\documentclass[a4paper,11pt]{article}	
\pdfoutput=1 
\usepackage{jheppub} 
                 \usepackage{tikz}    
\usepackage{autobreak}
\usepackage{enumerate}
\usepackage{ytableau}
\newcommand{\bitem}{\begin{itemize}}
\newcommand{\eitem}{\end{itemize}}
\newcommand{\be}{\begin{equation}}
\newcommand{\ee}{\end{equation}}
\newcommand{\ba}{\begin{aligned}}
\newcommand{\ea}{\end{aligned}}

\usepackage{enumerate}
\usepackage{ytableau}

\newcommand{\T}{\mathsf{T}}

\def\IZ{{\mathbb Z}}

\newcommand{\tc}{{\tilde c}}

\newcommand{\tilh}{{\tilde h}}
\newcommand{\tm}{{\tilde m}}

\newcommand{\nb}{\textcolor{red}}

\title{\boldmath Hecke Relations, Cosets and the Classification of 2d RCFTs}
\author{Zhihao Duan, Kimyeong Lee and Kaiwen Sun}
  
\affiliation{Korea Institute for Advanced Study,
85 Hoegiro, Dongdaemun-gu, Seoul 02455, Korea}

\preprint{KIAS-P22044}

\emailAdd{xduanz@kias.re.kr}
\emailAdd{klee@kias.re.kr}
\emailAdd{ksun@kias.re.kr}

\abstract{We systemically study the Hecke relations and the $c=8k$ coset relations among 2d rational conformal field theories (RCFTs) with up to seven characters. We propose that the characters of any 2d RCFT -- unitary or non-unitary -- satisfying a holomorphic modular linear differential equation (MLDE) can be realized as either a Hecke image or the coset of a Hecke image with respect to a $c=8k$ theory. Benefited from the recent results on holomorphic modular bootstrap, we check this proposal for all admissible theories with up to five characters. We also find many new interesting Hecke relations. For example, the characters of WZW models $(E_{6})_2,(E_7)_2,(E_{7\frac12})_2$ can be realized as the Hecke images $\mathsf{T}_{13},\mathsf{T}_{19},\mathsf{T}_{19}$ of Virasoro minimal models $M_{\rm sub}(7,6),M(5,4),M_{\rm eff}(13,2)$ respectively. Besides, we find the characters associated to the second largest Fisher group $Fi_{23}$ and the Harada-Norton group $HN$ can be realized as the Hecke images $\mathsf{T}_{23},\mathsf{T}_{19}$ of the product theories $M_{\rm eff}(5,2)\otimes M_{\rm eff}(7,2)$ and $M_{\rm eff}(7,2)^{\otimes 2}$ respectively. 
Mathematically, our study provides a great many interesting examples of vector-valued modular functions up to rank seven.}

\begin{document}

\setcounter{tocdepth}{2}
\maketitle
\flushbottom

\section{Introduction}
Two-dimensional conformal field theory is undoubtedly the basic working knowledge for modern theoretical physicists. It has wide applications in a variety of areas, ranging from describing the critical phenomenon in the second-order phase transition or boundary modes in the quantum Hall system in a lab, to the worldsheet theory of moving string in a higher dimensional spacetime. Moreover, it also has profound and beautiful geometric or algebraic structures, which have 
continuously sparkled fruitful interplay between mathematics and physics.

Two-dimensional rational conformal field theory (RCFT) is a special class that enjoys particularly nice properties. By definition, it means that there are only finitely many conformal primaries. Putting the theory on a torus, the partition function naturally decomposes in terms of finitely many characters labeled by those primaries. The large diffeomorphism group of torus, i.e., $\text{SL}(2,\mathbb{Z})$ naturally acts on the partition function and hence the characters. In this way, modularity plays an important role in the study of RCFTs. For example, combining with various other operations like fusion and braiding, it gives rise to an algebraic concept known as the modular tensor category (MTC). The study of MTC, especially its classification, is an active research topic in mathematics. See \cite{Rowell2007OnCO,Bruillard2015OnCO,Bruillard2019ClassificationOS, 2022arXiv220314829N} for a partial list.  

Despite several decades of extensive research, we are still witnessing new discoveries nowadays. Recently, Harvey and Wu proposed a striking relation among characters of RCFTs with different central charges via the so-called Hecke operator method \cite{Harvey:2018rdc}. Later, in \cite{Harvey:2019qzs} it was extended to an action among MTCs, which can be regarded as a “categorification" of the Galois action on the modular representations. A further exploration of this method is one of the main motivations for this paper.

Clearly, a natural playground to apply it is the RCFTs with a small number of characters. This links with another direction of research, namely, a complete list of classification for all of them. In the literature, Mathur, Mukhi and Sen first initiated a classification program for RCFTs based on two integral quantities: the number of characters and the index. In the original papers \cite{Mathur:1988na,Mathur:1988gt}, they completely classified all two-character cases with zero index. In \cite{Naculich,Hampapura:2015cea}, the authors further considered and even classified certain theories with positive index. Quasi-characters, which have integral but not necessarily positive $q$-expansion were considered in \cite{Chandra:2018pjq,Mukhi:2020gnj}. See \cite{Mukhi:2019xjy} for a review of the recent progress. What's more, they found an interesting duality phenomenon between theories with different indices, which they dubbed as the generalized coset construction. Recently, there is also an extension of their idea to fermionic RCFTs by demanding modular covariance only under suitable subgroups of $\text{SL}(2,\mathbb{Z})$ \cite{Bae:2020xzl,Bae:2021mej}.

Last year, motivated by results in the study of MTCs, \cite{Kaidi:2021ent} used the representation theory of finite groups to obtain a classification of consistent characters with number less than or equal to five.\footnote{Their classification actually only contains non-degenerate theories, which means that there does not exist any pair of conformal weights whose difference is an integer.} Greatly extending our previous knowledge, it is however not clear a priori if there is an underlying RCFT, perhaps of exotic type, for each of them. As a modest first step, we analyze systematically the possible relations among all the characters obtained in \cite{Kaidi:2021ent} and even beyond. The main tool we are using is the Hecke operator from \cite{Harvey:2018rdc}, but the modular linear differential equation (MLDE) also plays an important role. The highlights of our results are as follows:

\emph{1. We find a huge number of (generalized) Hecke relations among the characters of RCFT whose rank is not greater than seven.}

\emph{2. We find that each set of characters can be paired up to give a single and modular invariant character, which resembles a generalized coset relation w.r.t a putative single-character $c=8k$ theory. In particular, for $c = 24$ we recover 51 out of the 71 theories in the Schellekens' list \cite{Schellekens:1992db}, while we also find many new candidates which might describe certain non-conventional CFTs\,\footnote{One such example that we have in mind is the $(E_{7\frac{1}{2}})_1$ theory introduced in Section \ref{subsec:MLDE}.}. }

\emph{3. We propose that the characters of an arbitrary RCFT can be realized as either a Hecke image or the coset of a Hecke image w.r.t a putative $c=8k$ theory, starting from a handful of initial theories. As supporting evidence, for theories which satisfy holomorphic MLDEs with rank not greater than five, the initial ones are listed in Table \ref{table1-initial}.}

This paper is organized as follows. In Section \ref{sec:review}, we give an overview on some salient features of 2d RCFTs, including MLDEs, generalized cosets with respect to $c=8k$ and Hecke relations. A new ingredient we propose is the concept of generalized Hecke relations in Section \ref{sec:gHecke}, which describes Hecke image $\T_p$ for $p$ not coprime to the conductor $N$. 

In Section \ref{sec:cla}, we state our main results of this paper and elaborate them for the prototype -- RCFTs with two characters classified by Mathur-Mukhi-Sen (MMS) in the 1980s \cite{Mathur:1988na}. In Sections \ref{sec:3chi},\ref{sec:4chi},\ref{sec:5chi},\ref{sec:6chi},\ref{sec:7chi}, we discuss in detail the Hecke relations and $c=8k$ coset relations among 2d RCFTs with 3,4,5,6,7 characters which contain all the cases found in \cite{Kaidi:2021ent}. In particular we establish many new interesting Hecke and generalized Hecke relations for a large number of putative new 2d RCFTs. In Section \ref{sec:outlook}, we summarize our results and point out a possible generalization of our program to fermionic RCFTs.

Throughout the paper, we use $\T_p,\star$ to represent Hecke images with quasi-characters. We adopt the node order of affine Lie algebras as in SageMath \cite{sage}. Red color is used to highlight the weight-$3/2$ primaries which could be connected to emergent supersymmetry \cite{Bae:2021lvk,Kikuchi:2021qxz,Kikuchi:2022jbl}. 

\section{Review on 2d RCFTs}\label{sec:review}
\subsection{Basics of RCFTs}\label{sec:ex}
In this section, we give a brief overview of some basic facts of RCFTs that are important in this paper. We refer readers to \cite{Ginsparg:1988ui,DiFrancesco:1997nk} for more extensive reviews.

Consider the radial quantization in CFT with the Hilbert space of states $\mathcal{H}$ on the space slice $S^1$. It can be naturally decomposed into irreducible representations of the chiral algebra,
\be\label{eq:Hilbert}
\mathcal{H}_{S^1} = \bigoplus_{i,\bar\jmath} \mathcal{M}_{i, \bar\jmath} V_{h_i} \otimes V_{\bar{h}_{\bar\jmath} }\,,
\ee
where we label the irreducible representation by the conformal weight of lowest state $h_i, \bar{h}_{\bar\jmath}$, and $\mathcal{M}_{i, \bar\jmath}$ is the multiplicity. A nice way to encapsulate this piece of information is through the character, defined as
\be
\chi_i = \text{Tr}_{V_{h_i}} q^{L_0 - \frac{c}{24}},\quad  q = \exp(2\pi i\tau) \ \text{and}\ \tau = \tau_1 + i \tau_2 \in \mathbb{H}\,, 
\ee
with $c$ the central charge and $\mathbb{H}$ the upper half plane. Similarly one can also define its counterpart $\bar{\chi}_{\bar\jmath}$ as the trace over $V_{\bar{h}_{\bar\jmath}}$, such that the partition function $Z$ as a trace over $\mathcal{H}_{S^1}$ becomes
\be
Z(\tau,\bar{\tau}) = \text{Tr}_{\mathcal{H}_{S^1}}[\exp{(2\pi i \tau_1 P - 2\pi \tau_2 H)}] = \sum_{i,\bar\jmath} \mathcal{M}_{i, \bar\jmath} \chi_i\bar{\chi}_{\bar\jmath}\,,
\ee
where $P$ and $H$ are respectively the momentum and Hamiltonian of the system. On the other hand, $Z(\tau,\bar{\tau})$ has the interpretation as the partition function on the torus. In this paper, We will mainly focus on the holomorphic part and hence properties of $\chi_i$. 

RCFT, by definition, means that there are only finitely many summands in \eqref{eq:Hilbert}, assumed to be $d$ in total. This already entails that the central charge $c$ and all conformal weights $h_i$ are rational. In this paper, we label the set of characters as $\{\chi_i\}_{i = 0}^{d-1}$, and $\chi_0$ will denote the vacuum character with the following $q$-expansion,
\be\label{eq:vacuumchar}
\chi_0 = q^{-\frac{c}{24}} (1 + m_1 q + m_2 q^2 + \cdots)\,,
\ee
where the normalization is fixed by assuming a non-degenerate vacuum. In particular, $m_1$ counts the number of spin-1 currents. Meanwhile, for the non-vacuum primary with conformal weight $h_i$, we adopt the notation $(h_i)_{\mathcal{M}_i}$ to indicate its degeneracy $\mathcal{M}_i$. Crucially for us, the modular group $\text{SL}(2,\mathbb{Z})$ as a global diffeomorphism group on the torus, naturally acts on the whole set of $\chi_i$,
\be
\gamma  := \begin{pmatrix}
a&b\\
s&d
\end{pmatrix} \in \text{SL}(2,\mathbb{Z}),\quad  \chi_i\Big(\frac{a\tau + b}{c\tau +d}\Big) = \sum_{k = 0}^{d-1}\rho(\gamma)_{ik} \chi_{k}(\tau)\,,
\ee
where $\rho$ can be regarded as a $d$ dimensional representation of $\text{SL}(2,\mathbb{Z})$. $\rho$ also transforms $\bar{\chi}_{\bar\jmath}$ in such a way that $Z$ remains invariant. Mathematically speaking, all the $\chi_{i}$ form a vector-valued modular function. Since we know that $\text{SL}(2,\mathbb{Z})$ is generated by two elements,
\be
S =  \begin{pmatrix}
0&-1\\
1&0
\end{pmatrix}, \quad T =  \begin{pmatrix}
1&1\\
0&1
\end{pmatrix}\,,
\ee
their image under $\rho$ will play an important role. It is immediate that $\rho(T)$ is a diagonal and unitary matrix, but $\rho(S)$ in general is complicated.

In this part, we briefly review some well-known examples of 2d RCFTs for later convenience. In fact, we will mainly be concerned with two classes: minimal models and Wess-Zumino-Witten (WZW) models.

Minimal models are RCFTs with only Virasoro symmetry. They are classified by a pair of coprime positive integers $(p,q)$ with $p>q$, and we label them by $M(p,q)$. The central charge is given as
\be
c_{M(p,q)} = 1 - 6\frac{(p-q)^2}{pq}\,.
\ee
They are unitary with positive central charge if and only if $|p-q|=1$. Furthermore, we can label the conformal primaries by two integers $(r,s)$ which are constrained to lie inside the triangle 
\be
1\leq r < q, \quad 1\leq s < p, \quad pr > qs.
\ee
so there are altogether $(p-1)(q-1)/2$ fields. For reader's convenience, we give their conformal dimension
\be
h_{r,s} = \frac{(pr -qs)^2 - (p-q)^2}{4pq},
\ee
as well as the corresponding character
\be
\chi_{r,s}(\tau) = \frac{1}{\eta(\tau)} \sum_{n\in\mathbb{Z}}\Big(q^{\frac{(2pq\cdot n + pr-qs)^2}{4pq}} - q^{\frac{(2pq\cdot n + pr+qs)^2}{4pq}} \Big)\,.
\ee
Under the $S$-transformation, we have
\be\label{eq:Smatrix}
\chi_{r,s}(-\frac{1}{\tau}) = \sum \mathcal{S}_{(r,s),(\rho,\sigma)} \chi_{\rho,\sigma}(\tau), \quad \mathcal{S}_{(r,s),(\rho,\sigma)} = 2\sqrt{\frac{2}{pq}}(-1)^{1+s\rho+r\sigma}\sin(\pi\frac{p}{q}r\rho)\sin(\pi\frac{q}{p}s\sigma).
\ee
Last but not least, we remark that every primary is non-degenerate. Hence purely at the level of characters, we are free to choose any primary as the vacuum.\footnote{But surely only the correct vacuum gives rise to a well-defined fusion algebra.} In particular, even if $p - q > 1$, we can give an effective unitary description by declaring the character with the biggest exponent to be the vacuum character. This leads to the \emph{effective central charge} and \emph{effective conformal weights} as
\be\label{eq:effectivech}
c_\text{eff} = 1 - \frac{6}{pq}, \qquad h^{\text{eff}}_{r,s} = \frac{(pr -qs)^2 - 1}{4pq}\,.
\ee

On the other hand, WZW models are nonlinear sigma models whose fields $g$ are mappings from $S^2$ to a group manifold $G$ with Lie algebra $\mathfrak{g}$. It has conserved current $J^a(z)$, which when expanded in terms of modes,
$J^a(z) = \sum_{m \in \mathbb{Z}} J^a_m\, z^{m-1}\,,$
gives the commutation relation of affine Lie algebra,
\begin{equation}
[J^a_m, J^b_n] = i f^{ab}_c J^c_{m+n} + k m \delta^{a,b}\delta_{m+n,0}\,.
\end{equation}
Here $f^{ab}_c$ is the structure constant of $\mathfrak{g}$ and the integer $k$ is called the \textit{level}. From the Sugawara construction we have the central charge
\be\label{WZWc}
c = \frac{k\, \text{dim}\, \mathfrak{g}}{k + h^\vee}
\ee
with $h^{\vee}$ the dual Coxeter number.

In this paper, we will only consider WZW theories with positive level. To describe them, let us introduce some notations. First we name a given theory as $(G)_k$ where $G$ is the Lie group. Suppose the rank $r$ root system of its Lie algebra $\mathfrak{g}$ has a set of simple roots $\{\alpha_i\}_{i = 1}^{r}$ and the Killing form $\langle,\rangle$. We associate to each $\alpha_i$ a positive integer $a_{i}^{\vee}$ known as the comark. Moreover, a representation $\lambda$ can always be written as $\lambda = \sum \lambda_i \omega_i$, where $\{\omega_i\}_{i = 1}^{r}$ are fundamental weights and $\lambda_i$ are a set of non-negative integers.

In this way, the weight $\lambda$ which labels the primary  for a fixed level $k$ satisfies  an upper bound,
\begin{equation}
\lambda_0 := k - \sum\limits_{i = 1}^{r} \lambda_i\cdot a_{i}^{\vee} \geq 0\,,
\end{equation}
namely, we can associate to it an affine weight $\lambda_0 \omega_0 + \lambda$ (the definition of $\omega_0$ can be found in e.g., \cite{DiFrancesco:1997nk}). Its conformal dimension is given by the formula,
\begin{equation}\label{WZWweights}
h_{\lambda} = \frac{\langle\lambda, \lambda + \rho\rangle}{2( k + h^{\vee})}\,.
\end{equation}
Here $\rho$ is the Weyl vector, which is one half of the sum of all the positive roots.

For each conformal primary labeled by $\lambda$, the associated character $\chi_\lambda$ is given by the Weyl-Kac character formula, and we record the $S$-matrix of the characters below,
\be
\chi_{\lambda}(-\frac{1}{\tau}) = \sum S_{\lambda\mu}\chi_{\mu}(\tau),
\ee
with
\be
S_{\lambda\mu} = i^{|\Delta_+|} \left(\text{det}\,(\alpha^\vee_i, \alpha^\vee_j)\right)^{-\frac{1}{2}}(k+h^\vee)^{-\frac{r}{2}}\sum_{\omega \in W} \epsilon(\omega) \text{e}^{-2\pi i \frac{\langle\omega(\lambda + \rho), \mu+\rho\rangle}{k+h^\vee}}\,,
\ee
where $|\Delta_+|$ is the number of positive roots in $\mathfrak{g}$, $\{\alpha_i^\vee\}_{i = 1}^{r}$ are the coroots, $W$ is the Weyl group and $\epsilon(w)$ is the signature of an element $\omega \in W$ written as a product of transposes.

In later sections, we also need to take $G$ to be $U(1)$. For this somewhat degenerate situation, the WZW model actually describes a single compact boson and the level $k$ is replaced by $N$ with radius of the circle being $R = \sqrt{2N}$. It always has central charge one, and its conformal primaries are indexed by the finite group $\mathbb{Z}_{2N}$. For completeness we give their characters and $S$-matrix,
\begin{equation}
 \begin{aligned}
\chi_{k} (\tau) &= \frac{1}{\eta(\tau)} \sum\limits_{m \in \mathbb{Z}} q^{(k + m 2N)^2/4N}\,, \quad k \in \mathbb{Z}_{2N}\\
\chi_{k} (-\frac{1}{\tau}) &= \frac{1}{\sqrt{2N}}\sum\limits_{k^\prime \in \mathbb{Z}_{2N}} \text{e}^{-i \pi k k^\prime/N} \chi_{k^\prime}(\tau)\,,
 \end{aligned}
\end{equation}
where $\eta(\tau)$ is the Dedekind eta function.

Finally, we introduce the notion of level $k$ Lee-Yang model defined by
\be
(LY)_k=M_{\rm eff}(2k+3,2).
\ee
The nomenclature originates from the observation that the effective Lee-Yang model $(LY)_1$ can be seen as the first object in the Deligne exceptional series \cite{deligne}, which has dimension 1 and dual coxeter number $3/2$. Then by the central charge formula of WZW models \eqref{WZWc}, the level $k$ version should have central charge $c=k/(k+3/2)=\frac{2k}{2k+3}$. On the other hand, the effective central charge and effective conformal weights of $M(2k+3,2)$ minimal models follow from the general formula \eqref{eq:effectivech}
\begin{equation*}
    c_\mathrm{eff}=\frac{2k}{2k+3}, \qquad h^\mathrm{eff}_j=\frac{j(j+1)}{2(2k+3)}, \ j=0,1,2,\cdots, k \ .
\end{equation*}
The central charge indeed matches, which validates our proposal. The weights $h^\mathrm{eff}_j$ here actually can also be obtained from the weight formula \eqref{WZWweights} of WZW models, with $\lambda=j,\rho=1$ and bilinear form $\langle x,y \rangle=xy/2$. The characters of $(LY)_k$ theories can be nicely expressed as
\be
\ba\chi_j(\tau)  & = q^{-c_{\rm eff}/24+h^{\rm eff}_j} \prod_{n\neq 0, \pm (k-j+1) \mathrm{mod}(2k+3)} (1-q^n)^{-1} \\
& = q^{-c_{\rm eff}/24+h^{\rm eff}_j}\sum_{n_1,n_2\cdots n_k\ge 0}  \frac{q^{N_1^2+N_2^2+\cdots+N_k^2+N_{k-j+1}+\cdots+N_k}}{(q)_{n_1}  \cdots (q)_{n_k}},
\ea
\ee
where $N_i=n_i+\cdots+ n_k$ for $i\in\{1,2\cdots k\} $ and $(q)_n=(1-q)\cdots (1-q^n)$ with $(q)_0=1$. The expression in the second line is often called Nahm sums, while the equality between the first and the second line is famously known as the generalized Rogers-Ramanujan identities, or Andrews-Gordon identities, see e.g. \cite{Nahm:1992sx}. The $S$-matrix \eqref{eq:Smatrix} can be simplified as
\be S_{i,j} = \frac{2}{\sqrt{2k+3}} (-1)^{i+j+k} \sin \left( \frac{2\pi (k-i+1)(k-j+1)}{2k+3}  \right) .\ee


\subsection{Modular linear differential equations}\label{subsec:MLDE}

Based on the modularity or the existence of null states in the spectrum, it is well-known that the $d$ characters of an RCFT form solutions to an MLDE of the same order. Such differential equations were first proposed by Eguchi and Ooguri \cite{Eguchi:1987qd} in the context of CFT, and also by Kaneko and Zagier \cite{KZ} in mathematics. It is most convenient to present it based on the Wronskian method \cite{Mathur:1988na}, which we sketch here. 

First we define a covariant derivative called Serre derivative sending a modular form of weight $k$ to a new modular form of weight $k+2$ \cite{Don},
\be
D_k = \frac{1}{2\pi i}\frac{d}{d\tau} - \frac{k}{12}E_2(\tau)\,,
\ee
where $E_2(\tau)$ is the quasi-modular Eisenstein series of weight 2. Then we define the order $d$ modular derivative as
\be
{\cal D}^{d} := D_{2d-2} \circ D_{2d-4}\circ \cdots\circ D_0\,.
\ee
For an arbitrary function function $f(\tau)$ made of a linear combination of $d$ characters, the following determinant obviously vanishes,
\begin{align}
  {\rm det} \left( \begin{array}{cccc}
    \chi_0 & \cdots &\chi_{d-1} & f  \\
    {\cal D} \chi_0 & \cdots & {\cal D} \chi_{d-1} & {\cal D} f \\
    \vdots & & \vdots \\    {\cal D}^d \chi_0 & \cdots & {\cal D}^d \chi_{d-1} & {\cal D}^d f \\
\end{array}\right) = 0.
\end{align}
Expanding it according to the last column, we obtain the following equation,
\be\label{eq:MLDE}
  \left[ {\cal D}^d  + \sum_{k=0}^{d-1} \phi_k(\tau) {\cal D}^k \right] f(\tau) =0,
\ee
with
\be
\phi_k(\tau)= (-1)^{d-k} \frac{W_k}{W_d},\qquad  W_k = {\rm det} \left( \begin{array}{ccc}
    \chi_0 & \cdots &\chi_{d-1}  \\
    {\cal D} \chi_0 & \cdots & {\cal D} \chi_{d-1} \\
    \vdots & & \vdots \\
    {\cal D}^{k-1}\chi_0 & \cdots &{\cal D}^{k-1}\chi_{d-1}  \\
    {\cal D}^{k+1}\chi_0 & \cdots & {\cal D}^{k+1}\chi_{d-1} \\
    \vdots & & \vdots \\
    {\cal D}^d \chi_0 & \cdots & {\cal D}^d \chi_{d-1}  \\
\end{array}\right)\,.
\ee
Henceforth, we will call $W_d$ as the Wronskian. It turns out the equation \eqref{eq:MLDE} is precisely the MLDE that we are looking for. At the same time, due to possible zeros of the Wronskian, we also learn that the coefficients $\phi_k(\tau)$ can be meromorphic. In order to characterize the order of poles, one introduces a notion of index $l$ for $W_d$ as the sum of order of zeros multiplied by six in the fundamental domain. In other words, zero at the orbifold point $\tau = e^{2\pi i/3}$ contributes two, at the orbifold point $\tau = i$ contributes three while others give six. Furthermore, from the valence formula we are able to express $l$ as
\be\label{eq:index}
\frac{l}{6} = \frac{d(d-1)}{12} - \sum_{i = 0}^{d-1} \alpha_i,
\ee
where $\alpha_i = h_i - c/24$ is the leading exponent of each character. If $l = 0$, we call the corresponding MLDE holomorphic or monic. Clearly, in that situation all the coefficients $\phi_k(\tau)$ are polynomials in terms of Eisenstein series $E_4(\tau)$ and $E_6(\tau)$.

As mentioned in the introduction, Mathur, Mukhi and Sen first initiated a classification program for RCFTs based on the number of characters $d$ and the index $l$. Their strategy goes as follows: one first writes down the most general MLDE once given $d$ and $l$. Integrality of the characters requires all the solutions to have integral and non-negative $q$-expansion. For example, the solution corresponding to the vacuum character must take the form of \eqref{eq:vacuumchar}. Therefore, all the $m_i$ must be non-negative integers. Clearly, integrality imposes a very stringent constraint on the unknown parameters in the MLDE, and often renders the allowed choices of parameters finite.

In \cite{Mathur:1988na,Mathur:1988gt}, the authors completely classified all cases with $d = 2$ and $l = 0$, see also in mathematics literature \cite{Kaneko:2013uga}. There are only ten theories in total, and we provide some basic information in Table \ref{tab:MMS}.\footnote{If releasing the non-negativity constraint, a more general classification was given in \cite{Kiritsis:1988kq}.} Note from $A_1$ to $E_7$ and $E_8$ these are exactly the Deligne exceptional series \cite{deligne}. We remark that $(E_{7\frac12})_1$ is actually an intermediate vertex operator algebra (IVOA) \cite{Kawasetsu}, while $E_{7\frac12}$ is an intermediate Lie algebra filling a hole in the Deligne exceptional series between $E_7$ and $E_8$ \cite{LM}. The WZW $(E_8)_1$ only has a single-character but is included for completeness. Later, \cite{Naculich,Hampapura:2015cea} further considered theories with positive index, and they found an interesting duality phenomenon between theories in Table \ref{tab:MMS} and those with $l = 2$, which is the focus of the next section.

\begin{table}[ht]
\def\arraystretch{1.1}
\begin{center}
\begin{tabular}{c|cccccccccc}
 \hline
 & $(LY)_1$ & $(A_{1})_1$ & $(A_{2})_1$ & $(G_2)_1$ & $(D_4)_1$ & $(F_4)_1$  &  $(E_6)_1$ & $(E_7)_1$ & $(E_{7\frac12})_1$   & 
$(E_8)_1$ \\ \hline
$c$ &  $\frac{2}{5}$   &   1 &  2  &  
$\frac{14}{5}$ &  4&   $\frac{26}{5} $ &   6 & 
7 &   $\frac{38}{5} $&  8  \\
$h $ & $\frac15$ & $\frac14$ & $\frac13$ & $\frac25$ &  $\frac12$  & $\frac35$ & $\frac23$ & $\frac34$ & $\frac45$ & $-$ \\
$\mathcal{M}$ & 1 & 1  & 2 & 1 & 3 & 1 & 2 & 1 & 1 & $-$ \\
\hline
\end{tabular}
\caption{Central charge $c$ and non-vacuum conformal weight $h$ with its degeneracy $\mathcal M$ for Mathur-Mukhi-Sen series.}\label{tab:MMS}
\end{center}
\end{table} 

There have been many developments since the MMS classification of $d=2$ RCFTs. The classification of the $d=3$ cases was studied in \cite{kaneko3,Hampapura:2016mmz,Mukhi:2020gnj,Bae:2021mej,Das:2021uvd,Kaidi:2021ent}. In this work, We adopt the most recent results for $d=3,l=0$ in \cite{Kaidi:2021ent}. The classification of $d=4,5,l=0$ cases was also recently obtained in \cite{Kaidi:2021ent}, see also an earlier study on the $d=4$ case in \cite{ANS4}. Besides, 
the explicit values of $l$ for WZW models can be found in e.g. \cite{Das:2020wsi}. The MLDEs for Virasoro minimal models were discussed in \cite{Eguchi:1988wh}.

\subsection{Cosets with respect to $c=8k$}
In this section, we will explain the coset construction of RCFTs. It is by now well-known that many RCFTs admit coset presentation, the simplest which can be unitary minimal models in terms of diagonal coset of $SU(2)$ WZW models. In the classical textbook \cite{DiFrancesco:1997nk} it is restricted to taking coset within the class of WZW models, but here we actually need a more general setup. Therefore, we will adopt the terminology in \cite{Gaberdiel:2016zke} and call it as generalized coset.\footnote{Coset can also be understood from the more general procedure of gauging certain object in the fusion category \cite{Bhardwaj:2017xup}. We thank Ying-Hsuan Lin for discussion on this point.} We will also closely follow their treatment in the first part.

Suppose we are given an RCFT $\mathcal{G}$. The coefficient of $q$ in the expansion of the vacuum character contains the dimension of global symmetries. Assuming the affine currents can generate a subtheory $\mathcal{H}$, then the generalized coset,
\be\label{eq:coset}
\mathcal{C} = \mathcal{G}/\mathcal{H},
\ee
is formally defined as the subtheory in $\mathcal{G}$ whose generators of chiral algebra have trivial OPE with all those in $\mathcal{H}$. One can show that $\mathcal{C}$ is indeed well-defined, for instance it contains a genuine Virasoro algebra with central charge $c_{\mathcal{C}} = c_\mathcal{G} - c_\mathcal{H}$. This can be argued as follows. The stress-energy tensor of $\mathcal{H}$ that is given by the Sugawara construction, should still have the standard OPE with the affine currents $J^a (z)$. In other words, denoting their stress-energy tensor as $T^\mathcal{G}$ and $T^\mathcal{H}$ respectively, we have
\be
T^\mathcal{G}(z) J^a (w) \sim \frac{J^a(w)}{(z-w)^2} + \frac{\partial J^a(w)}{(z-w)}, \quad T^\mathcal{H}(z) J^a (w) \sim \frac{J^a(w)}{(z-w)^2} + \frac{\partial J^a(w)}{(z-w)}\,.
\ee
Their difference $T^\mathcal{C} := T^\mathcal{G} - T^\mathcal{H}$ hence has trivial OPE with all the generators of $\mathcal{H}$. By computing the OPE of $T^\mathcal{C}$ with itself, it is easy to see that the generalized coset $\mathcal{C}$ indeed has the desired central charge.

In fact, one can generalize the above picture even further, and take $\mathcal{H}$ to be a subtheory with a closed subchiral algebra generated by spin one or spin two currents. With a similar argument, one shows that the generalized coset is still a well-defined CFT. This kind of construction already appeared, say, in the decomposition of Monster CFT into theories with other sporadic group symmetries \cite{Bae:2018qfh,Bae:2020pvv}.

From the point of view of physical states, \eqref{eq:coset}
means that we decompose the Hilbert space of states in $\mathcal{G}$ into irreducible representations of commuting subchiral algebra for $\mathcal{C} \otimes \mathcal{H}$. At the level of characters this yields what is known as the \emph{bilinear relation},
\be\label{eq:bilinear}
\chi_0^{\mathcal{G}}(\tau) = \chi_0^{\mathcal{H}}(\tau)\cdot \chi_0^{\mathcal{C}}(\tau) + \sum_{i = 1}^{d-1} \mathcal{M}_i\, \chi_i^{\mathcal{H}}(\tau)\cdot \chi_i^{\mathcal{C}}(\tau)\,,
\ee
where $\chi_0^{\mathcal{H}} (\chi_0^{\mathcal{C}})$ is the vacuum character of $\mathcal{H} (\mathcal{C})$, while $\chi_i^{\mathcal{H}}  (\chi_0^{\mathcal{C}})$  for $1 \leq i \leq d-1$ are characters for the other primaries in $\mathcal{H}(\mathcal{C})$ with degeneracy $\mathcal{M}_i$.

In later sections, we shall discuss a lot of theories with various Wronskian indices $l$ that can be paired up to give a putative single-character theory with $c = 8k$. Some immediate constraints can already be deduced from the characters. Suppose theory $\mathcal{H} (\mathcal{C})$ has Wronskian index $l^{\mathcal{H}} (l^{\mathcal{C}})$ and conformal weights $h^\mathcal{H}_i (h^\mathcal{C}_i)$ for non-vacuum primaries labeled by $1\leq i \leq d-1$, then the bilinear relation \eqref{eq:bilinear} demands in particular
\be
h^\mathcal{H}_i + h^\mathcal{C}_i = n_i, \quad \text{with}\ n_i \in \mathbb{N} 
\ee
after a possible reordering of the indices. What's more, since the Wronskian index of each theory is constrained by the exponents \eqref{eq:index}, taking the sum of two equations and expressing $\alpha_i$ in terms of central charge and weights yield
\be\label{llrelation}
l^{\mathcal{H}}+l^{\mathcal{C}} =d^2+(2k-1)d-6\sum_{i=1}^{d-1}n_i\,.
\ee

Let us give some simple examples with $d=2$, $k=1$ in Table \ref{table1-2chi}, which are well-known decompositions of WZW $(E_8)_1$ theory into pairs of two-character theories. All theories here are exactly in the Mathur-Mukhi-Sen series, i.e., $l=l'=0$. Notice for each pair, $n_1=h_1+\tilh_1=1$, which satisfies the general relation \eqref{llrelation}.
\begin{table}[ht]
\def\arraystretch{1.1}
\centering
\begin{tabular}{|c|c|c|c|c|c|c|c|c|}
\hline
  \multicolumn{4}{|c|}{$l=0$} & \multicolumn{4}{c|}{$l'=0$} \\
\hline
 $c$ & $h$ & $m_1$ & remark & $\tc$ & $\tilh$ & $\tm_1$ &  remark \\
\hline
 $\frac25$ & $\frac15$ & 1 &  $(LY)_1$ & $\frac{38}{5}$ & $\frac45$ & 190 & $(E_{7\frac12})_1$ \\
\hline
 1 & $\frac14$ & 3 &  $(A_{1})_1$ &  7 & $\frac34$ & 133 & $(E_7)_1$\\
\hline
  2 & $\frac13$ & 8 & $(A_{2})_1$ & 6 & $\frac23$ & 78 & $(E_6)_1$ \\ 
\hline
  $\frac{14}{5}$ & $\frac25$ & 14 & $(G_2)_1$ & $\frac{26}{5}$ & $\frac35$ & 52 & $(F_4)_1$ \\
\hline
 4 & $\frac12$ & 28 & $(D_4)_1$ & 4 & $\frac12$ & 28 & $(D_4)_1$\\
\hline
\end{tabular}
\caption{Cosets of $c=8$ WZW $(E_8)_1$ theory with two characters.}
\label{table1-2chi}
\end{table}

More interesting examples are the $c=24$ pairs, i.e., $k=3$. In the case of RCFTs with two characters, i.e., $d=2$, such pairs have been studied in (\cite{Gaberdiel:2016zke}, Table 1), where each WZW theory from $A_1$ to $E_7$ in Mathur-Mukhi-Sen series has a dual $l'=2$ theory w.r.t $c=24$. These dual theories have $16<c<24$ and conformal weights $1<h_1<2$. In each pair, the conformal weights satisfy $h_1+\tilh_1=2$. Besides, each pair corresponds to some $c=24$ theories in the Schellekens' list \cite{Schellekens:1992db}. We can also extend it to the non-unitary cases $(LY)_1$ and $(E_{7\frac12})_1$. All together, we collect nine $c=24$ pairs in Table \ref{table2-2chi}.

\begin{table}[ht]
\def\arraystretch{1.1}
\centering
\begin{tabular}{|c|c|c|c|c|c|c|c|c|c|}
\hline
  \multicolumn{4}{|c|}{$l=0$} & \multicolumn{4}{c|}{$l'=2$} & \multicolumn{2}{c|}{duality}\\
\hline
 $c$ & $h$ & $m_1$ & remark & $\tc$ & $\tilh$ & $\tm_1$ & remark & $\!m_1+\tm_1\!$ & \!Schellekens No.\!\\
\hline
 $\frac25$ & $\frac15$ & 1 &  $(LY)_1$ & $\frac{118}{5}$ & $\frac95$ & 59 & $\T_{59}(LY)_1$ & 60 & $-$ \\
\hline
 1 & $\frac14$ & 3 &  $(A_{1})_1$ & 23 & $\frac74$ & 69& $\T_{23}(A_1)_1$ & 72 & $15-21$ \\
\hline
  2 & $\frac13$ & 8 & $(A_{2})_1$ & 22 & $\frac53$ & 88 & $\T_{11}(A_2)_1$ & 96  & $24, 26-28$ \\ 
\hline
  $\frac{14}{5}$ & $\frac25$ & 14 & $(G_2)_1,\T_{7}(LY)_1$ & $\frac{106}{5}$ & $\frac85$ & 106 & $\T_{53}(LY)_1$  & 120& $32,34$ \\
\hline
 4 & $\frac12$ & 28 & $(D_4)_1$ & 20 & $\frac32$ & 140 & $\T_5(D_4)_1$ & 168 & $42,43$\\
\hline
 $\frac{26}{5}$ & $\frac35$ & 52 & $(F_4)_1,\T_{13}(LY)_1$ & $\frac{94}{5}$ & $\frac75$ &  188 & $\T_{47}(LY)_1$ & 240 & $52,53$\\
\hline
 6 & $\frac23$ & 78 & $(E_6)_1$ & 18 & $\frac43$ & 234 & $-$ & 312 & $58,59$ \\
\hline
 7 & $\frac34$ & 133 & $(E_7)_1,\T_{7}(A_1)_1$ & 17 &  $\frac54$ & 323 & $\T_{17}(A_1)_1$  & 456 & $64,65$\\
\hline
 $\frac{38}{5}$ & $\frac45$ & 190 & $(E_{7\frac12})_1,\T_{19}(LY)_1$ & $\frac{82}{5}$ &  $\frac65$ & 410 & $\T_{41}(LY)_1$  & 600 & $-$\\
\hline
\end{tabular}
\caption{Cosets of $c=24$ theories with two characters. Hecke operation $\T_p$ will be defined in Section \ref{sec:Hecke}.}
\label{table2-2chi}
\end{table}

We can even form two-character RCFTs together as $c=40$ pairs, i.e., $k=5$. In such cases, the relation \eqref{llrelation} allows $n_1=h_1+\tilh_1=3$ and $l+l'=4$. There are two possibilities. We collect the $(l,l')=(2,2)$ pairs in Table \ref{table24-2chi} and $(l,l')=(0,4)$ pairs in Table \ref{table25-2chi}.\footnote{Although in Table \ref{table24-2chi} and \ref{table25-2chi} all $\mathcal{N}$'s are negative, the single character $(J(\tau)+\mathcal{N})j(\tau)^{2/3}$ always gives positive and integral $q$-expansion.}

\begin{table}[ht]
\def\arraystretch{1.1}
\centering
\begin{tabular}{|c|c|c|c|c|c|c|c|c|}
\hline
  \multicolumn{3}{|c|}{$l=2$} & \multicolumn{3}{c|}{$l'=2$} & {duality}\\
\hline
 $c$ & $h$ & $m_1$  & $\tc$ & $\tilh$ & $\tm_1$ &  $\mathcal{N}$ \\
\hline
 $\frac{82}{5}$ &  $\frac65$ & 410   & $\frac{118}{5}$ & $\frac95$ & 59 & $-27$ \\
\hline
 17 &  $\frac54$ & 323    & 23 & $\frac74$ & 69 & $-104$ \\
\hline
 18 & $\frac43$ & 234   & 22 & $\frac53$ & 88 & $-174$ \\ 
\hline
  $\frac{94}{5}$ & $\frac75$ &  188    & $\frac{106}{5}$ & $\frac85$ & 106  & $-202$\\
\hline
20 & $\frac32$ & 140  & 20 & $\frac32$ & 140 & $-216$\\
\hline
\end{tabular}
\caption{The $(l,l')=(2,2)$ coset relations of $c=40$ theories with two characters. The bilinear relation gives $(J(\tau)+\mathcal{N})j(\tau)^{2/3}$.}
\label{table24-2chi}
\end{table}

\begin{table}[ht]
\def\arraystretch{1.1}
\centering
\begin{tabular}{|c|c|c|c|c|c|c|c|c|}
\hline
  \multicolumn{4}{|c|}{$l=0$} & \multicolumn{3}{c|}{$l'=4$} & {duality}\\
\hline
 $c$ & $h$ & $m_1$ & remark & $\tc$ & $\tilh$ & $\tm_1$ &  $\mathcal{N}$ \\
\hline
 $\frac{38}{5}$ &  $\frac45$ & 190 & $(E_{7\frac12})_1$ & $\frac{162}{5}$ & $\frac{11}{5}$ & 4   & $-302$ \\
\hline
7 & $\frac34$ & 133 & $(E_7)_1$ & 33 & $\frac{9}{4}$ & 3    & $-360$\\
\hline
6 & $\frac23$ & 78 & $(E_6)_1$ & 34 & $\frac{7}{3}$ & 1    & $-417$\\
\hline
\end{tabular}
\caption{The $(l,l')=(0,4)$ coset relations of $c=40$ theories with two characters. The bilinear relation gives $(J(\tau)+\mathcal{N})j(\tau)^{2/3}$. Three $l'=4$ RCFTs of non-product type were conjectural to exist in Table 3 of \cite{Chandra:2018pjq} as IVOA.}
\label{table25-2chi}
\end{table}

\subsection{Hecke relations}\label{sec:Hecke}
Let us describe here an important method that will be used extensively in the core of this article to generate new RCFT characters, especially the candidate characters of a putative dual RCFT to the one character theory. We refer the reader to \cite{Harvey:2018rdc} for more details about this method.

In previous sections, we learn that the $n$ distinct characters of irreducible modules of an RCFT transform in an $n$-dimensional complex representation $\rho$ of the full modular group, hence forming a vector-valued modular function. In addition to that, according to the integrality conjecture which is recently proved by \cite{calegari2021unbounded}, there exists a smallest positive integer $N$ so that every character is already modular invariant under the principal congruence subgroup $\Gamma(N)$ of $\text{SL}(2,\mathbb{Z})$. Such a subgroup is defined as follows,
\be
\Gamma(N) := \Bigg\{\begin{pmatrix}
a&b\\
c&d
\end{pmatrix} \in \text{SL}(2,\mathbb{Z}),\ \  \begin{pmatrix}
a&b\\
c&d
\end{pmatrix} \equiv \begin{pmatrix}
1&0\\
0&1
\end{pmatrix} \text{mod}\ N\Bigg\}.
\ee
In \cite{Bantay:2001ni}, it is proved that for RCFT characters the integer $N$ coincides with the order of $\rho(T)$. Adopting the terminology in \cite{Harvey:2018rdc}, we henceforth refer it as the \emph{conductor} of the theory.

Now it is time to define the Hecke operator. In fact, the case for the full modular group and congruence subgroups can be found in the textbooks on number theory, e.g., \cite{Murty}. However, the novel point is that since the characters form a vector-valued modular function, it had better directly involve the representation $\rho$ instead of the Dirichlet characters. This kind of Hecke operator was first defined in \cite{Harvey:2018rdc} as follows. Given a prime number $p$ not dividing the conductor $N$, let us denote by $\bar p$ its multiplicative inverse in $\mathbb Z/N\mathbb Z$, and by $\sigma_p$ the pre-image of the diagonal matrix $\mathrm{diag}(\bar{p},p)$ under the mod $N$ map.\footnote{For example $\sigma_p$ can be chosen to be $T^{\bar p} S^{-1}T^p S T^{\bar p} S$.} Then one defines the action of the Hecke operator $\T_p$ on the $\alpha$-th component of the vector-valued modular function,
\be\label{Tp}
(\mathsf{T}_p \chi)_\alpha(\tau) =  \sum_\beta \rho_{\alpha \beta}(\sigma_p) \chi_\beta(p \tau) +\sum_{i=0}^{p-1} \chi_\alpha\left(\frac{\tau+iN}{p}\right) .
\ee
This leads to a formula in terms of the coefficients of the Fourier expansion
\be
\chi_\alpha(\tau)=\sum_n b_\alpha(n)\,q^\frac{n}{N},
\ee
as
\be
(\T_p\chi_\alpha)(\tau)=\sum_{n} b^{(p)}_\alpha(n)\,q^\frac{n}{N}\,,
\ee
with
	\begin{align}
		\begin{split}
			b^{(p)}_{\alpha}(n) =
			\left\{
			\begin{array}{ll}
			p b_{\alpha}(p n), &\ p \nmid n , \\
			p b_{\alpha}(p n) + \sum_{\beta} \rho_{\alpha}(\sigma_p) b_\beta
			\big( \frac{n}{p}\big), &\ p \ | \ n.
			\end{array}
			\right.
		\end{split}
	\end{align}

The above formula clearly indicates that the Fourier coefficients of each component of Hecke image $\T_p$ satisfy certain \emph{mod $p$ property}, i.e., for any Fourier coefficient not divisible by $p$, the $p-1$ Fourier coefficients behind or after it must be divisible by $p$.
Furthermore, if $p$ is no longer prime but still coprime to $N$, we can make use of the following formulas to define $\T_p$,
\begin{align}
		\begin{split}
			\left\{
			\begin{array}{ll}
			\T_{rs}=\T_r\circ \T_s,&\ (r,s)=1,  \\
			\T_{p^{n+1}}= \T_{p}\circ\T_{p^{n}}-p\sigma_p\circ\T_{p^{n-1}},&\ p\text{ prime.}
			\end{array}
			\right.
		\end{split}
	\end{align}
At the level of the action on the representation $\rho$, one can show that the new representation $\rho^{(p)}$ is given by
\begin{equation}
		\rho^{(p)}(T)=\rho(T^{\bar p})\,,\ \ \ \ \ \ \rho^{(p)}(S)=\rho(\sigma_p S)\,,
	\end{equation}
in terms of the generators of $\text{SL}(2,\mathbb{Z})$.

An easy consequence of the Hecke operator action is that is $p$ times the (effective) central charge and conformal weights of $\chi_\alpha$ should respectively give the central charge and conformal weights derived from $(\T_p\chi_\alpha)$, for the latter only up to integers. We may refer to it as the \emph{multiple $p$ requirement}.

Meanwhile, experimentally we observe that for all the cases considered, the characters $(\T_p\chi_\alpha)$ can always be decomposed into degree $p$ homogeneous polynomials with variables being $\chi_\alpha$. Despite a lack of proof, we believe it to hold in general and dub it the \emph{homogeneous property}.

Next we explain why Hecke operators are particularly suited for finding characters of dual RCFT. In \cite{Harvey:2018rdc}, it is shown that for given two Hecke images $\T_{p_1,p_2}\chi(\tau)$, the bilinear combination with the matrix $G_l=\rho(T^l S^{-1}T^{\bar l}S T^l S)$
\be\label{eq:heckepairing}
T_{p_1}\chi(\tau)^T \cdot G_l \cdot \T_{p_2}\chi(\tau)
\ee
is modular invariant if
\be
\bar p_2 +\bar p_1 l^2 \equiv 0\ \textrm{mod}\ N\,.
\ee
In other words, they combine to give the character of a putative single-character CFT of central charge $c=24k$.

For example, if one chooses $p_1 = 1, \ p_2 = N-1$, then setting $\bar p_1 = 1, \ \bar p_2 = -N-1$ we can simply take $l = 1$ to satisfy the above equation. If furthermore the charge conjugate matrix $C$ is the identity, then $G_{l=1}$ is also the identity matrix such that \eqref{eq:heckepairing} is reduced to the familiar diagonal pairing. The single-character for a $c=24k+8$ or $24k+16$ theory is not strictly speaking modular invariant but up to a phase $e^{\pm2\pi i/3}$. Two Hecke images $\T_{p_1,p_2}$ can also combine into a $c=24k+8$ or $24k+16$ theory. We will see many such examples in the main context.

Now we give some examples of Hecke relations which appear in \cite{Harvey:2018rdc} for RCFTs with two characters. First, consider the effective Lee-Yang model $(LY)_1$ which has central charge $c=2/5$, conformal weights $1/5$ and conductor $N=60$. The two characters are just the well-known Rogers–Ramanujan functions
\be
\phi_1=q^{-\frac{1}{60}}\prod_{n=0}^{\infty}\frac{1}{(1-q^{5n+1})(1-q^{5n+4})},\qquad
\phi_2=q^{\frac{11}{60}}\prod_{n=0}^{\infty}\frac{1}{(1-q^{5n+2})(1-q^{5n+3})},
\ee
with transformation of vector-valued modular form as
\be\label{YL}
\rho^{(YL)_1}(S) = \frac{2}{\sqrt{5}} 
\begin{pmatrix}  ~\sin(\frac{2 \pi}{5}) & ~\sin(\frac{\pi}{5}) \\  \sin(\frac{\pi}{5}) & -\sin(\frac{2 \pi}{5})  \end{pmatrix}, \qquad \rho^{(YL)_1}(T) = {\rm diag}( \xi_{60}^{-1}, \xi_{60}^{11}).
\ee
All Hecke images $\T_p$ of $(LY)_1$ with ordinary characters for $p<60$ are collected in Table \ref{table2-2chi}. For example, $\T_7(LY)_1=(G_2)_1$, $\T_{13}(LY)_1=(F_4)_1$ and $\T_{19}(LY)_1=(E_{7\frac12})_1$. Let us look at the $c=118/5$ theory in Table \ref{table2-2chi}, which can be realized as the $\T_{59}$ image of $(LY)_1$. We find the vacuum characters of this theory can be exactly written as 
\be\nonumber
\ba
\chi_0^{\T_{59}}&=\phi_{1}^{59}+162545 \phi_{1}^{49} \phi_{2}^{10}+8777430 \phi_{1}^{44} \phi_{2}^{15}+57609370 \phi_{1}^{39} \phi_{2}^{20}+48470919 \phi_{1}^{34} \phi_{2}^{25}\\
+\,&29482300 \phi_{1}^{29} \phi_{2}^{30}-34622085 \phi_{1}^{24} \phi_{2}^{35}+28804685 \phi_{1}^{19} \phi_{2}^{40}-2925810 \phi_{1}^{14} \phi_{2}^{45}+32509 \phi_{1}^9 \phi_{2}^{50},
\ea
\ee
while the non-vacuum character $\chi_{9/5}$ can be obtained from $\chi_0$ by replacing $\phi_1\to \phi_2,\phi_2\to -\phi_1$. Together with $(LY)_1$, they form a $c=24$ pair with bilinear relation of characters as
\be\label{118id}
\ba
J(\tau)+60=&\, \chi_0^{\T_{59}}\chi_0^{(LY)_1}+\chi_{9/5}^{\T_{59}}\chi_{1/5}^{(LY)_1}\\
=&\,\phi_{1}^{60}+195054 \phi_{1}^{50} \phi_{2}^{10}+11703240 \phi_{1}^{45} \phi_{2}^{15}+86414055 \phi_{1}^{40} \phi_{2}^{20}\\
&+83093004 \phi_{1}^{35} \phi_{2}^{25}+58964600 \phi_{1}^{30} \phi_{2}^{30}-83093004 \phi_{1}^{25} \phi_{2}^{35}\\
&+86414055 \phi_{1}^{20} \phi_{2}^{40}-11703240 \phi_{1}^{15} \phi_{2}^{45}+195054 \phi_{1}^{10} \phi_{2}^{50}+\phi_{2}^{60} \\
=&\,q^{-1}+60+196884 q+21493760 q^2+864299970 q^3+\dots.
\ea
\ee
On the other hand, consider the $c=82/5$ theory in Table \ref{table2-2chi}, which can be realized as the $\T_{41}$ image of $(LY)_1$. We find the vacuum characters of this theory can be exactly written as
\be
\ba
\chi_0^{\T_{41}}=&\,
\phi_{1}^{41}+369 \phi_{1}^{36} \phi_{2}^5+50594 \phi_{1}^{31} \phi_{2}^{10}+261580 \phi_{1}^{26} \phi_{2}^{15}+136735 \phi_{1}^{21} \phi_{2}^{20}\\
&-151003 \phi_{1}^{16} \phi_{2}^{25}+54858 \phi_{1}^{11} \phi_{2}^{30}-902 \phi_{1}^6 \phi_{2}^{35},
\ea
\ee
while the non-vacuum character $\chi_{6/5}$ again can be obtained from $\chi_0$ by replacing $\phi_1\to \phi_2,\phi_2\to -\phi_1$. Together with $(E_{7\frac12})_1$, they form a $c=24$ pair with bilinear relation of characters as
\be\label{82id}
\ba
J(\tau)+600=&\, \chi_0^{\T_{41}}\chi_0^{(E_{7+1/2})_1}+\chi_{6/5}^{\T_{41}}\chi_{4/5}^{(E_{7+1/2})_1}\\
=&\,\phi_{1}^{60}+540 \phi_{1}^{55} \phi_{2}^5+165354 \phi_{1}^{50} \phi_{2}^{10}+12353940 \phi_{1}^{45} \phi_{2}^{15}+79345455 \phi_{1}^{40} \phi_{2}^{20}\\
&+120668904 \phi_{1}^{35} \phi_{2}^{25}-13806340 \phi_{1}^{30} \phi_{2}^{30}-120668904 \phi_{1}^{25} \phi_{2}^{35}\\
&+79345455 \phi_{1}^{20} \phi_{2}^{40}-12353940 \phi_{1}^{15} \phi_{2}^{45}+165354 \phi_{1}^{10} \phi_{2}^{50}-540 \phi_{1}^5 \phi_{2}^{55}+\phi_{2}^{60}
 \\
=&\,q^{-1}+600+196884 q+21493760 q^2+864299970 q^3+\dots.
\ea
\ee 
The difference of the two identities \eqref{118id} and \eqref{82id} yields a simpler identity
\be
\phi_{1} \phi_{2} (\phi_{1}^{10}-11 \phi_{1}^5 \phi_{2}^5-\phi_{2}^{10})=1.
\ee
This famous identity was proposed by Ramanujan and later proved by Darling, Rogers and Mordell independently, see for example \cite{Bruce} for the history.

We also collect the Hecke relations involving $(A_1)_1,(A_2)_1,(D_4)_1$ in Table \ref{table2-2chi}. These relations were revealed in \cite{Harvey:2018rdc} and will be useful to us when discussing the Hecke images of product theory later. Lots of new Hecke relations involving RCFTs with $3,4,5,6,7$ characters will be given in the main context of this paper. Owing to the finite-generation property of $\Gamma(N)$ modular forms, it is always sufficient to verify a Hecke relation by checking the Fourier coefficients of the two sides to certain $q$ orders, called the Sturm bound \cite{sturm}.

\subsection{Generalized Hecke relations}\label{sec:gHecke}
In practice, we often encounter situations where two RCFTs seemingly have an analogous $\T_p$ relation but the suitable $p$ and $N$ are not coprime to each other. The simplest example perhaps is Ising model with $c=1/2,h_i=\frac{1}{16},\frac{1}{2}$ and WZW $(A_1)_2$ with $c=3/2,h_i=\frac{3}{16},\frac{1}{2}$ which appear to have a $\T_3$ relation but 3 is not coprime to the conductor $48$ of Ising model. At present, we are not able to find a similar formula as \eqref{Tp}, while its naive generalization does not make sense. Nevertheless, we propose that there should exist a \emph{generalized Hecke relation} $\T_p$ characterized by the following \emph{three} conditions when $p$ is not coprime to the conductor $N$: 
\begin{enumerate}
   \item The central charge and conformal weights satisfy the multiple $p$ requirement.
\item The degeneracy for each non-vacuum primary is inherited.
\item The homogeneous property holds.
\end{enumerate}
We regard these three conditions as the defining properties of generalized Hecke relation. 
By computing a large quantity of examples, we observe the following additional properties when a theory with conductor $N$ is mapped to its generalized Hecke image $\T_p$:
\begin{enumerate}
   \item Conductor $N$ becomes $N/p$.
\item Fourier coefficients of $\T_p$ satisfy the mod $p$ properties, just like ordinary Hecke image.
\item The classes of ordinary Hecke operation of $\T_p$ resemble those of the original theory. 
\end{enumerate}
We make some clarifications for the third point. A generalized Hecke image $\T_p$ can of course have its own ordinary Hecke images $\T_k$ with $k$ coprime to $N/p$. In these cases, we simply denote them as the generalized $\T_{kp}$ image of the original theory. This may brings in some uniqueness issues, but for all examples we encounter in the current paper this notion is indeed valid. The resemblance between the Hecke classes we declare in the third point will be shown for every example in the main context. Here let us just point out that the number of different $\rho(\sigma_k)$ matrices for the $\T_p$ image is always the same with the number of different $\rho(\sigma_p)$ matrices of the original theory. This is not obvious as the two cases have different conductors, different $T$ transformations, and in general different $S$ transformations.
As mentioned above, it would be desirable to give a concrete formula in terms of the coefficients in the $q$-expansion as \eqref{Tp}. For now, we can only establish generalized Hecke relations between existing theories. 
We give a summary of generalized Hecke relation appearing in the current work in Table \ref{tb:gHecke}. We will discuss these examples one by one in the main context, examine the defining conditions and check the above properties.
It is interesting to notice only generalized $\T_3,\T_2$ appearing in Table \ref{tb:gHecke}. We suspect these are the only two possibilities for generalized Hecke which may have something to do with the $24$ in vacuum $\alpha_0=-c/24$ only has factors 2 and 3.

\begin{table}[ht]
\def\arraystretch{1.1}
	\centering
	\begin{tabular}{|c|c|c|c|c|c|c|c|c|c|c|c|c|c|}
		\hline
Theory A	&	 $c$ & $h_i$ & $N$ & generalized $\T_p$\!\! & Theory B & $c'$ & $h_i'$ &  $N'$ \\
		\hline
  Ising  &    $\frac{1}{2}$  &  $ \frac{1}{16},\frac12 $ & $ 48$ & $\T_{3}$ & $ (A_1)_2$ &  $\frac32 $  &  $\frac{3}{16},\frac12  $ & $ 16$ \\        
		\hline
		$(LY)_2$ &    $\frac{4}{7}$  &  ${\frac{1}{7},\frac{3}{7}} $ & $ 42$ & $\T_{3}$ & $M_{12/7}$ &  $\frac{12}{7}$  &  ${\frac{2}{7},\frac{3}{7}} $ & $14$ \\
     \hline
$(LY)_1^{\otimes 2}$	   &    $\frac{4}{5}$  &  $ (\frac15)_2,\frac25 $ & $30 $ & $\T_{3}$ & $ (A_1)_{D_6} $ &  $\frac{12}{5} $  &  $ \frac15,(\frac35)_2 $ & $ 10$ \\        
		\hline
	$U(1)_2$	   &    $1$  &  $ (\frac{1}{8})_2,\frac{1}{2} $ & $24 $ & $\T_{3}$ & $(A_3)_1 $ &  $3 $  &  $ (\frac{3}{8})_2,\frac{1}{2} $ & $8 $ \\        
		\hline
		$(A_1)_1^{\otimes 2}$	   &    $2$  &  $(\frac14)_2,\frac12  $ & $12 $ & $\T_{3}$ & $(D_6)_1 $ &  $6 $  &  $ (\frac34)_2,\frac12 $ & $4 $ \\        	\hline
		   $M_{4/3}$ & $\frac43 $ &  $(\frac{2}{9})_3,\frac13,\frac23 $   & $18 $ & $\T_{2}$ & $M_{8/3} $ &  $ \frac83$  &  $ \frac13,(\frac49)_3,\frac23 $ & $ 9$ \\        
		\hline
 $(G_2)_1/(LY)_1$  &    $\frac{12}{5}$  &  ${\frac15,\frac25,\frac45} $ & $ 10$ & $\T_{2}$ & $(F_4)_1/(LY)_1$ &  $\frac{24}{5}$  &  ${\frac25,\frac35,\frac45} $ & $ 5$ \\        
		\hline

		\end{tabular}
			\caption{Examples of generalized Hecke relation $\T_p$ between theory A with central charge $c$, non-vacuum weights $h_i$, conductor $N$ and theory B with central charge $c'$, non-vacuum weights $h_i'$, conductor $N'$. Here $ (A_1)_{D_6} $ is the $D_6$ type modular invariant of $\hat{A}_1$. The $M_{12/7}$ theory will be discussed in Section \ref{sec:LY2}. The $M_{4/3}$ and $M_{8/3}$ theories will be discussed in Section \ref{sec:LY3}.}
			\label{tb:gHecke}
		\end{table}

We remark that the three defining conditions of generalized Hecke together are actually quite restricted. There exist situations where the first and third conditions are satisfied while the second is not. For example, the following two theories do \emph{not} satisfy a generalized $\T_2$ relation: 
\be
\ba
&(LY)_1^{\otimes 3},\ c=\frac{6}{5},\ h_i=0,\Big(\frac{1}{5}\Big)_3,\Big(\frac{2}{5}\Big)_3,\frac{3}{5},\ N=20,\\
&(G_2)_1/(LY)_1,\ c=\frac{12}{5},\ h_i=0,\frac15,\frac{2}{5},\frac45,\ N=10.
\ea
\ee

There also exist situations where the first two conditions are satisfied while the third is not. For example, the following two theories do \emph{not} satisfy a generalized $\T_2$ relation: 
\be
\ba
&M_{\rm sub}(6,5),\ c=\frac{4}{5},\ h_i=0,\Big(\frac{1}{15}\Big)_2,\frac{2}{5},\Big(\frac{2}{3}\Big)_2,\ N=30,\\
&{(A_2)_1}/{(LY)_1},\ c=\frac{8}{5},\ h_i=0,\Big(\frac{2}{15}\Big)_2,\frac{4}{5},\Big(\frac{1}{3}\Big)_2,\ N=15.
\ea
\ee
Here $M_{\rm sub}(6,5)$ describes the critical 3-state Potts model. 
One can easily check that the vacuum character of ${(A_2)_1}/{(LY)_1}$ is not the square of the vacuum character of $M_{\rm sub}(6,5)$.

\section{Classification of 2d RCFTs}\label{sec:cla}
The classification of 2d RCFTs we are interested in here is in more sense of number theory rather than category theory. One of the main reasons is that non-unitary RCFTs such as Lee-Yang theory and WZW $(E_{7\frac12})_1$ are of interest to us as much as unitary RCFTs. From the viewpoint of vector-valued modular forms, the characters of non-unitary and unitary RCFTs show no difference and deserve equal treatment. The classification of RCFTs exploiting the modularity of characters started from the seminal work of Mathur-Mukhi-Sen, and has now been achieved for the situation with up to five characters in \cite{Kaidi:2021ent}. 
One main aim of this paper is to better understand the classification results in \cite{Kaidi:2021ent}.

The 2d RCFTs having at least one pair of weights with integral difference are often called \emph{degenerate theories}, for example those with integral-weight non-vacuum primaries. Those theories bring more complexity in the holomorphic modular bootstrap, thus were not included in the classification in \cite{Kaidi:2021ent}. Let us consider the classification of non-degenerate theories appearing in \cite{Kaidi:2021ent}. We have the following main observation:
\emph{the characters of an arbitrary $l=0$ RCFT can be realized as either a Hecke image or the coset of a Hecke image w.r.t a $c=8k$ theory.}
We summarize all the initial theories for the (generalized) Hecke operations in Table \ref{table1-initial}. Each initial theory represents a \emph{type} which contains all its Hecke images and the $c=8k$ cosets of its Hecke images. Clearly all theories in one type share the same degeneracy. 
We will discuss all types in Table \ref{table1-initial} one by one in the following sections. 
If we further allow product or coset among initial theories, then the set of initial theories can be further reduced, which we collect in Table \ref{table1-pinitial} and call them primitive theories. \emph{One important consequence of our proposal is that it fixes the degeneracy and normalization of characters of all $l=0$ RCFTs} which are not explicit in the holomorphic modular bootstrap program.
\begin{table}[ht]
\def\arraystretch{1.1}
\centering
\begin{tabular}{|c|c|c|c|c|c|c|c|c|}
\hline

 $d$  & 2d RCFTs \\
\hline
 $2$  & $(LY)_1,(A_1)_1,(A_2)_1,{(D_4)_1}$\\
\hline
$3$   & $(LY)_2,(LY)_1^{\otimes 2},Spin(1,2,4)_1,(A_2)_1^{\otimes2},(A_4)_1$\\
\hline
$4$   & $\!\!(LY)_3,\!(LY)_1^{\otimes 3}\!,\!(A_1)_1^{\otimes 3}\!,M_{\rm eff}(5,3),M_{\rm sub}(6,5),M_{\rm sub}(14,3),M_{4/3,6/5,8/5,12/5,2},(A_6)_1,U(1)_3\!\!$\\
\hline
$5$  & $(LY)_4,(LY)_1^{\otimes 4}$\\
\hline
\end{tabular}
\caption{Initial 2d RCFTs for (generalized) Hecke operations and $c=8k$ cosets. The $Spin(1,2,4)_1$ theories are also denoted as $\textrm{Ising},U(1)_2,(A_1)_1^{\otimes2}$ respectively. $M_{8/5},M_{12/5}$ and $M_2$ can be realized as cosets $(A_2)_1/(LY)_1$, $(G_2)_1/(LY)_1$ and $(D_4)_1/(A_2)_1$ respectively. $M_{4/3}$ is a theory with central charge $c=4/3$ and weights with degeneracy $0,(\frac29)_3,\frac13,\frac23$ which will be discussed in Section \ref{sec:LY3}. $M_{6/5}$ is a subtheory of $\textrm{Ising}\otimes M(5,4)$ with central charge $c=6/5$ and weights with degeneracy $0,(\frac{1}{10})_3,(\frac{1}{2})_3,\frac35$ which will be discussed in Section \ref{sec:M53}. }
\label{table1-initial}
\end{table}

\begin{table}[ht]
\def\arraystretch{1.1}
\centering
\begin{tabular}{|c|c|c|c|c|c|c|c|c|}
\hline
 
 $d$  & 2d RCFTs \\
\hline
 $2$  & $(LY)_1,(A_1)_1,(A_2)_1,{(D_4)_1}$\\
\hline
$3$   & $(LY)_2,Spin(1,2)_1,(A_4)_1$\\
\hline
$4$   & $(LY)_3,M_{\rm sub}(6,5),M_{\rm sub}(14,3),M_{4/3,6/5},(A_6)_1$\\
\hline
$5$  & $(LY)_4$\\
\hline
\end{tabular}
\caption{Primitive 2d RCFTs for (generalized) Hecke operations and $c=8k$ cosets.}
\label{table1-pinitial}
\end{table}

Let us explain our proposal for RCFTs with $d=2$, for which we need initial theories $(LY)_1,(A_1)_1,(A_2)_1$ and ${(D_4)_1}$ from Table \ref{table1-initial}. We want to argue that the level 1 WZW $G_2,F_4,E_6,E_7,E_{7\frac12}$ models in the Mathur-Mukhi-Sen classification Table \ref{tab:MMS} can be generated by the level 1 $LY,A_1,A_2$ models from Hecke operations and $c=8$ cosets. This is actually obvious owing to $\T_5(LY)_1=(G_2)_1$ and the $c=8$ cosets in Table \ref{table1-2chi}. 

In the above discussion, we have limited ourselves to the non-degenerate RCFTs.  
For degenerate theories, one still need to add to the $l=0$ classification at least  with $(D_4)_1^{\otimes2}$ for $d=3$, $(D_4)_1^{\otimes3},(A_2)_1^{\otimes3}$ for $d=4$, and  $(D_4)_1^{\otimes4},(A_2)_1^{\otimes4},(A_1)_1^{\otimes4},U(1)_4,(A_1)_4$ for $d=5$. The Hecke images of degenerate theories are still degenerate and normally have $l>0$, thus are not the main interest of the current paper.

One interesting phenomenon we observe for $c=24$ pairs of Hecke images is that \emph{if a theory of central charge $c$ and conductor $N$ satisfies $cN=24$, then for any $c=24$ pair of its (generalized) Hecke images $(\T_p,\T_{N-p})$, the sum of spin-1 currents $m_1+\tm_1=\mathcal{N}$ is always divisible by the conductor $N$}. One can confirm this for all two-character RCFT pairs in Table \ref{table2-2chi}, knowing the conductor of $(LY)_1,(A_1)_1,(A_2)_1,{(D_4)_1}$ are $60,24,12,6$ respectively. We will demonstrate this intriguing phenomenon for more examples in later sections. In the cases where both images of a pair have ordinary characters, the bilinear relation of the characters is just equal to $J+\mathcal{N}$.  
It was recently suggested in \cite{Lin:2021bcp} that for $J+\mathcal{N}$ to be the single character of a consistent $c=24$ theory, the theory of topological modular forms constrains $\mathcal{N}$ to be divisible by $12$. However by computing the pairs of Hecke images, we do find some counter examples to this constraint. This  suggests that some Hecke images may not be physical as they may look. It would be interesting to further explore this point.

\section{RCFTs with three characters}\label{sec:3chi}
The potential RCFTs with three characters, non-integral weights and $l=0$ MLDEs have been recently classified by modular bootstrap in \cite{Kaidi:2021ent}, see also an early work \cite{kaneko3}. Apart from an infinite series $Spin(n)_1,n\neq 8k$, there are in total 65 theories in  (\cite{Kaidi:2021ent}, Table 5). We propose that all these theories including the infinite series can be merely generated by 7 simple theories which we will discuss individually in the following subsections. 
Therefore there are in total 7 types from the viewpoint of Hecke relation. 
We emphasize that RCFTs with three characters can have more than three primaries due to degeneracy. 
Among the 7 types we will discuss in the section, type Ising and $(LY)_2$ have three primaries, type $U(1)_2$, $(A_1)_1^{\otimes 2}$, $(LY)_1^{\otimes 2}$ and $(A_2)_1^{\otimes 2}$ have four primaries, while type $(A_4)_1$ have five primaries.

We remark that $l=0$ RCFTs with three characters can have integral weights. For example, the double product theory $(D_4)_1^{\otimes 2}$ has $c=8$ and conformal weights and degeneracy as
$h_i=0,(\frac{1}{2})_6,(1)_{9}.$ Its conductor $N=6$. We compute its Hecke images $\T_p$ for $p=5,7,11,...$ which all have $l\ge 6$ thus are not of our major interest here.

\subsection{Type Ising }
Critical Ising model $M(4,3)$ is the first of a series $Spin(n)_1$ unitary theory for $n=1$. It has central charge $c=\frac12$ and conformal weights $h_i=0,\frac{1}{16},\frac12$. The three characters are well-known to be
\be
\ba
\chi_0=\,&\frac{1}{2}\bigg(\sqrt{\frac{\theta_3(\tau)}{\eta(\tau)}}+\sqrt{\frac{\theta_4(\tau)}{\eta(\tau)}}\bigg)=q^{-\frac{1}{48}} ( 1 + q^2 + q^3 + 2 q^4 + 2 q^5 + 3 q^6  +\dots)
,\\
\chi_{\frac12}=\,&\frac{1}{2}\bigg(\sqrt{\frac{\theta_3(\tau)}{\eta(\tau)}}-\sqrt{\frac{\theta_4(\tau)}{\eta(\tau)}}\bigg)= q^{\frac{23}{48}} ( 1 + q + q^2 + q^3 + 2 q^4 + 2 q^5 + 3 q^6+\dots),\\
\chi_{\frac{1}{16}}\!=\,&\sqrt{\frac{\theta_2(\tau)}{2\eta(\tau)}}= q^{\frac{1}{24}} (1 + q + q^2 + 2 q^3 + 2 q^4 + 3 q^5 + 4 q^6 + 5 q^7 + 
 \dots),
\ea
\ee
where $\theta_i$ are Jacobi theta functions. Obviously the conductor $N=48$. The $S$-matrix for $(\chi_0,\chi_{\frac12},\chi_{\frac{1}{16}})$ is well-known to be
\be
\rho(S)^{\mathrm{Ising}}=\left(
\begin{array}{ccc}
 \frac{1}{2} & \frac{1}{2} & \frac{1}{\sqrt{2}} \\
 \frac{1}{2} & \frac{1}{2} & -\frac{1}{\sqrt{2}} \\
 \frac{1}{\sqrt{2}} & -\frac{1}{\sqrt{2}} & 0 \\
\end{array}
\right).
\ee
In fact, this is also the $S$-matrix of all WZW $(B_r)_1$ theories in the order of weights $0,\frac12,\frac{2r+1}{16}.$
The Hecke images $\T_p$ of Ising model for $p$ coprime to $48$ have been discussed in (\cite{Harvey:2018rdc}, Section 5.3). There are two classes for the Hecke operation: for $p=1, 7, 17, 23, 25, 31, 41, 47\textrm{ mod } 48$, i.e., $p^2\equiv 1 \textrm{ mod } 48$, $
\rho(\sigma_{p})=\mathrm{Id},$ while for $p=5, 11, 13, 19, 29, 35, 37, 43 \textrm{ mod } 48$, i.e., $p^2\equiv 25 \textrm{ mod } 48$, 
\be\label{sigmaIsing}
\rho(\sigma_p)=\left(
\begin{array}{ccc}
 0 & 1 & 0 \\
 1 & 0 & 0 \\
 0 & 0 & -1 \\
\end{array}
\right).
\ee
Notably, it was pointed out in \cite{Harvey:2018rdc} that the $\T_{31}$ image of Ising describe the WZW model $(E_8)_2$ and the $\T_{47}$ image describe the characters of the Baby Monster $\mathbb{B}$ vertex algebra \cite{hoehn}. We compute all admissible $\T_p$ for $p<48$ and collect the relevant information of all Hecke images in $c=24$ pairs in Table \ref{tb:HeckeIsing}. We observe more Hecke relations, such as the $\T_{35}$ image can describe a subtheory of $(C_{10})_1$, while the $\T_{37}$ image can describe a subtheory of $(E_{7})_2\otimes (F_{4})_1$.

The original Hecke operation in \cite{Harvey:2018rdc} does not include $\T_3$ as 3 is divisible by the conductor 48. However, we notice many well-defined RCFTs can be described by the generalized Hecke operation $\T_{3k}$ of Ising model, $k\in\IZ$. In particular, the generalized $\T_3$ image describes the WZW model $(A_1)_2$. The rest generalized Hecke images $\T_{3k}$ can be obtained simply by ordinary Hecke operation $\T_k(A_1)_2$. We include all  generalized Hecke images for $k=3,5,7,9,11,13,15$ in Table \ref{tb:HeckeIsing} as well. It is easy to check the sums of spin-1 currents $m_1+\tm_1$ for all pairs in Table \ref{tb:HeckeIsing} are divisible by the conductor 48. We also checked the bilinear relations of characters of all pairs which give $J(\tau)+m_1+\tm_1$. All theories in Table \ref{tb:HeckeIsing} that appeared in the Schellekens' list belong to the MTC classes $3^B_{\pm 1/2,\pm 3/2,\pm 5/2,\pm 7/2}$ in Table I of \cite{Schoutens:2015uia}.  Besides, all theories in the $l'=0$ column of Table \ref{tb:HeckeIsing} have appeared as solutions in the holomorphic modular bootstrap (\cite{Kaidi:2021ent}, Table 5), and were determined to have well-defined fusion algebras.
However, we notice that some $c=24$ pairs in Table \ref{tb:HeckeIsing} do not belong to the Schellekens’ list, for example the pair $(\T_{23},\T_{25})$.
This suggests that some Hecke images here such as $\T_{25}$ may not be unitary or physical theories. We also remark that the all theories in the $l = 0$ column can be fermionized to $2c$ copies of Majorana-Weyl free fermions, and all theories in the $l'=0$ column may have supersymmetric interpretations due to the presence of weight-$3/2$ primaries.

\begin{table}[ht]
\def\arraystretch{1.1}
	\centering
	\begin{tabular}{|c|c|c|c|c|c|c|c|c|c|c|c|c|c|}
		\hline
		 \multicolumn{4}{|c|}{$l=0$} & \multicolumn{4}{c|}{$l'=0$} & \multicolumn{2}{c|}{duality}\\
		\hline
		 $c$ & $h_i$  & $m_1$ & remark & $\tc$ & $\tilh_i$ &  $\tm_1$ & remark & $m_1+\tm_1$ & Sch\\
		\hline
$\frac{1}{2}$ & $\frac{1}{16},\frac12 $ & $0 $ & $\T_{1} $ & $\frac{47}{2}$ & ${\nb{\frac32},\frac{31}{16}} $ & $ 0$ & $\T_{47},\mathbb{B}$   &  $  0 $  & $0$\\
$\frac{3}{2}$ & $\frac{3}{16},\frac12 $ & $3 $ & $\T_{3},(A_1)_2 $ & $\frac{45}{2}$ & $ {\nb{\frac32},\frac{29}{16}}$ & $ 45$ & $\T_{45}$   &  $  48 $  & $5,7,8,10$\\
$\frac{5}{2}$ & $ \frac{5}{16},\frac12$ & $10 $ & $\T_{5},(B_2)_1 $ & $\frac{43}{2}$ & $ {\nb{\frac32},\frac{27}{16}}$ & $ 86$ & $\T_{43}$   &  $96   $  & $25,26,28$\\
$\frac{7}{2}$ & $ \frac{7}{16},\frac12$ & $ 21$ & $\T_{7},(B_3)_1 $ & $\frac{41}{2}$ & $ {\nb{\frac32},\frac{25}{16}}$ & $123 $ & $\T_{41}$   &  $  144 $  & $39,40$\\
$\frac{9}{2}$ & $\frac12,\frac{9}{16} $ & $ 36$ & $\T_{9},(B_4)_1 $ & $\frac{39}{2}$ & $ {\frac{23}{16},\nb{\frac32}}$ & $156 $ & $\T_{39}$   &  $  192 $  & $47,48$\\
$\frac{11}{2}$ & $\frac12,\frac{11}{16} $ & $ 55$ & $\T_{11},(B_5)_1 $ & $\frac{37}{2}$ & ${\frac{21}{16},\nb{\frac32}} $ & $ 185$ & $\T_{37}$   &  $  240 $  & $53$\\
$\frac{13}{2}$ & $\frac12,\frac{13}{16} $ & $ 78$ & $\T_{13},(B_6)_1 $ & $\frac{35}{2}$ & ${\frac{19}{16},\nb{\frac32}} $ & $210 $ & $\T_{35},(C_{10})_1$   &  $  288 $  & $56$\\
$\frac{15}{2}$ & $\frac12,\frac{15}{16} $ & $ 105$ & $\T_{15},(B_7)_1 $ & $\frac{33}{2}$ & $\frac{17}{16},\nb{\frac32} $ & $ 231$ & $\T_{33}$   &  $  336 $  & $-$\\
$\frac{17}{2}$ & ${\frac12,\frac{17}{16}} $ & $ 136$ & $\T_{17},(B_8)_1 $ & $\frac{31}{2}$ & $ {\frac{15}{16},\nb{\frac32}}$ & $ 248$ & $\T_{31},(E_8)_{2}$   &  $ 384  $  & $62$\\
$\frac{19}{2}$ & $\frac12,\frac{19}{16} $ & $171 $ & $\T_{19},(B_9)_1 $ & $\frac{29}{2}$ & $ {\frac{13}{16},\nb{\frac32}}$ & $261 $ & $\T_{29}$   &  $ 432  $  & $-$\\
$\frac{21}{2}$ & $\frac12,\frac{21}{16} $ & $ 210$ & $\T_{21},(B_{10})_1 $ & $\frac{27}{2}$ & $ {\frac{11}{16},\nb{\frac32}} $ & $270 $ & $\T_{27}$   &  $ 480  $  & $-$\\
$\frac{23}{2}$ & $\frac12,\frac{23}{16} $ & $253 $ & $\T_{23},(B_{11})_1 $ & $\frac{25}{2}$ & ${\frac{9}{16},\nb{\frac32}} $ & $ 275$ & $\T_{25}$   &  $  528 $  & $-$\\
		\hline
	
		\end{tabular}
			\caption{(Generalized) Hecke images of critical Ising model. Note for $p=3k,k\in\IZ$, we regard them as generalized Hecke images, i.e., Hecke images $\T_k$ of WZW $(A_1)_2$. Here and after, “Sch'' is short for the “Schellekens' list No." }
			\label{tb:HeckeIsing}
		\end{table}
In the following, we demonstrate in detail the basic generalized Hecke relation $(A_1)_2=\T_3M(4,3)$. 
WZW model $(A_1)_2$ has central charge $c=3/2$ and three primaries with characters:
\be
\ba
\chi_0^{2w_0}=&\,q^{-\frac{1}{16}}(1+3 q+9 q^2+15 q^3+30 q^4+54 q^5+94 q^6+\dots),\\
\chi_{1/2}^{2w_1}=&\,q^{\frac{7}{16}}(3+4 q+12 q^2+21 q^3+43 q^4+69 q^5+123 q^6+\dots),\\
\chi_{3/16}^{w_0+w_1}=&\,q^{\frac18}(2+6 q+12 q^2+26 q^3+48 q^4+84 q^5+146 q^6+\dots).
\ea
\ee
Here $w_0,w_1$ denote the affine weights of $\hat{A}_1$. Clearly the conductor of $(A_1)_2$ is $N=16$. We find the following exact relation between the $(A_1)_2$ characters and Ising characters:
\be\label{rel1}
\ba
\chi_0^{A_1}=\,\chi_0^3+3\chi_0\chi_{1/2}^2,\qquad
\chi_{1/2}^{A_1}=\,\chi_{1/2}^3+3\chi_{1/2}\chi_0^2,\qquad
\chi_{3/16}^{A_1}=\, 2\chi_{1/16}^3.
\ea
\ee
This means the $(A_1)_2$ characters are the degree 3 homogeneous polynomials of Ising characters. We can see that all three conditions of generalized Hecke operation $\T_3$ listed in Section \ref{sec:gHecke} are satisfied. Besides, it is easy to derive from \eqref{rel1} that the $S$-matrix of $(A_1)_2$ is identical to $\rho(S)^{\mathrm{Ising}}$. The Hecke operations of $(A_1)_2$ fall into two classes: for $p=1, 7, 9, 15\textrm{ mod } 16$, i.e., $p^2\equiv 1 \textrm{ mod } 16$, $
\rho(\sigma_{p})=\mathrm{Id},$ while for $p=3, 5, 11, 13 \textrm{ mod } 16$, i.e., $p^2\equiv 9 \textrm{ mod } 16$, $\rho(\sigma_p)$ is the same with \eqref{sigmaIsing}.
These highly resemble the two classes of original Hecke operation of Ising. Remark that although the $S$-matrices of Ising and $(A_1)_2$ are identical, their $T$-matrices are not. We further compute the Hecke images $\T_k (A_1)_2$ and observe the Hecke relations such as $\T_3 (A_1)_2=(B_4)_1$, $\T_5 (A_1)_2=(B_7)_1$ and $\T_7 (A_1)_2=(B_{10})_1$. Moreover, $\T_{15} (A_1)_2$ image can describe a subtheory of $(D_5)_8$. This subtheory pairs with $(A_1)_2$ to form a $c=24$ theory in the Schellekens' list No.10, of which the associated holomorphic VOA was constructed in \cite{lam2011}. 

\subsection{Type $U(1)_2$ }
The $U(1)_2$ theory is a well-known $c=1$ RCFT which can also be regarded as $Spin(n)_1$ theory with $n=2$. It has four primaries with conformal weights $h_i=0,(\frac{1}{8})_2,\frac12$. The three distinct characters are:
\be
\ba
\chi_0=&\,q^{-\frac{1}{24}}(1 + q + 4 q^2 + 5 q^3 + 9 q^4 + 13 q^5 + 21 q^6+\dots),\\
\chi_{1/8}=&\,q^{\frac{1}{12}}(1 + 2 q + 3 q^2 + 6 q^3 + 9 q^4 + 14 q^5 + 22 q^6+\dots),\\
\chi_{1/2}=&\,q^{\frac{11}{24}}(2 + 2 q + 4 q^2 + 6 q^3 + 12 q^4 + 16 q^5 + 26 q^6+\dots).
\ea
\ee
Obviously the conductor $N=24$. The $S$-matrix is well-known to be
\be\label{SU1_2}
\rho(S)=\left(
\begin{array}{ccc}
 \frac{1}{2} & 1 & \frac{1}{2} \\
 \frac{1}{2} & 0 & -\frac{1}{2} \\
 \frac{1}{2} & -1 & \frac{1}{2} \\
\end{array}
\right).
\ee
In fact, this is also the $S$-matrix of all WZW $(D_r)_1$ theories in the order of weights $0,\frac{r}{8},\frac12.$ Consider the Hecke images $\T_p$ of $U(1)_2$. 
It is easy to find for all $p$ coprime to $24$, $\rho(\sigma_p)=\mathrm{Id}$. We compute all admissible Hecke images for $p<24$ and summarize them in $c=24$ pairs in Table~\ref{tb:HeckeSpin2}. Interestingly, we notice many well-defined RCFTs can be described by the generalized Hecke operation $\T_{3k}$ of $U(1)_2$ theory, $k\in\IZ$. In particular, the generalized $\T_3$ image describes the WZW model $(A_3)_1$. In the following, we discuss the basic generalized Hecke relation $(A_3)_1=\T_3U(1)_2$.

WZW model $(A_3)_1$ has central charge $c=3$ and four primaries with conformal weights $h_i=0,(\frac{3}{8})_2,\frac12$. The three distinct characters are:
\be
\ba
\chi_0^{w_0}=&\,q^{-\frac{1}{8}}(1 + 15 q + 51 q^2 + 172 q^3 + 453 q^4 + 1128 q^5  +\dots),\\
\chi_{3/8}^{w_1}=&\,q^{\frac14}(4 + 24 q + 84 q^2 + 248 q^3 + 648 q^4 + 1536 q^5 +\dots),\\
\chi_{1/2}^{w_2}=&\,q^{\frac{3}{8}}(6 + 26 q + 102 q^2 + 276 q^3 + 728 q^4 + 1698 q^5  +\dots).
\ea
\ee
Here $w_{0,1,2,3}$ denote the affine weights of $\hat{A}_3^{(1)}$. 
The conductor of $(A_3)_1$ is $N=8$. 
We find the following exact relation between the $(A_3)_1$ characters and $U(1)_2$ characters:
\be\label{rel2}
\ba
\chi_0^{A_3}=\,\chi_0^3+3\chi_0\chi_{1/2}^2,\qquad
\chi_{3/8}^{A_3}=\, 4\chi_{1/8}^3,\qquad
\chi_{1/2}^{A_3}=\,\chi_{1/2}^3+3\chi_{1/2}\chi_0^2.
\ea
\ee
This means the $(A_3)_1$ characters are the degree 3 homogeneous polynomials of $U(1)_2$ characters. Note the relation \eqref{rel2} is almost identical to relation \eqref{rel1}. We checked all three conditions of generalized Hecke operation $\T_3$ are satisfied. From \eqref{rel2} and \eqref{SU1_2}, it is easy to find the $S$-matrix of $(A_3)_1$ is identical to the $S$-matrix \eqref{SU1_2} of $U(1)_2$. 

Consider the Hecke images $\T_k$ of $(A_3)_1$. We find for all $k$ coprime to the conductor 8, $\rho(\sigma_p)=\mathrm{Id}$, which resembles the situation of $U(1)_2$. We then compute all $\T_k(A_3)_1$ for $k=3,5,7$ and list them as generalized Hecke images in Table~\ref{tb:HeckeSpin2}. We observe Hecke relations such as $\T_3(A_3)_1=(D_9)_1$ and $\T_5(A_3)_1$ describes a subtheory of WZW $(A_{15})_1$. More precisely, the  $\T_5(A_3)_1$ image and  $(A_{15})_1$ characters satisfy the following relation
\be
\ba
\chi^{\T_5(A_3)_1}_0=&\,\chi_{0,1}^{(A_{15})_1,w_0}+\chi_{2,12870}^{(A_{15})_1,w_8}=q^{-\frac{5}{8}}( 1 + 255 q + 27525 q^2 + 713850 q^3+\dots),\\
\chi^{\T_5(A_3)_1}_{\frac78}=&\,\chi_{\frac78,120}^{(A_{15})_1,w_2}+\chi_{\frac{15}{8},8008}^{(A_{15})_1,w_6} =q^{\frac{1}{4}}(120 + 17104 q + 494040 q^2+ \dots),\\
\chi^{\T_5(A_3)_1}_{\frac32}=&\,2\chi_{\frac32, 1820}^{(A_{15})_1,w_4} =q^{\frac{7}{8}}( 3640 + 154056 q + 2878920 q^2  +\dots).
\ea
\ee
Here the primary with weight $7/8$ has degeneracy two. Together $\T_5(A_3)_1$  and $\T_3(A_3)_1$ form a $c=24$ theory in Schellekens' List No.63. Moreover, $\T_7(A_3)_1$ image can describe a subtheory of WZW $(C_7)_2$, which pairs with $(A_3)_1$ itself to form another $c=24$ theory in Schellekens' List No.35.
The holomorphic VOA associated to this $c=24$ theory was constructed in \cite{lam2011}.

We can see the sums of spin-1 currents $m_1+\tm_1$ for all pairs in Table \ref{tb:HeckeSpin2} are divisible by the conductor 24 of $U(1)_2$ theory. We also checked the bilinear relations of characters of all pairs which give $J(\tau)+m_1+\tm_1$. All theories in Table \ref{tb:HeckeSpin2} that appeared in the Schellekens' list belong to the MTC classes $4^B_{\pm1,\pm3}$ in Table I of \cite{Schoutens:2015uia}.
Besides, all theories in the $l'=0$ column of Table \ref{tb:HeckeSpin2} have appeared in the holomorphic modular bootstrap (\cite{Kaidi:2021ent}, Table 5), and were determined to have positive integer Verlinde formulas.
However, the last pair $(\T_{11},\T_{13})$ do not belong to the Schellekens’ list.
This suggests that the $\T_{13}$ image may not be unitary or physical theories.

\begin{table}[ht]
\def\arraystretch{1.1}
	\centering
	\begin{tabular}{|c|c|c|c|c|c|c|c|c|c|c|c|c|c|}
		\hline
		 \multicolumn{4}{|c|}{$l=0$} & \multicolumn{4}{c|}{$l'=0$} & \multicolumn{2}{c|}{duality}\\
		\hline
		 $c$ & $h_i$  & $m_1$ & remark & $\tc$ & $\tilh_i$ &  $\tm_1$ & remark & $\!\!m_1\!+\!\tm_1\!\!$ & Sch\\
		\hline
$1$ & ${\frac18,\frac12} $ & $ 1$ & $\T_{1} $ & $23$ & $\nb{\frac32},\frac{15}{8}$ & $ 23$ & $\T_{23}$   &  $  24 $  & $1$\\
$ 3 $ & ${\frac38,\frac12} $ & $ 15$ & $\T_{3 },(A_3)_1 $ & $21$ & $\frac{13}{8},\nb{\frac32}$ & $ 105$ & $\T_{21 }$   &  $  120 $  & $\!\!30,31,33-35\!\!$\\
$ 5 $ & ${\frac58,\frac12} $ & $45 $ & $\T_{ 5},(D_5)_1 $ & $19$ & $\frac{11}{8},\nb{\frac32} $ & $171 $ & $\T_{19 }$   &  $   216$  & $49$\\
$ 7 $ & ${\frac78,\frac12} $ & $ 91$ & $\T_{ 7},(D_7)_1 $ & $17$ & $\frac{9}{8},\nb{\frac32} $ & $221 $ & $\T_{17 }$   &  $  312 $  & $59$\\
$ 9 $ & ${\frac12,\frac98} $ & $153 $ & $\T_{ 9},(D_9)_1 $ & $15$ & $\frac{7}{8},\nb{\frac32} $ & $ 255$ & $\T_{15 },(A_{15})_1$   &  $ 408  $  & $63$\\
$ 11 $ & $ {\frac12,\frac{11}{8}}$ & $ 231$ & $\!\T_{ 11},(D_{11})_1\!$ & $13$ & $\frac{5}{8},\nb{\frac32} $ & $ 273$ & $\T_{ 13}$   &  $  504 $  & $-$\\

		\hline
	
		\end{tabular}
			\caption{(Generalized) Hecke images of $U(1)_2$. For $p=3k,k\in \IZ$, we regard them as generalized Hecke images as $\T_k(A_3)_1$.}
			\label{tb:HeckeSpin2}
		\end{table}
		
There is one more potential theory from holomorphic modular bootstrap (\cite{Kaidi:2021ent}, Table 5) falling into type $U(1)_2$, which has $c=23$, $h_i=0,\frac{7}{8},\frac{5}{2}$ and $m_1=2323$. We find it can form a $c=24$ pair with $U(1)_2$ which has bilinear relation of characters equal to $J+3474$.
		
\subsection{Type $(A_1)_1^{\otimes 2}$ }
Consider the double product theory $(A_1)_1^{\otimes 2}$ which has central charge
$c=2$ and conformal weights with degeneracy $h_i=0,(\frac14)_2,\frac12$. This can also be regarded as $Spin(4)_1$ theory. We use $h$ to denote $\chi^{(A_1)_1}$, then $h_0=\theta_3(2\tau)/\eta(\tau)$ and $h_{1/4}=\theta_2(2\tau)/\eta(\tau)$. The three distinct characters $(A_1)_1^{\otimes 2}$, i.e., $h_0^2,h_0h_{1/4},h_{1/4}^2$ have the same $S$-matrix as \eqref{SU1_2}. The conductor of $(A_1)_1^{\otimes 2}$ is $N=12$. Consider the Hecke images $\T_p$ of $(A_1)_1^{\otimes 2}$. We find for all $p$ coprime to $12$, $\rho(\sigma_p)=\mathrm{Id}$. We then compute all admissible Hecke images for $p<12$ and summarize them in $c=24$ pairs in Table~\ref{tb:HeckeSpin4}. Beside Hecke relation inherited from $\T_7(A_1)_1=(E_7)_1$, we observe new Hecke relations such as $\T_5(A_1)_1^{\otimes 2}=(D_{10})_1$, and $\T_{11}(A_1)_1^{\otimes 2}$ can describe a subtheory of $(D_6)_5$.

\begin{table}[ht]
\def\arraystretch{1.1}
	\centering
	\begin{tabular}{|c|c|c|c|c|c|c|c|c|c|c|c|c|c|}
		\hline
		 \multicolumn{4}{|c|}{$l=0$} & \multicolumn{4}{c|}{$l'=0$} & \multicolumn{2}{c|}{duality}\\
		\hline
		 $c$ & $h_i$  & $m_1$ & remark & $\tc$ & $\tilh_i$ &  $\tm_1$ & remark & $m_1+\tm_1$ & Sch\\
		\hline
$2$ & ${(\frac14)_2,\frac12} $ & $6 $ & $\T_{1} $ & $22$ & $\nb{\frac32},(\frac{7}{4})_2$ & $ 66$ & $\T_{11}$   &  $ 72  $  & $15-20$\\
$6$ & ${\frac12,(\frac34)_2} $ & $66 $ & $\T_{3},(D_6)_1 $ & $18 $ & $(\frac{5}{4})_2,\nb{\frac32}$ & $198 $ & $\T_{9}$   &  $ 264  $  & $54,55$\\
$10$ & ${\frac12,(\frac54)_2} $ & $190 $ & $\T_{5},(D_{10})_1 $ & $14$ & $(\frac{3}{4})_2,\nb{\frac32}$ & $ 266$ & $\T_{7},(E_7)_1^{\otimes 2}$   &  $  456 $  & $64$\\
		\hline
	
		\end{tabular}
			\caption{(Generalized) Hecke images of $(A_1)_1^{\otimes 2}$. All theories are unitary. For $p=3k$, we regard them as generalized Hecke images $\T_k(D_6)_1$.}
			\label{tb:HeckeSpin4}
		\end{table}

Interestingly, we notice some RCFTs can be described by the generalized Hecke operation $\T_{3k},k\in\IZ$ of $(A_1)_1^{\otimes 2}$. In particular, the generalized $\T_3$ image describes WZW model $(D_6)_1$. This model has central charge $c=6$ and four primaries with weights $h_i=0,(\frac{3}{4})_2,\frac12$. The three distinct characters associated to the affine, spinor and vector nodes of $\hat{D}_6^{(1)}$ are respectively
\be
\ba
\chi_0=&\,h_0^6 + 3 h_0^2 h_{1/4}^4=q^{-\frac{1}{4}}( 1 + 66 q + 639 q^2 + 3774 q^3 + 17283 q^4+\dots),\\
\chi_{3/4}=&\,4 h_{0}^3h_{1/4}^3=q^{\frac12}(32 + 384 q + 2496 q^2 + 12032 q^3 +48288 q^4+\dots),\\
\chi_{1/2}=&\,h_{1/4}^6 + 3 h_{1/4}^2 h_{0}^4=q^{\frac{1}{4}}(12 + 232 q + 1596 q^2 + 8328 q^3 +\dots).
\ea
\ee
Here the conductor $N=4$. It is easy to find the $S$-matrix of these three characters is still identical to \eqref{SU1_2}. Besides, the Hecke operations for $(D_6)_1$ always have $\rho(\sigma_p)=\mathrm{Id}$. We compute $\T_3(D_6)_1$ and regard it as generalized $\T_9$ Hecke image of $(A_1)_1^{\otimes 2}$. This image can be interpreted as a subtheory of $(A_9)_1^{\otimes 2}$ or a subtheory of $(D_6)_1^{\otimes 3}$. Besides, $(D_6)_1$ and $\T_3(D_6)_1$ form a $c=24$ pair which we add in Table~\ref{tb:HeckeSpin4}. Note the sums of spin-1 currents $m_1+\tm_1$ for all three pairs in Table \ref{tb:HeckeSpin4} are divisible by the original conductor 12 of $(A_1)_1^{\otimes 2}$ theory.

There is one more potential theory from holomorphic modular bootstrap (\cite{Kaidi:2021ent}, Table 5) falling into type $(A_1)_1^{\otimes 2}$, which has $c=22$, $h_i=0,\frac{3}{4},\frac{5}{2}$ and $m_1=1298$. We find it can form a $c=24$ pair with $(A_1)_1^{\otimes 2}$ which has bilinear relation of characters equal to $J+1920$.

\subsection{Type $(LY)_2$ }\label{sec:LY2}
Minimal model $M(7,2)$ is a non-unitary theory with central charge $-\frac{68}{7}$ and conformal weights $h_i={0,-\frac{2}{7},-\frac{3}{7}}$. The effective theory $(LY)_2$ has $c_{\rm eff}=\frac{4}{7}$ and $h_i^{\rm eff}=0,\frac{1}{7},\frac{3}{7}$. The three characters have the following Fourier expansion
\be\label{chiM37eff}
\ba
\chi_0=\,&q^{-\frac{1}{42}} (1 + q + 2 q^2 + 2 q^3 + 3 q^4 + 4 q^5 + 6 q^6 +\dots)
,\\
\chi_{\frac17}=\,& q^{\frac{5}{42}} (1 + q + q^2 + 2 q^3 + 3 q^4 + 3 q^5 + 5 q^6 + \dots),\\
\chi_{\frac37}=\,& q^{\frac{17}{42}} (1 + q^2 + q^3 + 2 q^4 + 2 q^5 + 3 q^6 + 3 q^7 +\dots).
\ea
\ee
Clearly, the conductor $N=42$. The $S$-matrix for $(LY)_2$ is
\be
\rho(S)=\frac{2}{\sqrt{7}}\left(
\begin{array}{ccc}
 \cos \left(\frac{\pi }{14}\right) & \cos \left(\frac{3 \pi }{14}\right) & \sin \left(\frac{\pi }{7}\right) \\
 \cos \left(\frac{3 \pi }{14}\right) & -\sin \left(\frac{\pi }{7}\right) & -\cos \left(\frac{\pi }{14}\right) \\
 \sin \left(\frac{\pi }{7}\right) & -\cos \left(\frac{\pi }{14}\right) & \cos \left(\frac{3 \pi }{14}\right) \\
\end{array}
\right).
\ee
Many Hecke images $\T_p$ of $(LY)_2$ for $p$ coprime to $42$ have been discussed in \cite{Harvey:2018rdc}. In particular, there exist three classes for the Hecke operation: for $p=1,13,29,41\textrm{ mod } 42$, i.e., $p^2\equiv 1 \textrm{ mod } 42$, $\rho(\sigma_{p})=\mathrm{Id},$ 
\be
\textrm{for }p=5,19,23,37 \textrm{ mod } 42,\textrm{ i.e., } p^2\equiv 25 \textrm{ mod } 42,\quad
\rho(\sigma_p)=\left(
\begin{array}{ccc}
 0 & 1 & 0 \\
 0 & 0 & -1 \\
 -1 & 0 & 0 \\
\end{array}
\right),
\ee
\be
\textrm{for }p=11,17,25,31 \textrm{ mod } 42,\textrm{ i.e., } p^2\equiv 37 \textrm{ mod } 42,\quad
\rho(\sigma_p)=\left(
\begin{array}{ccc}
 0 & 0 & -1 \\
 1 & 0 & 0 \\
 0 & -1 & 0 \\
\end{array}
\right).
\ee
It was also pointed out in \cite{Harvey:2018rdc} that the $\T_5$ image is related to the Witten-Reshetikhin–Turaev invariant of the Brieskorn homology sphere $\Sigma(2, 3, 7)$ \cite{Hikami}. We compute all admissible Hecke images for $p<42$ and list them in pairs w.r.t $c=24$ in Table \ref{tb:HeckeM27}. The pairs $(\T_p,\T_{p'})$ satisfy $p+p'=42$. We also use the $\T_{-17}$ as a formal notation to denote the original $M(7,2)$ minimal model, which is dual to the Hecke image $\T_{59}$. This $c=24$ pair $(M(7,2),\T_{59})$ has the following bilinear relation of characters\footnote{We notice it is a typical phenomenon that when a non-unitary minimal model and a $c>24$ Hecke image pair up to a putative $c=24$ theory, the bilinear relation of characters always involves some negative signs. We do not have a clear physical understanding of this phenomenon at the moment.}
\be\label{1647id}
\ba
J(\tau)=&\, \chi_0^{\T_{59}}\chi_0^{M(7,2)}-\chi_{\frac{16}{7}}^{\T_{59}}\chi_{-\frac27}^{M(7,2)}+\chi_{\frac{17}{7}}^{\T_{59}}\chi_{-\frac37}^{M(7,2)}.
\ea
\ee

Interestingly, we find many theories appeared in the holomorphic modular bootstrap (\cite{Kaidi:2021ent}, Table 5) can be described by the generalized Hecke operation $\T_{3k}$ of $(LY)_2$, $k\in\IZ$, which are ordinary $\T_k$ of a $c=\frac{12}{7}$ theory with weights $h_i=0,\frac{2}{7},\frac{3}{7}$. We compute all such generalized Hecke images and list them in pairs in Table \ref{tb:HeckeM27} as well. We can see the sums of spin-1 currents $m_1+\tm_1$ for all pairs in Table \ref{tb:HeckeM27} are divisible by the conductor 42 of $(LY)_2$. In the following, we discuss in detail the intriguing $c=\frac{12}{7}$ theory and its Hecke operations.

\begin{table}[ht]
\def\arraystretch{1.1}
	\centering
	\begin{tabular}{|c|c|c|c|c|c|c|c|c|c|c|c|c|c|}
		\hline
		 \multicolumn{4}{|c|}{$l=0$} & \multicolumn{4}{c|}{$l'=0$} & \multicolumn{2}{c|}{duality}\\
		\hline
		 $c$ & $h_i$  & $m_1$ & remark & $\tc$ & $\tilh_i$ &  $\tm_1$ & remark & $m_1+\tm_1$ & Sch\\
		\hline
$-\frac{68}{7}$  &  ${-\frac{2}{7},-\frac{3}{7}} $ & $ 0$ & $\T_{-17}$ & $\frac{236}{7}$ & ${\frac{16}{7},\frac{17}{7} }$ & $ 0$ & $\T_{59} $ & $ 0$ & $0$\\ 
    $\frac{4}{7}$  &  ${\frac{1}{7},\frac{3}{7}} $ & $ 1$ & $\T_{1}$ & $\frac{164}{7}$ & ${\frac{11}{7},\frac{13}{7} }$ & $ 41$ & $\T_{41} $ & $ 42$ & $-$\\
    $\frac{12}{7}$  &  ${\frac{2}{7},\frac{3}{7}} $ & $ 6$ & $\T_{3}$ & $\frac{156}{7}$ & ${\frac{11}{7},\frac{12}{7} }$ & $ 78$ & $\T_{39} $ & $ 84$ & $-$\\
        $\frac{20}{7}$  &  ${\frac{1}{7},\frac{5}{7}} $ & $ 10$ & $\T_{5},\star,\Sigma(2,3,7)$ & $\frac{148}{7}$ & ${ \frac{9}{7},\frac{13}{7}}$ & $74 $ & $\T_{37},\star $ & $ 84$ & $-$ \\
   $\frac{36}{7}$  &  ${\frac{2}{7},\frac{6}{7}} $ & $ 36$ & $\T_{9},\star$ & $\frac{132}{7}$ & ${\frac{8}{7},\frac{12}{7} }$ & $132 $ & $\T_{33},\star $ & $168 $ & $-$\\
            $\frac{44}{7}$  &  ${\frac{4}{7},\frac{5}{7}} $ & $88 $ & $\T_{11}$ & $\frac{124}{7}$ & ${\frac{9}{7},\frac{10}{7} }$ & $ 248$ & $\T_{31} $ & $ 336$ & $-$\\
                $\frac{52}{7}$  &  ${\frac{4}{7},\frac{6}{7}} $ & $156 $ & $\T_{13}$ & $\frac{116}{7}$ & ${ \frac{8}{7},\frac{10}{7}}$ & $ 348$ & $\T_{29} $ & $ 504$ & $-$ \\
        $\frac{60}{7}$  &  ${\frac{3}{7},\frac{8}{7}} $ & $ 210$ & $\T_{15}$ & $\frac{108}{7}$ & ${ \frac{6}{7},\frac{11}{7}}$ & $378 $ & $\T_{27} $ & $ 588$ & $-$ \\
                    $\frac{68}{7}$  &  ${\frac{3}{7},\frac{9}{7}} $ & $221 $ & $\T_{17}$ & $\frac{100}{7}$ & ${ \frac{5}{7},\frac{11}{7}}$ & $ 325$ & $\T_{25} $ & $ 546$ & $-$\\
            $\frac{76}{7}$  &  ${\frac{5}{7},\frac{8}{7}} $ & $ 152$ & $\T_{19},\star$ & $\frac{92}{7}$ & ${ \frac{6}{7},\frac{9}{7}}$ & $184 $ & $\T_{23},\star $ & $ 336$ & $-$\\
            
		\hline
	
		\end{tabular}
			\caption{(Generalized) Hecke images of $(LY)_2$. All theories are non-unitary. For $p=3k,k\in\IZ$, we regard them as generalized Hecke images.}
			\label{tb:HeckeM27}
		\end{table}
		
Consider the $c=\frac{12}{7}$ theory with weights $h_i=0,\frac{2}{7},\frac{3}{7}$ appeared in the holomorphic modular bootstrap \cite{Kaidi:2021ent}. We compute from $l=0$ MLDE that the three characters are
\be\label{chi127}
\ba
\chi_0^M=\,&q^{-\frac{1}{14}} (1 + 6 q + 12 q^2 + 28 q^3 + 57 q^4 + 108 q^5 + 191 q^6 +  
\dots)
,\\
\chi_{\frac37}^M=\,& q^{\frac{5}{14}} (2 + 3 q + 12 q^2 + 20 q^3 + 42 q^4 + 75 q^5 + 140 q^6 +\dots),\\
\chi_{\frac27}^M=\,& q^{\frac{3}{14}} (3 + 8 q + 21 q^2 + 42 q^3 + 87 q^4 + 156 q^5 + 285 q^6 + \dots).
\ea
\ee
The conductor $N=14$. We checked that the degeneracy of this theory is $(1,1,1)$, and notice that this is actually a subtheory of $(LY)_2^{\otimes3}$. The exact character relations are
\be
\ba
\chi_0^M= \chi_0^3 + 3\chi_{\frac17} \chi_{\frac37}^2 
,\quad
\chi_{\frac27}^M= 3\chi_0 \chi_{\frac17}^2  - \chi_{\frac37}^3,\quad
\chi_{\frac37}^M=3\chi_0^2  \chi_{\frac37} - \chi_{\frac17}^3.
\ea
\ee
These suggest that the $c=\frac{12}{7}$ theory satisfies the generalized Hecke  $\mathsf{T}_3$ relation with $(LY)_2$. Note the Fourier coefficients in \eqref{chi127} satisfy the mod 3 properties. 
The $S$ matrix of this theory can be computed from the $S$-matrix of $(LY)_2$ as 
\be
\rho(S)=\frac{2}{\sqrt{7}}\left(
\begin{array}{ccc}
 \cos \left(\frac{3 \pi }{14}\right) & \cos \left(\frac{\pi }{14}\right) & \sin \left(\frac{\pi }{7}\right) \\
 \cos \left(\frac{\pi }{14}\right) & -\sin \left(\frac{\pi }{7}\right) & -\cos \left(\frac{3 \pi }{14}\right) \\
 \sin \left(\frac{\pi }{7}\right) & -\cos \left(\frac{3 \pi }{14}\right) & \cos \left(\frac{\pi }{14}\right) \\
\end{array}
\right).
\ee
We then compute all admissible Hecke images of this $c=\frac{12}{7}$ theory. We find there exist three classes for its Hecke operation: for $p=1, 13\textrm{ mod } 14$, i.e., $p^2\equiv 1 \textrm{ mod } 14$, $
\rho(\sigma_{p})=\mathrm{Id},$ 
\be
\textrm{for }p=3, 11 \textrm{ mod } 14,\textrm{ i.e., } p^2\equiv 9 \textrm{ mod } 14,\quad
\rho(\sigma_p)=\left(
\begin{array}{ccc}
 0 & 1 & 0 \\
 0 & 0 & -1 \\
 -1 & 0 & 0 \\
\end{array}
\right),
\ee
\be
\textrm{for }p=5, 9 \textrm{ mod } 14,\textrm{ i.e., } p^2\equiv 11 \textrm{ mod } 14,\quad
\rho(\sigma_p)=\left(
\begin{array}{ccc}
 0 & 0 & -1 \\
 1 & 0 & 0 \\
 0 & -1 & 0 \\
\end{array}
\right).
\ee
These resemble the three classes of original Hecke operation of $(LY)_2$.


There are three more potential theories from holomorphic modular bootstrap (\cite{Kaidi:2021ent}, Table 5, the forth column) falling into type $(LY)_2$. One has $c=\frac{100}{7}$, $h_i=0,\frac{4}{7},\frac{12}{7}$ and $m_1=380$. We find it can form a $c=24$ pair with $\T_{17}$ which has bilinear relation of characters equal to $J+1536$. The second has $c=\frac{108}{7}$, $h_i=0,\frac{4}{7},\frac{13}{7}$ and $m_1=456$, which can form a $c=24$ pair with $\T_{15}$ which has bilinear relation of characters equal to $J+1056$. The third has $c=\frac{156}{7}$, $h_i=0,\frac{5}{7},\frac{18}{7}$ and $m_1=1248$, which can form a $c=24$ pair with $\T_{3}$ which has bilinear relation of characters equal to $J+1644$.

\subsection{Type $(LY)_1^{\otimes 2}$ }\label{sec:LY12}
Consider the double product of effective Lee-Yang model $(LY)_1^{\otimes 2}$ which has central charge $c=\frac45$ and conformal weights with degeneracy $h_i=0,(\frac{1}{5})_2,\frac{2}{5}$. This theory can also be regarded as a subtheory of effective minimal model $M_{\rm eff}(10,3)$. 
The conductor of $(LY)_1^{\otimes 2}$ is $N=30$. The full $S$-matrix is $(\rho(S)^{(LY)_1})^{\otimes 2}$, or for the three distinct characters  $\phi_1^2,\phi_1\phi_2,\phi_2^2$ as
\be
\rho(S)=\frac{1}{\sqrt{5}}\left(
\begin{array}{ccc}
 \frac{1}{2} \left(\sqrt{5}+1\right) & 2 & \frac{1}{2} \left(\sqrt{5}-1\right) \\
 1 & -1 & -1 \\
 \frac{1}{2} \left(\sqrt{5}-1\right) & -2 & \frac{1}{2} \left(\sqrt{5}+1\right) \\
\end{array}
\right).
\ee
Consider the ordinary Hecke images $\T_p$ of $(LY)_1^{\otimes 2}$ for $p$ coprime to 30. 

We find there exist two classes for the Hecke operation: for $p=1, 11, 19, 29 \textrm{ mod } 30$, i.e., $p^2\equiv 1 \textrm{ mod } 30$, $\rho(\sigma_p)=\mathrm{Id}$, while for $p= 7, 13, 17, 23 \textrm{ mod } 30$, i.e., $p^2\equiv 19 \textrm{ mod } 30$,
\be\label{rholy12}
\rho(\sigma_p)=\left(
\begin{array}{ccc}
 0 & 0 & 1 \\
 0 & -1 & 0 \\
 1 & 0 & 0 \\
\end{array}
\right).
\ee
We compute all admissible Hecke images of $(LY)_1^{\otimes 2}$ for $p<30$ and list them in $c=24$ pairs in Table \ref{tb:HeckeYL12}. The pairs satisfy $p+p'=30$. 
Beside the Hecke relations inherit from those of $(YL)_1$, we observe some new ones such as $\T_{17}$ describes a subtheory of WZW $(C_8)_1$, and $\T_{23}$ describes a subtheory of WZW $(G_2)_1\otimes (E_6)_3$. For example, we find the following
relation between the $\T_{17}$ image and $(C_8)_1$ characters:
\be
\ba
\chi^{\T_{17}}_0=&\,\chi_{0,1}^{(C_8)_1,w_0}+\chi_{2,4862}^{(C_8)_1,w_8}=q^{-\frac{17}{30}}( 1 + 136 q + 10438 q^2 + 216920 q^3+\dots),\\
\chi^{\T_{17}}_{\frac45}=&\,\chi_{\frac45,119}^{(C_8)_1,w_2}+\chi_{\frac95,6188}^{(C_8)_1,w_6}=q^{\frac{7}{30}}( 119 + 13328 q + 326026 q^2 +\dots),\\
\chi^{\T_{17}}_{\frac75}=&\,\chi_{\frac75,1700}^{(C_8)_1,w_4}=q^{\frac{5}{6}}(1700 + 61625 q + 1009000 q^2 +\dots).
\ea
\ee
Here the primary with weight ${7}/{5}$ has degeneracy 2. 
The two pairs $(\T_7,\T_{23})$ and $(\T_{13},\T_{17})$ have appeared in the Schellekens' list No.32 and No.52 \cite{Schellekens:1992db}.

\begin{table}[ht]
\def\arraystretch{1.1}
	\centering
	\begin{tabular}{|c|c|c|c|c|c|c|c|c|c|c|c|c|c|}
		\hline
		 \multicolumn{4}{|c|}{$l=0$} & \multicolumn{4}{c|}{$l'=0$} & \multicolumn{2}{c|}{duality}\\
		\hline
		 $c$ & $h_i$  & $m_1$ & remark & $\tc$ & $\tilh_i$ &  $\tm_1$ & remark & $\!\!m_1\!+\!\tm_1\!\!$ & Sch\\
		\hline
   $\!\!-\frac{44}{5}\!\!$  &  $\!\!-\{(\frac{1}{5})_2,\frac{2}{5}\} \!\!$ & $ 0$ & $\T_{-11}$ & $\frac{164}{5}$ & ${ (\frac{11}{5})_2,\frac{12}{5}}$ & $0 $ & $\T_{41} $ & $0 $ & $0$ \\
$\frac{4}{5}$  &  ${(\frac15)_2,\frac25} $ & $ 2$ & $\T_{1}$ & $\frac{116}{5}$ & ${\frac85,(\frac95)_2 }$ & $ 58$ & $\T_{29} $ & $60 $ & $-$\\
$\frac{12}{5}$  &  ${\frac15,(\frac35)_2} $ & $3 $ & $\T_{3},(A_1)_8$ & $\frac{108}{5}$ & ${(\frac75)_2,\frac95 }$ & $ 27$ & $\T_{27} $ & $ 30$ & $-$ \\
$\frac{28}{5}$  &  ${(\frac25)_2,\frac45} $ & $28 $ & $\T_{7},(G_2)_1^{\otimes 2}$ & $\frac{92}{5}$ & ${\frac65,(\frac85 )_2}$ & $ 92$ & $\T_{23} $ & $ 120$ & $32$\\
$\frac{36}{5}$  &  ${\frac35,(\frac45)_2} $ & $144 $ & $\T_{9}$ & $\frac{84}{5}$ & ${(\frac65)_2,\frac75 }$ & $336 $ & $\T_{21} $ & $480 $ & $-$ \\
$\frac{44}{5}$  &  ${\frac25,(\frac65)_2} $ & $220 $ & $\T_{11}$ & $\frac{76}{5}$ & ${(\frac45)_2,\frac85 }$ & $380 $ &  $\T_{19},(E_{7\frac12})_1^{\otimes2} $ & $600 $ & $-$\\
$\frac{52}{5}$  &  ${(\frac35)_2,\frac65} $ & $104 $ & $\T_{13},(F_4)_1^{\otimes 2}$ & $\frac{68}{5}$ & ${\frac45,(\frac75)_2 }$ & $ 136$ & $\T_{17},(C_8)_1 $ & $240 $ & $52$\\
		\hline
	
		\end{tabular}
			\caption{(Generalized) Hecke images of $(YL)_1^{\otimes 2}$. For $p=3k,k\in\IZ$, we regard them as generalized Hecke images $\T_k$ of a subtheory of WZW $(A_1)_8$.}
			\label{tb:HeckeYL12}
		\end{table}
		
Consider a unitary $c=12/5$ RCFT which is a subtheory of WZW $(A_1)_8$, -- the $D_6$ type in the ADE classification of $\hat{A}_1$ modular invariant, see e.g. \cite{DiFrancesco:1997nk}. The three distinct characters are
\be
\ba
\chi_0^{8w_0}=\,&q^{-\frac{1}{10}} (1 + 3 q + 18 q^2 + 38 q^3 + 99 q^4 + 207 q^5 + 438 q^6 +\dots)
,\\
\chi_{3/5}^{4w_0+4w_1}=\,& q^{\frac{1}{2}} (5 + 15 q + 45 q^2 + 110 q^3 + 255 q^4 + 525 q^5 + 1060 q^6 +\dots),\\
\chi_{1/5}^{2w_0+6w_1}=\,& q^{\frac{1}{10}} (3 + 16 q + 48 q^2 + 129 q^3 + 294 q^4 + 642 q^5 + 1302 q^6 + 
 \dots).
\ea
\ee
The primary with weight $3/5$ has degeneracy two.
Clearly the conductor $N=10$. We find 
the exact formulas for the characters as
\be
\ba
\chi_0=\phi_1^6-3\phi_1\phi_2^5
,\quad 
\chi_{3/5}=5\phi_1^3\phi_2^3,\quad
\chi_{1/5}=\phi_2^6+3\phi_1^5\phi_2.
\ea
\ee
It is easy to see these are the degree 3 homogeneous polynomials of $(LY)_1^{\otimes 2}$ characters. 
We regard this theory as the generalized Hecke image $\mathsf{T}_3$ of $(LY)_1^{\otimes 2}$. 
From $\rho(S)^{(LY)_1}$, it is easy to find the $S$-matrix for $\chi_0,\chi_{3/5},\chi_{1/5}$ is
\be
\rho(S)=\frac{1}{\sqrt{5}}\left(
\begin{array}{ccc}
 \frac{1}{2} \left(\sqrt{5}-1\right) & 2 & \frac{1}{2} \left(\sqrt{5}+1\right) \\
 1 & 1 & -1 \\
 \frac{1}{2} \left(\sqrt{5}+1\right) & -2 & \frac{1}{2} \left(\sqrt{5}-1\right) \\
\end{array}
\right).
\ee

Consider the Hecke images $\T_p$ of this $c=\frac{12}{5}$ theory. 
We find there exist two classes for the Hecke operation: for $p=1, 9 \textrm{ mod } 10$, i.e., $p^2\equiv 1 \textrm{ mod } 10$, $\rho(\sigma_p)=\mathrm{Id}$, while for $p= 3,7 \textrm{ mod } 10$, i.e., $p^2\equiv 9 \textrm{ mod } 10$, $\rho(\sigma_p)$ is the same with \eqref{rholy12}. These resemble the two classes of Hecke operation of $(LY)_1^{\otimes 2}$. We compute Hecke images $\T_k,k=3,7,9$ of the $c=\frac{12}{5}$ theory and regard them as generalized $\T_{3k}$ images of $(LY)_1^{\otimes 2}$. We add these theories in $c=24$ pairs in Table \ref{tb:HeckeYL12}. Clearly, the sums of spin-1 currents $m_1+\tm_1$ for all pairs in Table \ref{tb:HeckeYL12} are divisible by the conductor 30 of $(LY)_1^{\otimes 2}$.

We remark that the $c=24$ pair between the double product of original $M(5,2)$ model and the $\T_{41}$ Hecke image has the following bilinear relation of characters:
\be\label{1645id}
\ba
J(\tau)=&\, \chi_0^{\T_{41}}\chi_0^{M(5,2)^{\otimes 2}}-2\chi_{\frac{11}{5}}^{\T_{41}}\chi_{-\frac15}^{M(5,2)^{\otimes 2}}+\chi_{\frac{12}{5}}^{\T_{41}}\chi_{-\frac25}^{M(5,2)^{\otimes 2}}.
\ea
\ee
There are two more potential theories from holomorphic modular bootstrap (\cite{Kaidi:2021ent}, Table 5, the third column) falling into type $(LY)_1^{\otimes 2}$. One has $c=\frac{76}{5}$, $h_i=0,\frac{3}{5},\frac{9}{5}$ and $m_1=437$. We find it can form a $c=24$ pair with $\T_{11}$ which has bilinear relation of characters equal to $J+1284$. The second has $c=\frac{108}{5}$, $h_i=0,\frac{4}{5},\frac{12}{5}$ and $m_1=1404$, which can form a $c=24$ pair with $\T_{3}$ which has bilinear relation of characters equal to $J+2784$.

\subsection{Type $(A_2)_1^{\otimes 2}$ }
Consider the double product theory $(A_2)_1^{\otimes 2}$ which has central charge
$c=4$ and conformal weights with degeneracy $h_i=0,(\frac13)_2,\frac23$. The modular form expression of $(A_2)_1$ characters can be found in e.g. \cite{Kaneko:2013uga}. The $S$-matrix of $(A_2)_1^{\otimes 2}$ is 
\be 
\rho(S)= \left(
\begin{array}{ccc}
 \frac{1}{3} & \frac{4}{3} & \frac{4}{3} \\
 \frac{1}{3} & \frac{1}{3} & -\frac{2}{3} \\
 \frac{1}{3} & -\frac{2}{3} & \frac{1}{3} \\
\end{array}
\right).
\ee
The conductor of $(A_2)_1^{\otimes 2}$ is $N=6$. Consider the Hecke images of $(A_2)_1^{\otimes 2}$. We find for all $p$ coprime to 6, $\rho(\sigma_p)=\mathrm{Id}$. We compute all admissible Hecke images $\T_p$ for $p<12$. Obviously the only one $c=24$ pair of $(\T_p,\T_{p'})$ is $(p,p')=(1,5)$, which we list in Table \ref{tb:HeckeA212}. 
We find the $\T_5$ image can be interpreted as a subtheory of WZW $(A_8)_3$. For example, the vacuum character of $\T_5$ can be expressed by
\be
\ba
\chi^{\T_{5}}_0=&\,\chi_{0,1}^{(A_8)_3,3w_0}+2\chi_{2,17325}^{(A_8)_3,w_1+w_2+w_6}+\chi_{2,8820}^{(A_8)_3,w_4+w_5+w_0}+2\chi_{3,41580}^{(A_8)_3,3w_3}\\
=&\,q^{-\frac{5}{6}}(1 + 80 q + 46790 q^2 + 2654800 q^3 + 68308625 q^4+\dots).
\ea
\ee
This pair gives a $c=24$ theory appearing in the Schellekens’ List No.27, with the associated holomorphic VOA constructed in \cite{Lam:2015pjc} by certain $\IZ_2$ orbifold. 
Besides, we find the $\T_{7}$ image has $h_i=0,\frac{4}{3},\frac{5}{3}$ and $\T_{11}$ image has $h_i=0,\frac{7}{3},\frac{8}{3}$. Both $\T_{7}$ and $\T_{11}$ images have $m_1=0$ and $l=6$ MLDEs.

\begin{table}[ht]
\def\arraystretch{1.1}
	\centering
	\begin{tabular}{|c|c|c|c|c|c|c|c|c|c|c|c|c|c|}
		\hline
		 \multicolumn{4}{|c|}{$l=0$} & \multicolumn{4}{c|}{$l'=0$} & \multicolumn{2}{c|}{duality}\\
		\hline
		 $c$ & $h_i$  & $m_1$ & remark & $\tc$ & $\tilh_i$ &  $\tm_1$ & remark & $m_1+\tm_1$ & Sch\\
		\hline
$4$  &  ${(\frac13)_2,\frac23} $ & $16 $ & $\T_{1}$ & $20$ & ${\frac 43,(\frac53)_2}$ & $80 $ & $\T_{5} $ & $96 $ & $24,26,27$\\
		\hline
	
		\end{tabular}
			\caption{Hecke images of $(A_2)_1^{\otimes 2}$.}
			\label{tb:HeckeA212}
		\end{table}

We find there exist six more potential theories from holomorphic modular bootstrap (\cite{Kaidi:2021ent}, Table 5) falling into the type $(A_2)_1^{\otimes 2}$ beside the $\T_1,\T_5$ images. The $(E_6)_1^{\otimes 2}$ which has central charge $c=12$ and conformal weights with degeneracy $h_i=0,(\frac{2}{3})_2,\frac{4}{3}$, naturally is dual to $\T_1$ w.r.t $(E_8)_1^{\otimes 2}$. The $c=12,h_i=0,\frac{1}{3},\frac53$ theory is also dual to $\T_1$ w.r.t $(E_8)_1^{\otimes 2}$. The $c=20,h_i=\frac13,\frac{8}{3}$ theory can pair with $(A_2)_1^{\otimes 2}$ with bilinear relation of characters equal to $J+960$. The $c=20,h_i=\frac23,\frac{7}{3}$ theory can pair with $(A_2)_1^{\otimes 2}$ with bilinear relation of characters equal to $J+1716$. Besides, 
we find the $c=28,h_i=\frac23,\frac{10}{3}$ theory is the $c=32$ dual to $(A_2)_1^{\otimes 2}$ with bilinear relation of characters equal to $j^{1/3}(J+3066)$. 
We also find that the $c=36,h_i=\frac23,\frac{13}{3}$ theory is the $c=40$ dual to $(A_2)_1^{\otimes 2}$ with bilinear relation of characters equaling $j^{2/3}(J+4848)$.

\subsection{Type $(A_4)_1$ }
WZW model $(A_4)_1$ has central charge $c=4$ and conformal weights with degeneracy as $h_i=0,(\frac{2}{5})_2,(\frac{3}{5})_2.$
The three distinct characters are:
\be
\ba
\chi_0^{w_0}=&\,q^{-\frac{1}{6}}(1 + 24 q + 124 q^2 + 500 q^3 + 1625 q^4 + 4752 q^5+\dots),\\
\chi_{2/5}^{w_1}=&\,q^{\frac{7}{30}}(5 + 50 q + 220 q^2 + 820 q^3 + 2525 q^4 + 7070 q^5 +\dots),\\
\chi_{3/5}^{w_2}=&\,q^{\frac{13}{30}}( 10 + 65 q + 300 q^2 + 1025 q^3 + 3140 q^4 + 8565 q^5+\dots).
\ea
\ee
Here $w_{0,1,2,3,4}$ denote the affine weights of $\hat{A}_4^{(1)}$. Clearly the conductor $N=30$. WZW $(A_4)_1$ is self-dual w.r.t $c=8$, which means $\chi_0^2+4\chi_{2/5}\chi_{3/5}=j^{1/3}$. The Hecke images $\T_p$ of $(A_4)_1$ for $p\ge 5$ all have $l\ge 6$ MLDEs, thus we are only brief here. We find there are two classes for the Hecke operation: for $p=1, 11, 19, 29 \textrm{ mod } 30$, $\rho(\sigma_p)=\mathrm{Id}$, while for $p= 7, 13, 17, 23 \textrm{ mod } 30$,
\be 
\rho(\sigma_p)=\left(
\begin{array}{ccc}
 -1 & 0 & 0 \\
 0 & 0 & -1 \\
 0 & -1 & 0 \\
\end{array}
\right).
\ee

We find there are two more potential theories with $c=12,20$ from holomorphic modular bootstrap (\cite{Kaidi:2021ent}, Table 5) falling into the type $(A_4)_1$. The $c=20$ theory has $h_i=0,\frac{7}{5},\frac{8}{5}$ and $m_1=120$. We find this theory is dual to  $(A_4)_1$ w.r.t $c=24$, thus the degeneracy is also $(1,2,2)$. 
This allows us to determine the multiplicities of the primaries of weights $\frac{7}{5},\frac{8}{5}$ to be 2500 and 8125. The bilinear relation of characters gives $J+144$.
We also find the $c=12$ theory with $h_i=0,\frac{3}{5},\frac{7}{5}$ and $m_1=222$ is the dual of $(A_4)_1$ w.r.t $c=16$. Thus it again has degeneracy $(1,2,2)$. This coset allows us to determine the multiplicities of the primaries of weights $\frac{3}{5},\frac{7}{5}$ to be 25 and 1275.

\section{RCFTs with four characters}\label{sec:4chi}
The potential RCFTs with four characters, non-integral weights and $l=0$ MLDEs have been recently classified by modular bootstrap in  \cite{Kaidi:2021ent}. There are in total 72 theories, see Table 6 therein. We propose that all these 72 theories can be merely generated by 13 simple theories which will be discussed individually in the following subsections. In other words, there are in total 13 types from the viewpoint of Hecke relation. When two of the 13 initial theories satisfy the \emph{first} condition of generalized Hecke relation, we put them in the same subsection. Among the 13 types, types $(LY)_3,M_{12/5},M_{\rm eff}(5,3)$ have four primaries, type $M_{\rm eff}(14,3)$ has five primaries, types $M_{4/3},M_{\rm sub}(6,5),M_{8/5},U(1)_3$ have six primaries, type $(A_6)_1$ has seven primaries, types $(LY)_1^{\otimes 3},(A_1)_1^{\otimes 3},M_{6/5}$ have eight primaries, type $M_2$ has 12 primaries.

RCFTs with four characters and integral weights include $(A_2)_1^{\otimes 3}$ which has central charge $c=6$, conductor $N=12$ and weights and degeneracy $h_i=0,(\frac{1}{3})_6,(\frac{2}{3})_{12},(1)_{8},$ and $(D_4)_1^{\otimes 3}$ theory with $c=12$, $N=2$ and $h_i=0,(\frac{1}{2})_9,(1)_{27},(\frac{3}{2})_{27}.$ We also compute the Hecke images of these two theories and find all $\T_p,p>1$ have $l\ge 6$ MLDEs, thus are not of our main interest here.

\subsection{Type $(LY)_3$ and type $M_{4/3}$ }\label{sec:LY3}
Minimal model $M(9,2)$ is a non-unitary RCFT with central charge $c=-\frac{46}{3}$ and conformal weights $h_i=0,-\frac13,-\frac59,-\frac23$. The effective theory $M_{\rm eff}(9,2)$, i.e., $(LY)_3$ has $c_{\rm eff}=\frac23$ and $h^{\rm eff}_i=0,\frac{1}{9},\frac{1}{3},\frac{2}{3}$. The conductor of $(LY)_3$ is $N=36$. The $S$-matrix of $(LY)_3$ is
\be 
\rho(S)= \frac{1}{3}\left(
\begin{array}{cccc}
 2 \cos \left(\frac{\pi }{18}\right) & \sqrt{3} & 2 \sin \left(\frac{2 \pi }{9}\right) & 2 \sin \left(\frac{\pi }{9}\right) \\
 \sqrt{3} & 0 & -\sqrt{3} & -\sqrt{3} \\
 2 \sin \left(\frac{2 \pi }{9}\right) & -\sqrt{3} & -2 \sin \left(\frac{\pi }{9}\right) & 2 \cos \left(\frac{\pi }{18}\right) \\
 2 \sin \left(\frac{\pi }{9}\right) & -\sqrt{3} & 2 \cos \left(\frac{\pi }{18}\right) & -2 \sin \left(\frac{2 \pi }{9}\right) \\
\end{array}
\right).
\ee
Various Hecke images $\T_p$ of $(LY)_3$ for $p$ coprime to 36 have been discussed in \cite{Harvey:2019qzs}. For example, $\T_7(LY)_3=(G_2)_2$ and $\T_{29}(LY)_3$ describes a subtheory of $(C_5)_3\otimes(A_1)_1$. Together these two images $\T_7$ and $\T_{29}$ form a $c=24$ theory in the Schellekens' list No.21, of which the associated holomorphic VOA was constructed in \cite{vanEkeren:2017scl}.
We compute all admissible Hecke images of $(LY)_3$ for $p<36$ and summarize them in $c=24$ pairs in Table \ref{tb:HeckeM29}. We also consider the dual of original minimal model $M(9,2)$ w.r.t $c=24$, which is the $\T_{59}$ Hecke image with weights $0,\frac{7}{3},\frac{23}{9},\frac{8}{3}$. We find the vacuum of the $\T_{59}$ image has quasi-character,
while the rest three primaries have ordinary characters. We find the bilinear relation of characters between $M(9,2)$ and $\T_{59}$ is
\be\label{1183id}
\ba
J(\tau)=&\, -\chi_0^{\T_{59}}\chi_0^{M(9,2)}+\chi_{\frac{7}{3}}^{\T_{59}}\chi_{-\frac13}^{M(9,2)}-\chi_{\frac{23}{9}}^{\T_{59}}\chi_{-\frac59}^{M(9,2)}+\chi_{\frac{8}{3}}^{\T_{59}}\chi_{-\frac{2}{3}}^{M(9,2)}.
\ea
\ee

Beside those in Table \ref{tb:HeckeM29}, we find there are two more theories from holomorphic modular bootstrap (\cite{Kaidi:2021ent}, Table 6) falling into type $(LY)_3$. One has $c=\frac{58}{3}$, $h_i=0,\frac{2}{3},\frac{4}{3},\frac{20}{9}$ and $m_1=638$. We find it pairs with $\T_7(LY)_3$ with bilinear relation of characters equal to $J+1464 $. The other one has $c=\frac{70}{3}$, $h_i=0,\frac{2}{3},\frac{4}{3},\frac{26}{9}$ and $m_1=2730$. We find it pairs with $(LY)_3$ with bilinear relation of characters equal to $J+2976$.

\begin{table}[ht]
\def\arraystretch{1.1}
	\centering
	\begin{tabular}{|c|c|c|c|c|c|c|c|c|c|c|c|c|c|}
		\hline
		 \multicolumn{4}{|c|}{$l=0$} & \multicolumn{4}{c|}{$l'=0$} & \multicolumn{2}{c|}{duality}\\
		\hline
		 $c$ & $\!h_i\!$  & $m_1$ & remark & $\tc$ & $\tilh_i$ &  $\tm_1$ & remark & $\!m_1\!+\!\tm_1\!$ & Sch\\
		\hline
    $\!\!-\frac{46}{3}\!\!$  &  $\!\!-\{\frac23,\frac59,\frac13\}\!\!$ & $ 0$ & $\T_{-23}$ & $\!\frac{118}{3}\!$ & $ \frac73,\frac{23}{9},\frac{8}{3}   $ & $ 0$ & $\T_{59},\star $ & $0 $ & $0$\\
     $\frac{2}{3}$  &  $ \frac{1}{9},\frac{1}{3},\frac{2}{3} $ & $ 1$ & $\T_{1}$ & $\frac{ 70}{3}$ & $  \frac{4}{3},\frac{5}{3},\frac{17}{9}  $ & $ 35$ & $\T_{35} $ & $36 $ & $-$\\
$\frac{10}{3}$  &  $ \frac{1}{3},\frac{5}{9},\frac{2}{3} $ & $15 $ & $\T_{5}$ & $\frac{ 62}{3}$ & $  \frac{4}{3},\frac{13}{9},\frac{5}{3}  $ & $ 93$ & $\T_{31} $ & $ 108$ & $-$\\
                $\frac{14}{3}$  &  $ \frac{1}{3},\frac{2}{3},\frac{7}{9} $ & $14 $ & $\!\T_{7},(G_2)_2\!$ & $\frac{58 }{3}$ & $ \frac{11}{9},\frac{4}{3},\frac{5}{3}   $ & $ 58$ & $\!\T_{29}, (C_5)_3\otimes(A_1)_1\!$ & $ 72$ & $21$\\
    $\frac{22}{3}$  &  $ \frac{1}{3},\frac{2}{3},\frac{11}{9} $ & $44 $ & $\T_{11},\star$ & $\frac{ 50}{3}$ & $  \frac{7}{9},\frac{4}{3},\frac{5}{3}  $ & $ 100$ & $\T_{25},\star $ & $ 144$ & $-$\\
    $\frac{26}{3}$  &  $ \frac{4}{9},\frac{2}{3},\frac{4}{3} $ & $ 130$ & $\T_{13},\star$ & $\frac{46 }{3}$ & $ \frac{2}{3},\frac{4}{3},\frac{14}{9}   $ & $230 $ & $\T_{23},\star $ & $360 $ & $-$\\
        $\frac{34}{3}$  &  $\frac{2}{3},\frac{8}{9},\frac{4}{3}  $ & $306 $ & $\T_{17},\star$ & $\frac{38 }{3}$ & $  \frac{2}{3},\frac{10}{9},\frac{4}{3}  $ & $342 $ & $\T_{19},\star $ & $648 $ & $-$\\
		\hline
	
		\end{tabular}
			\caption{Hecke images of $(LY)_3$.}
			\label{tb:HeckeM29}
		\end{table}
		
Consider the $c=\frac{4}{3}$ theory with weights $h_i=0,\frac{2}{9},\frac13,\frac23$ appeared in holomorphic modular bootstrap (\cite{Kaidi:2021ent}, Table 6). We denote this theory as $M_{4/3}$ and find it is possible to express its four characters as degree 2 homogeneous polynomials of $(LY)_3$ characters:
\be\label{rel4}
\ba
\chi_0^{M_{4/3}}=&\,\chi_0^2+2\chi_{\frac13}\chi_{\frac23}=q^{-\frac{1}{18}}(1 + 4 q + 7 q^2 + 14 q^3 + 26 q^4 + 44 q^5 +\dots),\\
\chi_{2/9}^{M_{4/3}}=&\, \chi_{\frac19}^2=q^{\frac16}(1 + 2 q + 5 q^2 + 8 q^3 + 16 q^4 + 26 q^5+44 q^6+\dots),\\
\chi_{1/3}^{M_{4/3}}=&\,2\chi_{0}\chi_{\frac13}- \chi_{\frac23}^2=q^{\frac{5}{18}}(2 + 3 q + 8 q^2 + 14 q^3 + 26 q^4 + 41 q^5+\dots),\\
\chi_{2/3}^{M_{4/3}}=&\, 2\chi_{0}\chi_{\frac23}- \chi_{\frac13}^2=q^{\frac{11}{18}}(1 + 3 q^2 + 4 q^3 + 7 q^4 + 10 q^5 + 20 q^6+\dots).
\ea
\ee
Note the conductor is $N=18$. The $S$-matrix of these four characters is
\be 
\rho(S)^{M_{8/3}}= \frac{1}{3} \left(
\begin{array}{cccc}
 2 \cos \left(\frac{\pi }{9}\right) & 3 & 2 \cos \left(\frac{2 \pi }{9}\right) & 2 \sin \left(\frac{\pi }{18}\right) \\
 1 & 0 & -1 & -1 \\
 2 \cos \left(\frac{2 \pi }{9}\right) & -3 & -2 \sin \left(\frac{\pi }{18}\right) & 2 \cos \left(\frac{\pi }{9}\right) \\
 2 \sin \left(\frac{\pi }{18}\right) & -3 & 2 \cos \left(\frac{\pi }{9}\right) & -2 \cos \left(\frac{2 \pi }{9}\right) \\
\end{array}
\right).
\ee 
We find the primary with weight $2/9$ actually has degeneracy 3. Since the degeneracy of $M_{4/3}$ is different from the one of $(LY)_3$, we do \emph{not} regard it as the generalized $\T_2$ of $(LY)_3$. Consider all admissible Hecke images of $M_{4/3}$ theory. We find there exist three classes for the Hecke operation: for $p=1, 17\textrm{ mod } 18$, i.e., $p^2\equiv 1 \textrm{ mod } 18$, $
\rho(\sigma_{p})=\mathrm{Id},$ 
\be
\textrm{for }p=5,13 \textrm{ mod } 18,\textrm{ i.e., } p^2\equiv 7 \textrm{ mod } 18,\quad
\rho(\sigma_p)= \left(
\begin{array}{cccc}
 0 & 0 & -1 & 0 \\
 0 & 1 & 0 & 0 \\
 0 & 0 & 0 & 1 \\
 -1 & 0 & 0 & 0 \\
\end{array}
\right),
\ee
\be
\textrm{for }p=7, 11 \textrm{ mod } 18,\textrm{ i.e., } p^2\equiv 13 \textrm{ mod } 18,\quad
\rho(\sigma_p)=\left(
\begin{array}{cccc}
 0 & 0 & 0 & -1 \\
 0 & 1 & 0 & 0 \\
 -1 & 0 & 0 & 0 \\
 0 & 0 & 1 & 0 \\
\end{array}
\right) .
\ee
We collect all admissible $\T_p$ of $M_{4/3}$ for $p<18$ in $c=24$ pairs in Table \ref{tb:HeckeM29gen}. 

\begin{table}[ht]
\def\arraystretch{1.1}
	\centering
	\begin{tabular}{|c|c|c|c|c|c|c|c|c|c|c|c|c|c|}
		\hline
		 \multicolumn{4}{|c|}{$l=0$} & \multicolumn{4}{c|}{$l'=0$} & \multicolumn{2}{c|}{duality}\\
		\hline
		 $c$ & $\!h_i\!$  & $m_1$ & remark & $\tc$ & $\tilh_i$ &  $\tm_1$ & remark & $\!m_1\!+\!\tm_1\!$ & Sch\\
		\hline
  $\frac{4}{3}$  &  $(\frac{2}{9})_3,\frac13,\frac23  $ & $ 4$ & $\T_{1}$ & $\frac{68 }{3}$ & $\frac43,\frac53,(\frac{16}{9} )_3   $ & $ 68$ & $\T_{17} $ & $72 $ & $-$\\
$\frac{8}{3}$  &  $\frac13,(\frac49)_3,\frac23  $ & $ 12$ & $\T_{2}$ & $\frac{64 }{3}$ & $   \frac43,(\frac{14}{9})_3,\frac53 $ & $ 96$ & $\T_{16} $ & $108 $ & $-$\\
$\frac{16}{3}$  &  $\frac13,\frac23,(\frac89)_3  $ & $ 8$ & $\T_{4},(A_2)_6$ & $\frac{56 }{3}$ & $   (\frac{10}{9})_3,\frac43,\frac53 $ & $28 $ & $\T_{14},(D_4)_{12} $ & $36 $ & $3$\\
$\frac{20}{3}$  &  $ \frac{1}{3},\frac{2}{3},(\frac{10}{9})_3 $ & $ 20$ & $\T_{5},\star$ & $\frac{52 }{3}$ & $  (\frac{8}{9})_3,\frac{4}{3},\frac{5}{3}  $ & $52 $ & $\T_{13},\star $ & $ 72$ & $-$\\
$\frac{28}{3}$  &  $ (\frac{5}{9})_3,\frac{2}{3},\frac{4}{3} $ & $ 182$ & $\T_{7},\star$ & $\frac{44 }{3}$ & $  \frac{2}{3},\frac{4}{3},(\frac{13}{9})_3  $ & $ 286$ & $\T_{11},\star $ & $ 468$ & $-$\\
$\frac{32}{3}$  &  $ \frac{2}{3},(\frac{7}{9})_9,\frac{4}{3} $ & $ 272$ & $\T_{8},\star$ & $\frac{ 40}{3}$ & $   \frac{2}{3},(\frac{11}{9})_9,\frac{4}{3} $ & $340 $ & $\T_{10},\star $ & $612 $ & $-$\\		\hline
	
		\end{tabular}
			\caption{(Generalized) Hecke images of $M_{4/3}$. For $p=2k$, we regard them as generalized Hecke images $\T_kM_{8/3}$.}
			\label{tb:HeckeM29gen}
		\end{table}

Consider the $c=\frac{8}{3}$ theory with weights $h_i=0,\frac13,\frac{4}{9},\frac23$ appeared in holomorphic modular bootstrap (\cite{Kaidi:2021ent}, Table 5), which we denote as $M_{8/3}$. We compute the four distinct characters and find it is possible to express them as degree 2 homogeneous polynomials of $M_{4/3}$ characters:
\be
\ba
\chi_0^{M_{8/3}}=&\,(\chi_0^{M_{4/3}})^2+2\chi_{\frac13}^{M_{4/3}}\chi_{\frac23}^{M_{4/3}}=q^{-\frac{1}{9}}( 1 + 12 q + 36 q^2 + 112 q^3 + 275 q^4 +\dots),\\
\chi_{4/9}^{M_{8/3}}=&\,3 (\chi_{\frac29}^{M_{4/3}})^2=q^{\frac13}( 3 + 12 q + 42 q^2 + 108 q^3 + 267 q^4 + 588 q^5+\dots),\\
\chi_{1/3}^{M_{8/3}}=&\,2\chi_{0}^{M_{4/3}}\chi_{\frac13}^{M_{4/3}}- (\chi_{\frac23}^{M_{4/3}})^2=q^{\frac{2}{9}}(4 + 21 q + 68 q^2 + 184 q^3 + 456 q^4 +\dots),\\
\chi_{2/3}^{M_{8/3}}=&\,  (\chi_{\frac13}^{M_{4/3}})^2-2\chi_{0}^{M_{4/3}}\chi_{\frac23}^{M_{4/3}}=q^{\frac{5}{9}}(2 + 4 q + 21 q^2 + 44 q^3 + 112 q^4 +\dots).
\ea
\ee
Here the conductor is $N=9$. Note the character relation here resembles the character relation \eqref{rel4} between $(LY)_3$ and $M_{4/3}$, except the last line has an opposite sign. It is easy to find the $S$-matrix of $M_{8/3}$ is
\be 
\rho(S)^{M_{8/3}}= \frac{1}{3}  \left(
\begin{array}{cccc}
 2 \cos \left(\frac{2 \pi }{9}\right) & 3 & 2 \cos \left(\frac{\pi }{9}\right) & 2 \sin \left(\frac{\pi }{18}\right) \\
 1 & 0 & -1 & 1 \\
 2 \cos \left(\frac{\pi }{9}\right) & -3 & 2 \sin \left(\frac{\pi }{18}\right) & -2 \cos \left(\frac{2 \pi }{9}\right) \\
 2 \sin \left(\frac{\pi }{18}\right) & 3 & -2 \cos \left(\frac{2 \pi }{9}\right) & -2 \cos \left(\frac{\pi }{9}\right) \\
\end{array}
\right).
\ee 
The weight-$4/9$ primary of this theory has degeneracy 3. Given the same degeneracy and the homogeneous expressions, we
regard $M_{8/3}$ as the generalized $\T_2$ Hecke image of $M_{4/3}$. 
Next consider all admissible Hecke images of $M_{8/3}$ theory. We find there exist three classes for the Hecke operation: for $p=1, 8\textrm{ mod } 9$, i.e., $p^2\equiv 1 \textrm{ mod } 9$, $
\rho(\sigma_{p})=\mathrm{Id},$ 
\be
\textrm{for }p=2,7 \textrm{ mod } 9,\textrm{ i.e., } p^2\equiv 4 \textrm{ mod } 9,\quad
\rho(\sigma_p)= \left(
\begin{array}{cccc}
 0 & 0 & -1 & 0 \\
 0 & 1 & 0 & 0 \\
 0 & 0 & 0 & -1 \\
 1 & 0 & 0 & 0 \\
\end{array}
\right) ,
\ee
\be
\textrm{for }p=4, 5 \textrm{ mod } 9,\textrm{ i.e., } p^2\equiv 7 \textrm{ mod } 9,\quad
\rho(\sigma_p)=\left(
\begin{array}{cccc}
 0 & 0 & 0 & 1 \\
 0 & 1 & 0 & 0 \\
 -1 & 0 & 0 & 0 \\
 0 & 0 & -1 & 0 \\
\end{array}
\right)  .
\ee
We compute all admissible $\T_p$ of $M_{8/3}$ for $p<9$ and find $\T_2M_{8/3} $ describe a subtheory of WZW $(A_2)_6$, while $\T_7M_{8/3} $ describe a subtheory of WZW $(D_4)_{12}$. For example, we find the following relation between the $\T_2M_{8/3} $ image and $(A_2)_6$ characters:
\be
\ba
\chi^{\T_2M_{8/3}}_0=&\,\chi_0^{(A_2)_6,6w_0}+2\chi_2^{(A_2)_6,6w_1}=q^{-\frac{2}{9}}(1 + 8 q + 100 q^2 + 480 q^3 + 2020 q^4 +\dots),\\
\chi^{\T_2M_{8/3}}_{\frac13}=&\,\chi_{\frac13}^{(A_2)_6,w_1+w_2+4w_0}+2\chi_{\frac43}^{(A_2)_6,4w_1+w_2+w_0}=q^{\frac{1}{9}}(8 + 134 q + 912 q^2 + \dots),\\
\chi^{\T_2M_{8/3}}_{\frac23}=&\,2\chi_{\frac23}^{(A_2)_6,3w_1+3w_0}+\chi_{\frac53}^{(A_2)_6,3w_1+3w_2}=q^{\frac{4}{9}}(20 + 224 q + 1267 q^2   +\dots),\\
\chi^{\T_2M_{8/3}}_{\frac89}=&\,\chi_{\frac89}^{(A_2)_6,2w_1+2w_2+2w_0}=q^{\frac{2}{3}}( 27 + 216 q + 1188 q^2 + 4968 q^3+\dots).
\ea
\ee
Here the primary with weight $8/9$ has degeneracy three. This subtheory is actually known as the $D_6$ type modular invariant of $\hat{A}_2$ \cite{Gannon:1992ty}. Together $\T_2M_{8/3} $ and $\T_7M_{8/3}$ form a $c=24$ theory in Schellekens' List No.3, with the associated holomorphic VOA  constructed in \cite{vanEkeren:2017scl} by a $\IZ_6$ orbifold of the Niemeier
lattice $A_2^{12}$. We regard the Hecke images of $M_{8/3}$ as the generalized $\T_{2k}$ images of $M_{4/3}$ and list them in $c=24$ pairs in Table \ref{tb:HeckeM29gen}. Note the sums of spin-1 currents $m_1+\tm_1$ for all  pairs in Table \ref{tb:HeckeM29gen} are divisible by the conductor 18 of $M_{4/3}$.

Beside the Hecke images with characters in Table \ref{tb:HeckeM29gen}, we find there are two more theories from holomorphic modular bootstrap (\cite{Kaidi:2021ent}, Table 6) falling into type $M_{4/3}$. One has $c=\frac{56}{3}$, $h_i=0,\frac{2}{3},\frac{4}{3},\frac{19}{9}$ and $m_1= 420$. We find it pairs with $\T_4M_{4/3}$ with bilinear relation of characters equal to $J+1212$. The other one has $c=\frac{68}{3}$, $h_i=0,\frac{2}{3},\frac{4}{3},\frac{25}{9}$ and $m_1= 2278$. We find it pairs with $M_{4/3}$ with bilinear relation of characters equal to $J+2724$. Note these two theories should have degeneracy $(1,3,1,1)$ automatically.

\subsection{Type $(LY)_1^{\otimes 3}$  and type $M_{12/5}$ }\label{sec:LY13}
The product theory $(LY)_1^{\otimes 3}$ has central charge $\frac{6}{5}$ and conformal weights with degeneracy $0,(\frac{1}{5})_3,(\frac25)_3,\frac35$. 
The four distinct characters can be easily computed from those of $(LY)_1$.
The conductor $N=20$. We compute the Hecke images $\T_p$ of $(LY)_1^{\otimes 3}$ like in Section \ref{sec:LY12} and
find Hecke relations such as the $\T_7$ image describes WZW $(G_2)_1^{\otimes 3}$, while the $\T_{13}$ image describes a subtheory of WZW $(E_6)_3$. More precisely, we observe the following relation between the $\T_{13}$ image and $(E_6)_3$ characters:
\be
\ba
\chi^{\T_{13}}_0=&\,\chi_0^{(E_6)_3,3w_0}+2\chi_2^{(E_6)_3,3w_1}=q^{-\frac{13}{20}}(1 + 78 q + 9165 q^2+263926 q^3+\dots),\\
\chi^{\T_{13}}_{\frac85}=&\,\chi_{\frac85}^{(E_6)_3,w_4}=q^{\frac{19}{20}}(2925 + 122550 q + 2340325 q^2+\dots),\\
\chi^{\T_{13}}_{\frac65}=&\,\chi_{\frac65}^{(E_6)_3,w_1+w_6+w_0}=q^{\frac{19}{20}}(650 + 50700 q + 1241175 q^2+\dots),\\
\chi^{\T_{13}}_{\frac45}=&\,\chi_{\frac45}^{(E_6)_3,w_2+w_0}+2\chi_{\frac95}^{(E_6)_3,w_1+w_3}=q^{\frac{3}{20}}(78 + 17732 q + 606684 q^2 +\dots).
\ea
\ee
Here the primaries with weights ${8}/{5}$ and ${6}/{5}$ have degeneracy 3 as required by the Hecke operation. Together $\T_7$ and $\T_{13}$ form a $c=24$ theory appearing in the Schellekens' List No.32. The construction of holomorphic VOA of this $c=24$ theory was given in \cite{Miyamoto} by a $\IZ_3$ orbifold from the Niemeier lattice $E_6^4$, see also \cite{Sagaki}. 
There are in total four pairs of Hecke images $(\T_p,\T_{p'})$ w.r.t $c=24$ which satisfy $p+p'=20$. We summarize all admissible Hecke images $\T_p$ for $p<20$ in pairs in Table \ref{tb:HeckeYL3}. Note the sums of spin-1 currents $m_1+\tm_1$ for all  pairs in Table \ref{tb:HeckeYL3} are divisible by the conductor 20.

\begin{table}[ht]
\def\arraystretch{1.1}
	\centering
	\begin{tabular}{|c|c|c|c|c|c|c|c|c|c|c|c|c|c|}
		\hline
		 \multicolumn{4}{|c|}{$l=0$} & \multicolumn{4}{c|}{$l'=0$} & \multicolumn{2}{c|}{duality}\\
		\hline
		 $c$ & $h_i$  & $m_1$ & remark & $\tc$ & $\tilh_i$ &  $\tm_1$ & remark & $\!\!m_1\!+\!\tm_1\!\!$ & \!Sch\!\\
		\hline
 $\frac{6}{5}$  &  ${(\frac{1}{5})_3,(\frac{2}{5})_3,\frac{3}{5}} $ & $3 $ & $\T_1$ & $\frac{144}{5}$ & ${ \frac{7}{5},(\frac{8}{5})_3,(\frac{9}{5})_3}$ & $ 57$ & $\T_{19} $ & $60 $ & $ -$ \\
 $\frac{18}{5}$  &  ${(\frac{1}{5})_3,(\frac{3}{5})_3,\frac{4}{5}} $ & $ 9$ & $\T_{3},\star$ & $\frac{102}{5}$ & ${\frac{6}{5},(\frac{7}{5})_3,(\frac{9}{5})_3 }$ & $51 $ & $\T_{17},\star $ & $ 60$ & $-$\\
  $\frac{42}{5}$  &  ${(\frac{2}{5})_3,(\frac{4}{5})_3,\frac{6}{5}} $ & $42 $ & $\!\T_{7},(G_2)_1^{\otimes 3}\!$ & $\frac{78}{5}$ &  ${\frac{4}{5},(\frac{6}{5})_3,(\frac{8}{5})_3 }$ & $78 $ & $\!\T_{13},(E_6)_3\!$ & $ 120$ & $32$ \\
   $\frac{54}{5}$  &  ${(\frac{3}{5})_3,(\frac{4}{5})_3,\frac{7}{5}} $ & $ 270$ & $\T_{9},\star$ & $\frac{66}{5}$ & ${\frac{3}{5},(\frac{6}{5})_3,(\frac{7}{5})_3 }$ & $ 330$ & $\T_{11},\star $ & $ 600$ & $-$\\
		\hline
	
		\end{tabular}
			\caption{Hecke images of $(YL)_1^{\otimes 3}$.}
			\label{tb:HeckeYL3}
		\end{table}
		
Beside the Hecke images with ordinary characters in Table \ref{tb:HeckeYL3},
we notice there exists two more theories from holomorphic modular bootstrap (\cite{Kaidi:2021ent}, Table 6) falling into type $(YL)_1^{\otimes 3}$. 
The $c=\frac{78}{5}$ theory with $h_i=0,\frac{3}{5},\frac{6}{5},\frac{9}{5}$ and $m_1=156$ is just $(F_{4})_1^{\otimes 3} $. 
Naturally it forms a $c=24$ pair with $\T_7(YL)_1^{\otimes 3} $, of which the bilinear relation of characters gives $J+ 744$.
The $c=\frac{114}{5}$ theory with $h_i=0,\frac{4}{5},\frac{8}{5},\frac{12}{5}$ and $m_1=570$ is just $(E_{7\frac12})_1^{\otimes 3} $ which  
similarly forms a $c=24$ pair with $(YL)_1^{\otimes 3} $.
		
Now let us consider a $c=\frac{12}{5}$ RCFT appeared in holomorphic modular bootstrap (\cite{Kaidi:2021ent}, Table 6) which has weights $h_i=0,\frac{1}{5},\frac{2}{5},\frac{4}{5}$. We denote this theory as $M_{{12}/{5}}$. Naively, it looks like a generalized $\T_2$ image of $(LY)_1^{\otimes 3}$, as the first condition of generalized Hecke relation is satisfied. However, we will show that their degeneracies do not match. It is easy to compute the four characters from MLDE and we determine the following exact formulas:
\be
\ba
&\chi_0^{M_{12/5}}=\phi_1^6+2\phi_1\phi_2^5 ,\quad
\chi_{1/5}^{M_{12/5}}=2\phi_1^5\phi_2-\phi_2^6 ,\\
&\chi_{2/5}^{M_{12/5}}=5\phi_1^4\phi_2^2 ,\qquad
\chi_{4/5}^{M_{12/5}}=5\phi_1^2\phi_2^4 .
\ea
\ee
The conductor of this theory is $N=10$. 
From the characters, we find this theory can actually be realized as the coset $(G_2)_1/(LY)_1$. Indeed, we checked the following character relations
\be
\chi^{(G_2)_1}_0=\chi^{M_{12/5}}_0\chi^{(LY)_1}_0+\chi^{M_{12/5}}_{4/5}\chi^{(LY)_1}_{1/5},\quad \chi^{(G_2)_1}_{2/5}=\chi^{M_{12/5}}_{2/5}\chi^{(LY)_1}_0+\chi^{M_{12/5}}_{1/5}\chi^{(LY)_1}_{1/5}.
\ee
This suggests that the degeneracy of $M_{{12}/{5}}$ theory is actually $(1,1,1,1)$. From $\rho(S)^{(LY)_1}$, we compute the $S$-matrix of $M_{{12}/{5}}$ as
\be \label{S152}
\rho(S)^{M_{{12}/{5}}}= \frac{1}{\sqrt{5}}  \left(
\begin{array}{cccc}
 1 & 1 & \frac{1}{2} \left(\sqrt{5}+1\right) & \frac{1}{2} \left(\sqrt{5}-1\right) \\
 1 & 1 & \frac{1}{2} \left(1-\sqrt{5}\right) & \frac{1}{2} \left(-\sqrt{5}-1\right) \\
 \frac{1}{2} \left(\sqrt{5}+1\right) & \frac{1}{2} \left(1-\sqrt{5}\right) & -1 & 1 \\
 \frac{1}{2} \left(\sqrt{5}-1\right) & \frac{1}{2} \left(-\sqrt{5}-1\right) & 1 & -1 \\
\end{array}
\right).
\ee 
With $S$-matrix at hand, we can further study its Hecke images. We find there exist two classes for the Hecke operation of $M_{{12}/{5}}$: for $p=1, 9\textrm{ mod } 10$, i.e., $p^2\equiv 1 \textrm{ mod } 10$, $
\rho(\sigma_{p})=\mathrm{Id},$ while for $p=3,7\textrm{ mod } 10$, i.e., $p^2\equiv 9 \textrm{ mod } 10$,
\be\label{sigma152}
\rho(\sigma_p)= \left(
\begin{array}{cccc}
 0 & -1 & 0 & 0 \\
 -1 & 0 & 0 & 0 \\
 0 & 0 & 0 & 1 \\
 0 & 0 & 1 & 0 \\
\end{array}
\right).
\ee
We compute Hecke images $\T_pM_{{12}/{5}}$ for $p=3,7,9$ and list them in $c=24$ pairs in Table \ref{tb:Heckec125}. All of them are non-unitary. 

\begin{table}[ht]
\def\arraystretch{1.1}
	\centering
	\begin{tabular}{|c|c|c|c|c|c|c|c|c|c|c|c|c|c|}
		\hline
		 \multicolumn{4}{|c|}{$l=0$} & \multicolumn{4}{c|}{$l'=0$} & \multicolumn{2}{c|}{duality}\\
		\hline
		 $c$ & $h_i$  & $m_1$ & remark & $\tc$ & $\tilh_i$ &  $\tm_1$ & remark & $m_1+\tm_1$ & Sch\\
		\hline
    $\frac{12}{5}$  &  ${\frac15,\frac25,\frac45} $ & $ 8$ & $\T_{1}$ & $\frac{108}{5}$ & ${\frac65,\frac{8}{5},\frac{9}{5} }$ & $ 72$ & $\T_{9} $ & $ 80$ & $-$ \\
$\frac{24}{5}$  &  ${\frac25,\frac35,\frac45} $ & $ 36$ & $\T_{2}$ & $\frac{96}{5}$ & ${\frac65,\frac75,\frac85 }$ & $ 144$ & $\T_{8} $ & $180 $ & $-$\\
$\frac{36}{5}$  &  $ \frac{2}{5},\frac{3}{5},\frac{6}{5} $ & $96 $ & $\T_{3},\star$ & $\frac{84}{5}$ & ${\frac{4}{5},\frac{7}{5},\frac{8}{5} }$ & $224 $ & $\T_{7},\star $ & $320 $ & $-$\\
$\frac{48}{5}$  &  ${\frac{3}{5},\frac{4}{5},\frac{6}{5}} $ & $ 168$ & $\T_{4},\star$ & $\frac{72}{5}$ & ${\frac{4}{5},\frac{6}{5},\frac{7}{5} }$ & $252 $ & $\T_{6},\star $ & $420 $ & $-$\\
		\hline
	
		\end{tabular}
			\caption{(Generalized) Hecke images of $M_{12/5}$ theory. For $\T_{2k}$, we regard them as generalized Hecke images $\T_kM_{24/5}$.}
			\label{tb:Heckec125}
		\end{table}

Let us further consider a $c=\frac{24}{5}$ RCFT appeared in holomorphic modular bootstrap (\cite{Kaidi:2021ent}, Table 6) which has weights $h_i=0,\frac{2}{5},\frac{3}{5},\frac{4}{5}$. We denote this theory as $M_{{24}/{5}}$. We find it can be regarded as generalized Hecke image $\T_2$ of $M_{{12}/{5}}$. 
We compute the four characters of $M_{{24}/{5}}$ from MLDE and determine the exact formulas are
\be
\ba
&\chi_0^{M_{24/5}}=\phi_1^{12} + 24 \phi_1^7 \phi_2^5 - 6 \phi_1^2 \phi_2^{10}
  ,\quad
\chi_{2/5}^{M_{24/5}}= 6 \phi_1^{10} \phi_2^2 + 24 \phi_1^5 \phi_2^7 - \phi_2^{12},\\
&\chi_{3/5}^{M_{24/5}}=5( 4\phi_{1}^9\phi_{2}^3+3\phi_{1}^4\phi_{2}^8) ,\qquad\ \ \ \ 
\chi_{4/5}^{M_{24/5}}= 5(3 \phi_1^8 \phi_2^4 - 4 \phi_1^3 \phi_2^9).
\ea
\ee
Here the conductor $N=5$. We observe this theory can be realized as the coset $(F_4)_1/(LY)_1$, with character relations
\be
\chi^{(F_4)_1}_0=\chi^{M_{24/5}}_0\chi^{(LY)_1}_0+\chi^{M_{24/5}}_{4/5}\chi^{(LY)_1}_{\frac15},\quad \chi^{(F_4)_1}_{3/5}=\chi^{M_{24/5}}_{3/5}\chi^{(LY)_1}_0+\chi^{M_{24/5}}_{\frac25}\chi^{(LY)_1}_{1/5}.
\ee
Therefore, the degeneracy is $(1,1,1,1)$, the same with $M_{{12}/{5}}$. We also find the $S $-matrix of $M_{{24}/{5}}$ is exactly the same with the one \eqref{S152} of $M_{{12}/{5}}$.
The Hecke operation of $M_{{24}/{5}}$ also resembles the one of $M_{{12}/{5}}$. We find for $p=1, 4\textrm{ mod } 5$, i.e., $p^2\equiv 1 \textrm{ mod } 5$, $
\rho(\sigma_{p})=\mathrm{Id},$ while for $p=2,3\textrm{ mod } 5$, i.e., $p^2\equiv 4 \textrm{ mod } 10$, $\rho(\sigma_{p})$ is the same with \eqref{sigma152}. We collect the Hecke images $\T_k$ of $M_{{24}/{5}}$ as the generalized $\T_{2k}$ images of $M_{{12}/{5}}$, and list them in Table \ref{tb:Heckec125} as well. We can see the sums of spin-1 currents $m_1+\tm_1$ for all  pairs in Table \ref{tb:Heckec125} are divisible by the conductor 10 of $M_{12/5}$.

Beside the Hecke images with ordinary characters in Table \ref{tb:Heckec125},
we find there exist two more theories from holomorphic modular bootstrap (\cite{Kaidi:2021ent}, Table 5) falling into type $M_{{12}/{5}}$. The $c=\frac{96}{5}$ theory with $h_i=0,\frac{3}{5},\frac{7}{5},\frac{11}{5}$ and $m_1=276$ can pair with $M_{24/5}$, of which the bilinear relation of characters gives $J+444$. The $c=\frac{108 }{5}$ theory with $h_i=0,\frac{4}{5},\frac85,\frac{11}{5} $ and $m_1=204$ can pair with $ M_{12/5}$, of which the bilinear relation of characters gives $J+344$.

\subsection{Type $M_{\rm eff}(5,3)$ and type $M_{6/5}$ }\label{sec:M53}
Minimal model $M(5,3)$ is non-unitary with central charge $c=-\frac{3}{5}$ and conformal weights $h_i=0,-\frac{1}{20},\frac15,\frac34$. While the effective theory $M_{\rm eff}(5,3)$ has
$c_{\rm eff}=\frac{3}{5}$ and $h_i^{\rm eff}=0,\frac{1}{20},\frac{1}{4},\frac{4}{5}$. The conductor $N=40$. Both $M(5,3)$ and $M_{\rm eff}(5,3)$ are well known as the coset theories between WZW models $(LY)_1$ and $(A_1)_1$ or between WZW models $(E_7)_1$ and $(E_{7\frac{1}{2}})_1$, see for example \cite{Kawasetsu}. To be precise,
\be
\frac{(E_{7})_1}{(E_{7\frac{1}{2}})_1}=\frac{(LY)_1}{(A_1)_1}=M_{(5,3)},\quad\mathrm{and}\quad \frac{(E_{7\frac{1}{2}})_1}{(E_7)_1}=\frac{(A_1)_1}{(LY)_1}=M_{\rm eff}(5,3).
\ee
Various Hecke images $\T_p$ of $M_{\rm eff}(5,3)$ have been discussed in \cite{Harvey:2019qzs} for $p$ coprime to $40$, for example $\T_3=(A_1)_3$ and $\T_{7}=(C_3)_1$. We compute all Hecke images for admissible $p\le 41$ and collect them in $c=24$ pairs in Table \ref{tb:HeckeM35}. All these Hecke images have $l=0$ MLDEs. The $c=24$ pairs have $p+p'=40$. We can see from Table \ref{tb:HeckeM35} that the sum of spin-1 currents $m_1+\tm_1$ for all pairs are divisible by the conductor $40$.

\begin{table}[ht]
\def\arraystretch{1.1}
	\centering
	\begin{tabular}{|c|c|c|c|c|c|c|c|c|c|c|c|c|c|}
		\hline
		 \multicolumn{4}{|c|}{$l=0$} & \multicolumn{4}{c|}{$l'=0$} & \multicolumn{2}{c|}{duality}\\
		\hline
		 $c$ & $h_i$  & $m_1$ & remark & $\tc$ & $\tilh_i$ &  $\tm_1$ & remark & $m_1+\tm_1$ & Sch\\
		\hline
    $-\frac{3}{5}$  &  $-\frac{1}{20},\frac15,\frac34  $ & $ 0$ & $\T_{-1}$ & $\frac{ 123}{5}$ & $  \frac{5}{4},\frac{9}{5},\frac{41}{20}  $ & $0 $ & $\T_{41} $ & $ 0$ & $0$\\
$\frac{3}{5}$  &  $ \frac{1}{20},\frac{1}{4},\frac{4}{5} $ & $ 1$ & $\T_{1}$ & $\frac{117}{5}$ & $  \frac{6}{5},\frac{7}{4},\frac{39}{20}  $ & $39 $ & $\T_{39} $ & $ 40$ & $-$\\
$\frac{9}{5}$  &  $ \frac{3}{20},\frac{2}{5},\frac{3}{4} $ & $3 $ & $\T_{3},(A_1)_3$ & $\frac{111 }{5}$ & $  \frac{5}{4},\frac{8}{5},\frac{37}{20}  $ & $37 $ & $\T_{37} $ & $ 40$ & $-$\\
$\frac{21}{5}$  &  $ \frac{7}{20},\frac{3}{5},\frac{3}{4} $ & $21 $ & $\T_{7},(C_3)_1$ & $\frac{99 }{5}$ & $   \frac{5}{4},\frac{7}{5},\frac{33}{20} $ & $ 99$ & $\T_{33} $ & $ 120$ & $33$\\
$\frac{27}{5}$  &  $ \frac{1}{4},\frac{9}{20},\frac{6}{5} $ & $ 63$ & $\T_{9},\star$ & $\frac{ 93}{5}$ & $ \frac{4}{5},\frac{31}{20},\frac{7}{4}   $ & $217 $ & $\T_{31},\star $ & $280 $ & $-$\\
$\frac{33}{5}$  &  $  \frac{11}{20},\frac{3}{4},\frac{4}{5}$ & $99 $ & $\T_{11}$ & $\frac{87 }{5}$ & $ \frac{6}{5},\frac{5}{4},\frac{29}{20}   $ & $261 $ & $\T_{29} $ & $ 360$ & $-$\\
$\frac{39}{5}$  &  $\frac{2}{5},\frac{13}{20},\frac{5}{4}  $ & $ 91$ & $\T_{13},\star$ & $\frac{81 }{5}$ & $  \frac{3}{4},\frac{27}{20},\frac{8}{5}  $ & $189 $ & $\T_{27},\star $ & $ 280$ & $-$\\
$\frac{51}{5}$  &  $ \frac{3}{5},\frac{17}{20},\frac{5}{4} $ & $153 $ & $\T_{17},\star$ & $\frac{69 }{5}$ & $ \frac{3}{4},\frac{23}{20},\frac{7}{5}   $ & $207 $ & $\T_{23},\star $ & $ 360$ & $-$\\
$\frac{57}{5}$  &  $ \frac{3}{4},\frac{19}{20},\frac{6}{5} $ & $ 285$ & $\T_{19},\star$ & $\frac{ 63}{5}$ & $ \frac{4}{5},\frac{21}{20},\frac{5}{4}   $ & $315 $ & $\T_{21},\star $ & $ 600$ & $-$\\
		\hline
	
		\end{tabular}
			\caption{Hecke images of $M_{\rm eff}(5,3)$. We remark that although the $\T_{37}$ image has a positive integer Verlinde formula \cite{Kaidi:2021ent}, the pair $(\T_3,\T_{37})$ yields bilinear relation of characters equal to $J+40$, which does not appear in the Schellekens' list, nor satisfies the divisibility condition proposed in \cite{Lin:2021bcp} for $c=24$ theories. This suggests that the $\T_{37}$ image may not be unitary or physical theories.}
			\label{tb:HeckeM35}
		\end{table}

Consider a $c=\frac{123}{5}>24$ theory proposed in (\cite{Kaidi:2021ent}, Table 6) from holomorphic modular bootstrap, which has conformal weights $h_{1,2,3}=\frac{5}{4},\frac{9}{5},\frac{41}{20}$. We find this theory is actually just the $\T_{41}$ Hecke image of $M_{\rm eff}(5,3)$. 
This theory can pair with the original ${M(5,3)}$. Indeed we find the following bilinear relation of characters:
\be\label{1235id}
\ba
J(\tau)=&\, \chi_0^{\T_{41}}\chi_0^{M(5,3)}-\chi_{\frac{5}{4}}^{\T_{41}}\chi_{\frac34}^{M(5,3)}-\chi_{\frac{9}{5}}^{\T_{41}}\chi_{\frac15}^{M(5,3)}+\chi_{\frac{41}{20}}^{\T_{41}}\chi_{-\frac{1}{20}}^{M(5,3)}.
\ea
\ee

There exists one more theory from holomorphic modular bootstrap (\cite{Kaidi:2021ent}, Table 6) falling into type $M_{\rm eff}(5,3)$. We find the $c=\frac{117}{5}$ theory with $h_i=0,\frac{19}{20},\frac{7}{4},\frac{11}{5}$ and $m_1=325$ can pair with $M_{\rm eff}(5,3)$ with $n_i=0,1,2,3$, of which the bilinear relation of characters gives $J+ 612$.

An interesting $c=\frac{6}{5}$ RCFT was studied in \cite{lam2000z2} which has weights and degeneracy $h_i=0,(\frac{1}{10})_3,(\frac{1}{2})_3,\frac{3}{5} $.
This theory which we denote as $M_{6/5}$ can be viewed as a subtheory of $\textrm{Ising}\otimes M(5,4)$ and associated to the $D_{2\rm A}$ conjugacy class of the Monster group \cite{lam2000z2}. Although the central charge and conformal weights of $M_{6/5}$ satisfy the generalized Hecke $\T_2$ condition of $M_{\rm eff}(5,3)$, the degeneracy does not match. Therefore, we do \emph{not} regard $M_{6/5}$ as a type $M_{\rm eff}(5,3)$ theory. We compute the Fourier expansion of the four characters to be
\be
\ba
\chi_0=\,&q^{-\frac{1}{20}} (1 + 3 q^2 + 4 q^3 + 9 q^4 + 12 q^5 + 22 q^6 + 30 q^7  +\dots)
,\\
\chi_{\frac{1}{10}}=\,& q^{\frac{1}{20}} (  1 + 2 q + 4 q^2 + 8 q^3 + 13 q^4 + 22 q^5 + 35 q^6+\dots),\\
\chi_{\frac{1}{2}}=\,& q^{\frac{9}{20}} ( 1 + 2 q + 3 q^2 + 6 q^3 + 10 q^4 + 16 q^5 + 26 q^6+\dots),\\
\chi_{\frac{3}{5}}=\,& q^{\frac{11}{20}} ( 2 + 3 q + 6 q^2 + 9 q^3 + 18 q^4 + 27 q^5 + 44 q^6 +\dots).
\ea
\ee
Clearly, the conductor $N=20$. The modular $S$-matrix as well as the relation between these characters with those of Ising and $M(5,4)$ can be found in e.g. (\cite{Bae:2020pvv}, Section 3.2.2). We find there exist 4 classes for the Hecke operation: for $p$ mod 20,
\be
\rho(\sigma_{1,19})=-\rho(\sigma_{9,11})=\mathrm{Id},\ \textrm{ or }\  \rho(\sigma_{3,17})=-\rho(\sigma_{7,13})=\left(
\begin{array}{cccc}
 0 & 0 & 0 & -1 \\
 0 & 0 & 1 & 0 \\
 0 & -1 & 0 & 0 \\
 1 & 0 & 0 & 0 \\
\end{array}
\right).
\ee
We then compute the Hecke images $\T_p$ for all admissible $p<20$ and summarize them in $c=24$ pairs in Table \ref{tb:HeckeMc65}. Notably, it was observed in \cite{Bae:2020pvv} that the $\T_{19}$ Hecke image produces the characters of the RCFT associated to the Steinberg group $2^2.^2E_6(2)$. We find the $\T_{11},\T_{13},\T_{17},\T_{19}$ Hecke images may have emergent supersymmetry owing to the presence of weight-$3/2$ primaries. In particular, the fermionization of the $2^2.^2E_6(2)$ RCFT has been studied in \cite{Bae:2020xzl}. Note all theories in Table \ref{tb:HeckeMc65} automatically have degeneracy $(1,3,3,1)$ as required by the Hecke operation.

\begin{table}[ht]
\def\arraystretch{1.1}
	\centering
	\begin{tabular}{|c|c|c|c|c|c|c|c|c|c|c|c|c|c|}
		\hline
		 \multicolumn{4}{|c|}{$l=0$} & \multicolumn{4}{c|}{$l'=0$} & \multicolumn{2}{c|}{duality}\\
		\hline
		 $c$ & $h_i$  & $m_1$ & \!remark\! & $\tc$ & $\tilh_i$ &  $\tm_1$ & remark & $\!\!m_1\!+\!\tm_1\!\!$ & \!\!Sch\!\!\\
		\hline
$\frac{6}{5}$  &  $(\frac{1}{10})_3,(\frac{1}{2})_3,\frac{3}{5}  $ & $ 0$ & $\!\!\!\T_{1},D_{2\rm A}\!\!\!$ & $\frac{ 114}{5}$ & $\frac{7}{5},(\nb{\frac{3}{2}})_3,(\frac{19}{10})_3    $ & $ 0$ & $\!\!\T_{19},2^2.^2E_6(2) \!\!$ & $0 $ & $0$\\
$\frac{18}{5}$  &  $(\frac{3}{10})_3,(\frac{1}{2})_3,\frac{4}{5}  $ & $18 $ & $\T_{3}$ & $\frac{102 }{5}$ & $ \frac{6}{5},(\nb{\frac32})_3,(\frac{17}{10} )_3  $ & $102 $ & $\T_{17} $ & $ 120$ & $-$\\
$\frac{42}{5}$  &  $ (\frac{1}{2})_3,(\frac{7}{10})_3,\frac{6}{5} $ & $ 126$ & $\T_{7},\star$ & $\frac{ 78}{5}$ & $  \frac{4}{5},(\frac{13}{10})_3,(\nb{\frac{3}{2}})_3  $ & $234 $ & $\T_{13},\star $ & $360 $ & $-$\\
$\frac{54}{5}$  &  $ (\frac{1}{2})_3,(\frac{9}{10})_3,\frac{7}{5} $ & $108 $ & $\T_{9},\star$ & $\frac{66 }{5}$ & $  \frac{3}{5},(\frac{11}{10})_3,(\nb{\frac{3}{2} })_3 $ & $ 132$ & $\T_{11},\star $ & $ 240$ & $-$\\
		\hline
	
		\end{tabular}
			\caption{Hecke images of $M_{6/5}$ with degeneracy $(1,3,3,1)$.}
			\label{tb:HeckeMc65}
		\end{table}
		
There exists one more theory from holomorphic modular bootstrap (\cite{Kaidi:2021ent}, Table 6) falling into type $M_{{6}/{5}}$. We find the $c=\frac{114}{5}$ theory with $h_i=0,\frac{9}{10},\frac{3}{2},\frac{12}{5}$ and $m_1=1938$ can pair with $M_{6/5}$ with $n_i=0,1,2,3$, of which the bilinear relation of characters gives $J+3876$.

\subsection{Type $M_{\rm sub}(14,3)$ }
Minimal model $M(14,3)$ is non-unitary with central charge $c=-\frac{114}{7}$ and 13 primaries. The effective theory $M_{\rm eff}(14,3)$ has $c_{\rm eff}=\frac{6}{7}$ and $h_{\rm eff}=0,\frac{1}{56},\frac{5}{56},\frac{1}{7},\frac{2}{7},\frac{3}{8},\frac{33}{56},\frac{5}{7},1,\frac{85}{56},\frac{15}{7},\frac{23}{8},\frac{26}{7}$.
It is known a subset of the 13 primaries can form a block-diagonal modular invariant known as $(A_2,D_8)$ theory \cite{DiFrancesco:1997nk} with weights and degeneracy as 
$h_i=0,\frac{1}{7},(\frac{2}{7})_2,\frac{5}{7}.$
We compute the four characters to be
\be
\ba
\chi_0=&\,q^{-\frac{1}{28}}( 1 + 2 q + 3 q^2 + 5 q^3 + 8 q^4 + 11 q^5 + 17 q^6 +  \dots),\\
\chi_{\frac{1}{7}}=&\,q^{\frac{3}{28}}( 1 + q + 3 q^2 + 3 q^3 + 6 q^4 + 8 q^5 + 13 q^6 + \dots),\\
\chi_{\frac{2}{7}}=&\,q^{\frac{1}{4}}(1 + q + 2 q^2 + 3 q^3 + 5 q^4 + 7 q^5 + 11 q^6 + \dots),\\
\chi_{\frac{5}{7}}=&\,q^{\frac{19}{28}}(1 + q^2 + 2 q^3 + 3 q^4 + 3 q^5 + 6 q^6 + 7 q^7  +\dots).
\ea
\ee
Clearly, the conductor $N=28$. The $S$ matrix for the four characters is
\be
\rho(S)=\frac{1}{\sqrt{7}}\left(
\begin{array}{cccc}
 2 \cos \left(\frac{\pi }{7}\right) & 2 \sin \left(\frac{3 \pi }{14}\right) & 2 & 2 \sin \left(\frac{\pi }{14}\right) \\
 2 \sin \left(\frac{3 \pi }{14}\right) & 2 \sin \left(\frac{\pi }{14}\right) & -2 & -2 \cos \left(\frac{\pi }{7}\right) \\
 1 & -1 & -1 & 1 \\
 2 \sin \left(\frac{\pi }{14}\right) & -2 \cos \left(\frac{\pi }{7}\right) & 2 & -2 \sin \left(\frac{3 \pi }{14}\right) \\
\end{array}
\right).
\ee

Now let us consider the Hecke images $\T_p$ of this subtheory. 
We find there exist 6 classes for the Hecke operation. 
For $p^2\equiv 1 \textrm{ mod } 28$, $
\rho(\sigma_{1,27})=-\rho(\sigma_{13,15})=\mathrm{Id}.$ 
While for  $p^2\equiv 9 \textrm{ mod } 28$ and  $p^2\equiv 25 \textrm{ mod } 28$, the corresponding $\rho(\sigma_{p})$ matrices are
\be
\rho(\sigma_{3,25})=-\rho(\sigma_{11,17})=\left(
\begin{array}{cccc}
 0 & -1 & 0 & 0 \\
 0 & 0 & 0 & -1 \\
 0 & 0 & 1 & 0 \\
 1 & 0 & 0 & 0 \\
\end{array}
\right),\  \rho(\sigma_{5,23})=-\rho(\sigma_{9,19})=\left(
\begin{array}{cccc}
 0 & 0 & 0 & -1 \\
 1 & 0 & 0 & 0 \\
 0 & 0 & -1 & 0 \\
 0 & 1 & 0 & 0 \\
\end{array}
\right).
\ee
Using these $\rho(\sigma_{p})$ matrices, we compute the Hecke images for all admissible $p<28$ and collect them in $c=24$ pairs in Table \ref{tb:HeckeM143}. We can see the sums of spin-1 currents $m_1+\tm_1$ for all pairs in Table \ref{tb:HeckeM143} are divisible by the conductor 28. We find the $\T_3$ image describe a subtheory of WZW $(A_1)_{12}$, which is actually the $D_8$ type invariant of $\hat{A}_1$ in the ADE classification. Note the $\T_{1,3,5,23,25,27}$ images appear in the holomorphic modular bootstrap (\cite{Kaidi:2021ent}, Table 6, the fifth column). One extra theory there has $c=\frac{150}{7}$, $h_i=0,\frac{6}{7},\frac{11}{7},\frac{15}{7}$ and $m_1=300$. We find this theory can form a $c=24$ pair with the $\T_3$ image with $n_i=0,1,2,3$. This pair has bilinear relation of characters equal to $J+1128$. 

\begin{table}[ht]
\def\arraystretch{1.1}
	\centering
	\begin{tabular}{|c|c|c|c|c|c|c|c|c|c|c|c|c|c|}
		\hline
		 \multicolumn{4}{|c|}{$l=0$} & \multicolumn{4}{c|}{$l'=0$} & \multicolumn{2}{c|}{duality}\\
		\hline
		 $c$ & $h_i$  & $m_1$ & remark & $\tc$ & $\tilh_i$ &  $\tm_1$ & remark & $\!\!m_1\!+\!\tm_1\!\!$ & \!Sch\!\\
		\hline
$\!\!-\frac{114}{7}\!\!$  &  $\!\!-\{\frac57,(\frac47)_2,\frac37 \}\!\! $ & $0 $ & $\T_{-19}$ & $\frac{282}{7}$ & ${ \frac{17}{7},(\frac{18}{7})_2,\frac{19}{7} }$ & $ 0$ & $\T_{47},\star$ & $0 $ & $0$\\
$\frac{6}{7}$  &  ${\frac{1}{7},(\frac{2}{7})_2,\frac{5}{7}} $ & $2 $ & $\T_{1}$ & $\frac{162}{7}$ & ${ \frac{9}{7},(\frac{12}{7})_2,\frac{13}{7}}$ & $54 $ & $\T_{27} $ & $56 $ & $-$\\
$\frac{18}{7}$  &  ${\frac{1}{7},\frac{3}{7},(\frac{6}{7})_2} $ & $ 3$ & $\T_{3},(A_1)_{12}$ & $\frac{150}{7}$ & ${ (\frac{8}{7})_2,\frac{11}{7},\frac{13}{7}}$ & $ 25$ & $\T_{25} $ & $ 28$ & $-$\\
$\frac{30}{7}$  &  ${(\frac{3}{7})_2,\frac{4}{7},\frac{5}{7}} $ & $30 $ & $\T_{5}$ & $\frac{138}{7}$ & ${\frac{9}{7},\frac{10}{7},(\frac{11}{7})_2}$ & $138 $ & $\T_{23} $ & $168 $ & $-$\\
$\frac{54}{7}$  &  ${\frac{3}{7},(\frac{4}{7})_2,\frac{9}{7}} $ & $ 117$ & $\T_{9},\star$ & $\frac{114}{7}$ & ${ \frac{5}{7},(\frac{10}{7})_2,\frac{11}{7}}$ & $ 247$ & $\T_{19} ,\star$ & $ 364$ & $-$\\
$\frac{66}{7}$  &  ${\frac{4}{7},\frac{6}{7},(\frac{8}{7})_2} $ & $66 $ & $\T_{11},\star$ & $\frac{102}{7}$ & ${(\frac{6}{7})_2,\frac{8}{7},\frac{10}{7} }$ & $ 102$ & $\T_{17},\star $ & $168 $ & $-$\\
$\frac{78}{7}$  &  ${(\frac{5}{7})_2,\frac{6}{7},\frac{9}{7}} $ & $312 $ & $\T_{13},\star$ & $\frac{90}{7}$ & ${ \frac{5}{7},\frac{8}{7},(\frac{9}{7})_2}$ & $ 360$ & $\T_{15},\star$ & $ 672$ & $-$\\
		\hline
	
		\end{tabular}
			\caption{Hecke images of $M_{\rm sub}(14,3)$.}
			\label{tb:HeckeM143}
		\end{table}

We also study the $c=24$ dual of the subtheory of original $M(14,3)$ which we formally denote as $\T_{-19}$. The dual theory has central charge $c=\frac{282}{7}$ and weights $0,\frac{17}{7},\frac{18}{7},\frac{19}{7}$. We find it can be realized as Hecke image $\T_{47}$ but with quasi-characters. We find this pair has the following bilinear relation of characters
\be
\ba
J(\tau)=&\, \chi_{0}^{\T_{-19}}\chi_0^{\T_{47}}-\chi_{-\frac57}^{\T_{-19}}\chi_{\frac{19}{7}}^{\T_{47}}-\chi_{-\frac47}^{\T_{-19}}\chi_{\frac{18}{7}}^{\T_{47}}+2\chi_{-\frac37}^{\T_{-19}}\chi_{\frac{17}{7}}^{\T_{47}}.
\ea
\ee
 
\subsection{Type $M_{\rm sub}(6,5),$ 3-states Potts model and type $M_{8/5}$ }
Unitary minimal model $M(6,5)$ has central charge $c=\frac45$ and 10 primary fields with weights $h_i=0,\frac{1}{40},\frac{1}{15},\frac{1}{8},\frac{2}{5},\frac{21}{40},\frac{2}{3},\frac{7}{5},\frac{13}{8},3$. It is well-known  a subtheory of $M(6,5)$ describes the three-states Potts model which has four characters with weights and degeneracy 
$h_i=0,(\frac{1}{15})_2,\frac{2}{5},(\frac{2}{3})_2$ \cite{DiFrancesco:1997nk}.
The four characters have the following Fourier expansion
\be
\ba
\chi_0=&\,q^{-\frac{1}{30}}(1 + q^2 + 2 q^3 + 3 q^4 + 4 q^5 + 7 q^6  +\dots),\\
\chi_{\frac{1}{15}}=&\,q^{\frac{1}{30}}(1 + q + 2 q^2 + 3 q^3 + 5 q^4 + 7 q^5 + 10 q^6 +\dots),\\
\chi_{\frac{2}{5}}=&\,q^{\frac{11}{30}}( 1 + 2 q + 2 q^2 + 4 q^3 + 5 q^4 + 8 q^5 + 11 q^6 + \dots),\\
\chi_{\frac{2}{3}}=&\,q^{\frac{19}{30}}(1 + q + 2 q^2 + 2 q^3 + 4 q^4 + 5 q^5 + 8 q^6 + \dots).
\ea
\ee
Clearly the conductor $N=30$. 
The $S$ matrix for the four characters is
\be
S=\left(
\begin{array}{cccc}
 a_{-}& 2 a_{+}& a_{+}& 2 a_{-}\\
 a_{+}& a_{-}& -a_{-}& -a_{+}\\
 a_{+}& -2 a_{-}& -a_{-}& 2 a_{+}\\
 a_{-}& -a_{+}& a_{+}& -a_{-}\\
\end{array}
\right)
,\textrm{ with } a_{\pm}=2 \sqrt{\frac{1}{15} \left(\frac{5}{8}\pm\frac{\sqrt{5}}{8}\right)}.
\ee

Let us consider the Hecke images $\T_p$ of this subtheory $M_{\rm sub}(6,5)$. 
We find the following $\rho(\sigma_{p})$ matrices with $p$ modulo 30:
\be
\rho(\sigma_{1,29})=-\rho(\sigma_{11,19})=\mathrm{Id},\ \textrm{   and   }\  \rho(\sigma_{7,23})=-\rho(\sigma_{13,17})=\left(
\begin{array}{cccc}
 0 & 0 & -1 & 0 \\
 0 & 0 & 0 & 1 \\
 1 & 0 & 0 & 0 \\
 0 & -1 & 0 & 0 \\
\end{array}
\right).
\ee
Using these $\rho(\sigma_{p})$ matrices, we compute its Hecke images for all admissible $p\le 29$ and collect them in $c=24$ pairs in Table \ref{tb:HeckeM65}. Notably, it was observed in \cite{Bae:2020pvv} that the $\T_{29}$ image produces the characters of the RCFT associated to the largest Fischer group $Fi_{24}$. We remark that the $\T_7,\T_{23}$ images are non-unitary.

\begin{table}[ht]
	\centering
	\begin{tabular}{|c|c|c|c|c|c|c|c|c|c|c|c|c|c|}
		\hline
		 \multicolumn{4}{|c|}{$l=0$} & \multicolumn{4}{c|}{$l'=0$} & \multicolumn{2}{c|}{duality}\\
		\hline
		 $c$ & $h_i$  & $m_1$ & remark & $\tc$ & $\tilh_i$ &  $\tm_1$ & remark & $\!\!m_1\!+\!\tm_1\!\!$ & Sch\\
		\hline
    $\frac{4}{5}$  &  ${(\frac{1}{15})_2,\frac{2}{5},(\frac{2}{3})_2} $ & $ 0$ & $\T_{1}$ & $\frac{116}{5}$ & ${ (\frac{4}{3})_2,\frac{8}{5},(\frac{29}{15})_2}$ & $ 0$ & $\T_{29},Fi_{24} $ & $0 $ & $0$\\
$\frac{28}{5}$  &  ${(\frac{7}{15})_2,(\frac{2}{3})_2,\frac{4}{5}} $ & $56 $ & $\T_{7}$ & $\frac{92}{5}$ & ${\frac{6}{5},(\frac{4}{3})_2,(\frac{23}{15})_2 }$ & $184 $ & $\T_{23} $ & $ 240$ &$-$ \\
$\frac{44}{5}$  &  ${\frac{2}{5},(\frac{11}{15})_2,(\frac{4}{3})_2} $ & $88 $ & $\T_{11},\star$ & $\frac{76}{5}$ & ${ (\frac{2}{3})_2,(\frac{19}{15})_2,\frac{8}{5}}$ & $152 $ & $\T_{19},\star $ & $240 $ & $-$\\
$\frac{52}{5}$  &  ${(\frac{2}{3})_2,(\frac{13}{15})_2,\frac{6}{5}} $ & $208 $ & $\T_{13},\star$ & $\frac{68}{5}$ & ${ \frac{4}{5},(\frac{17}{15})_2,(\frac{4}{3})_2}$ & $272 $ & $\T_{17},\star $ & $480 $ & $-$\\
		\hline
	
		\end{tabular}
			\caption{Hecke images of $M_{\rm sub}(6,5)$.}
			\label{tb:HeckeM65}
		\end{table}

There exists one more theory from holomorphic modular bootstrap (\cite{Kaidi:2021ent}, Table 6) falling into type $M_{\rm sub}(6,5)$. We find the $c=\frac{116}{5}$ theory with $h_i=0,\frac{14}{15},\frac{8}{5},\frac{7}{3}$ and $m_1= 1566$ can pair with $M_{\rm sub}(6,5)$ with $n_i=0,1,2,3$, of which the bilinear relation of characters gives $J+ 3132$.

A non-unitary theory with central charge $c=\frac{8}{5}$ and weights and degeneracy $h_i=0,(\frac{2}{15})_2,\frac{4}{5},(\frac{1}{3})_2$ was studied in \cite{Harvey:2019qzs}, which can be realized as a coset ${(A_2)_1}/{(LY)_1}$. 
We denote this theory as $M_{8/5}$. 
The conductor is $N=15$. The Fourier expansion can be found in (\cite{Harvey:2019qzs}, Section 2.4). Although the weights and degeneracy of $M_{8/5}$ satisfy the generalized $\T_2$ conditions of $M_{\rm sub}(6,5),$ unfortunately we checked the characters of $M_{8/5}$ can not be written as degree two homogeneous polynomial of the characters of $M_{\rm sub}(6,5)$. Therefore, we do \emph{not} regard $M_{8/5}$ as the generalized $\T_2$ image of $M_{\rm sub}(6,5)$. 
Still, it is worthy to explore the Hecke images of $M_{8/5}$ itself. We determine the $S$-matrix of $M_{8/5}$ as 
\be
\rho(S)=\left(
\begin{array}{cccc}
 a_{+}& 2 a_{-}& a_{-}& 2 a_{+}\\
 a_{-}& a_{+}& -a_{+}& -a_{-}\\
 a_{-}& -2 a_{+}& -a_{+}& 2 a_{-}\\
 a_{+}& -a_{-}& a_{-}& -a_{+}\\
\end{array}
\right).
\ee
Then we find the following $\rho(\sigma_{p})$ matrices of $M_{8/5}$ with $p$ modulo 15:
\be
\rho(\sigma_{1,14})=-\rho(\sigma_{4,11})=\mathrm{Id},\ \textrm{   and   }\  \rho(\sigma_{2,13})=-\rho(\sigma_{7,8})=\left(
\begin{array}{cccc}
 0 & 0 & -1 & 0 \\
 0 & 0 & 0 & 1 \\
 1 & 0 & 0 & 0 \\
 0 & -1 & 0 & 0 \\
\end{array}
\right).
\ee
Note these exactly resemble the $\rho(\sigma_{p})$ matrices of $M_{\rm sub}(6,5)$. 
We compute the Hecke images of $M_{8/5}$ for all admissible $p\le 14$ and collect the results in Table \ref{tb:HeckeM85}. It was observed in \cite{Harvey:2019qzs} that the $\T_2$ and $\T_{13}$ images describe the WZW $(A_2)_2$ and a subtheory of $(F_4)_6$. Together they form a $c=24$ theory appearing in the Schellekens' list No.14,  with holomorphic VOA construction recently given in \cite{Lam}.

\begin{table}[ht]
\def\arraystretch{1.1}
	\centering
	\begin{tabular}{|c|c|c|c|c|c|c|c|c|c|c|c|c|c|}
		\hline
		 \multicolumn{4}{|c|}{$l=0$} & \multicolumn{4}{c|}{$l'=0$} & \multicolumn{2}{c|}{duality}\\
		\hline
		 $c$ & $h_i$  & $m_1$ & remark & $\tc$ & $\tilh_i$ &  $\tm_1$ & remark & $\!\!m_1\!+\!\tm_1\!\!$ & \!Sch\!\\
		\hline
$\frac{8}{5}$  &  ${(\frac{2}{15})_2,(\frac{1}{3})_2,\frac{4}{5}} $ & $ 4$ & $\T_{1}$ & $\frac{112}{5}$ & ${\frac65,(\frac53)_2,(\frac{28}{15})_2 }$ & $56 $ & $\T_{14} $ & $60 $ & $-$\\
$\frac{16}{5}$  &  ${(\frac{4}{15})_2,\frac{3}{5},(\frac{2}{3})_2} $ & $8 $ & $\!\T_{2},(A_2)_2\!$ & $\frac{104}{5}$ & $(\frac{4}{3})_2,\frac{7}{5},(\frac{26}{15})_2$ & $ 52$ & $\!\T_{13},(F_4)_6\! $ & $ 60$ & $14$\\
$\frac{32}{5}$  &  ${(\frac{1}{3})_2,(\frac{8}{15})_2,\frac{6}{5}} $ & $ 80$ & $\T_{4},\star$ & $\frac{88}{5}$ & ${\frac{4}{5},(\frac{22}{15})_2,(\frac{5}{3})_2 }$ & $220 $ & $\T_{11},\star $ & $ 300$ & $-$ \\
$\frac{56}{5}$  &  ${\frac{3}{5},(\frac{14}{15})_2,(\frac{4}{3})_2} $ & $ 140$ & $\T_{7},\star$ & $\frac{64}{5}$ & ${ (\frac{2}{3})_2,(\frac{16}{15})_2,\frac{7}{5}}$ & $160 $ & $\T_{8},\star $ & $300 $ & $-$ \\
		\hline
	
		\end{tabular}
			\caption{Hecke images of $M_{8/5}$.}
			\label{tb:HeckeM85}
		\end{table}

There exist one more theory from holomorphic modular bootstrap (\cite{Kaidi:2021ent}, Table 6) falling into type $M_{8/5}$. We find the $c=\frac{112}{5}$ theory with $h_i=0,\frac{13}{15},\frac{5}{3},\frac{11}{5}$ and $m_1= 210$ can pair with $M_{8/5}$ with $n_i=0,1,2,3$, of which the bilinear relation of characters gives $J+ 368$.

\subsection{Type $U(1)_{3}$ and type $M_2$ } 

Consider the $U(1)_{3}$ theory with central charge $c = 1$ and conformal weights and degeneracy $h_i=0,(\frac{1}{12})_2,(\frac{1}{3})_2,\frac34.$ This theory can also be regarded as coset $(A_2)_1/(A_1)_1$. It is easy to compute the four distinct characters as
\be
\ba
\chi_0=&\,q^{-\frac{1}{24}}( 1 + q + 2 q^2 + 5 q^3 + 7 q^4 + 11 q^5 + 17 q^6   +\dots),\\
\chi_{\frac{1}{12}}=&\,q^{\frac{1}{24}}(1 + q + 3 q^2 + 4 q^3 + 8 q^4 + 11 q^5 + 18 q^6   +\dots),\\
\chi_{\frac{1}{3}}=&\,q^{\frac{7}{24}}( 1 + 2 q + 3 q^2 + 5 q^3 + 8 q^4 + 13 q^5 + 19 q^6  +\dots),\\
\chi_{\frac{3}{4}}=&\,q^{\frac{17}{24}}( 2 + 2 q + 4 q^2 + 6 q^3 + 10 q^4 + 14 q^5 + 24 q^6   +\dots).
\ea
\ee
Note the conductor $N=24$. 
Consider the Hecke images $\T_p$ of $U(1)_3$ theory. We find for $p=1, 5,19,23\textrm{ mod } 24$, $\rho(\sigma_{p})=\mathrm{Id},$ while for $p=7, 11,13,17\textrm{ mod } 24$, $\rho(\sigma_{p})=-\mathrm{Id}$. 
We collect all admissible Hecke images in $c=24$ pairs in Table \ref{tb:HeckeSpin2_3}. Note the sums of spin-1 currents $m_1+\tm_1$ for all pairs in Table \ref{tb:HeckeSpin2_3} are divisible by the conductor 24.

\begin{table}[ht]
\def\arraystretch{1.1}
	\centering
	\begin{tabular}{|c|c|c|c|c|c|c|c|c|c|c|c|c|c|}
		\hline
		 \multicolumn{4}{|c|}{$l=0$} & \multicolumn{4}{c|}{$l'=0$} & \multicolumn{2}{c|}{duality}\\
		\hline
		 $c$ & $h_i$  & $m_1$ & remark & $\tc$ & $\tilh_i$ &  $\tm_1$ & remark & $\!\!m_1\!+\!\tm_1\!\!$ &\!Sch\!\\
		\hline
$1$ & ${(\frac{1}{12})_2,(\frac13)_2, \frac{3}{4}} $ & $ 1$ & $\T_{1} $ & $23$ & $\frac{5}{4},(\frac{5}{3})_2, (\frac{23}{12})_2$ & $ 23$ & $\T_{23}$   &  $  24 $  & $1$\\
$ 5 $ & ${(\frac{5}{12})_2,(\frac{2}{3})_2,\frac{3}{4} } $ & $35 $ & $\!\!\T_{ 5},(A_5)_1\!\!$ & $19$ & $\frac{5}{4},(\frac{4}{3})_2,(\frac{19}{12})_2 $ & $133 $ & $\T_{19 }$   &  $   168$  & $\!43-45\!$\\
$ 7 $ & ${(\frac13)_2,(\frac{7}{12})_2,\frac{5}{4}} $ & $77$ & $\T_{ 7},\star $ & $17$ & $\frac{3}{4},(\frac{17}{12})_2,(\frac{5}{3})_2 $ & $187 $ & $\T_{17 }, \star$   &  $ 264 $  & $-$\\
$ 11 $ & $ {(\frac23)_2,(\frac{11}{12})_2, \frac{5}{4}}$ & $ 187$ & $\T_{ 11} ,\star $ & $13$ & $\frac{3}{4},(\frac{13}{12})_2,(\frac{4}{3} )_2$ & $ 221$ & $\T_{ 13} ,\star$   &  $  408 $  & $-$\\
		\hline
	
		\end{tabular}
			\caption{Hecke images of $U(1)_{3}$.}
			\label{tb:HeckeSpin2_3}
		\end{table}

We find the Hecke relations such as $\T_5U(1)_3=(A_5)_1$ and $\T_{19}U(1)_3$ can describe a subtheory of WZW $(E_7)_3$. More precisely, we have the following relation between the $\T_{19}$ Hecke image and $(E_7)_3$ characters:
\be
\ba
\chi^{\T_{19}}_0=&\,\chi_0^{(E_7)_3,3w_0}+\chi_2^{(E_7)_3,w_2+w_7}=q^{-\frac{19}{24}}( 1 + 133 q + 49799 q^2 + 2414976 q^3+\dots),\\
\chi^{\T_{19}}_{\frac{19}{12}}=&\,\chi_{\frac{19}{12}}^{(E_7)_3,w_1+w_7}=q^{\frac{19}{24}}(6480 + 541728 q + 16030224 q^2+\dots),\\
\chi^{\T_{19}}_{\frac43}=&\,\chi_{\frac43}^{(E_7)_3,w_6+w_0}=q^{\frac{13}{24}}( 1539 + 204687 q + 7299477 q^2+\dots),\\
\chi^{\T_{19}}_{\frac54}=&\,\chi_{\frac54}^{(E_7)_3,w_2+w_0}+\chi_{\frac94}^{(E_7)_3,3w_7}=q^{\frac{11}{24}}( 912 + 145616 q + 5572928 q^2 +\dots).
\ea
\ee
The primaries with weights ${19}/{12}$ and ${4}/{3}$ have degeneracy 2 as required by the Hecke operation. Together these two images $\T_5$ and $\T_{19}$ can form a $c=24$ theory in Schellekens’ List No.45, with the associated holomorphic VOA  constructed in \cite{Lam:2015pjc} by certain $\IZ_2$ orbifold.

There exists one more theory from holomorphic modular bootstrap (\cite{Kaidi:2021ent}, Table 6) falling into type $U(1)_3$. We find the $c=23$ theory with $h_i=0,\frac{11}{12},\frac{5}{3},\frac{9}{4}$ and $m_1= 575$ can pair with $U(1)_3$ with $n_i=0,1,2,3$, of which the bilinear relation of characters gives $J+ 1128$.

Consider a $c=2$ theory with conformal weights $h_i=0,\frac16,\frac12,\frac23$ appeared in holomorphic modular bootstrap (\cite{Kaidi:2021ent}, Table 5), which we denote as $M_{2}$. Although the central charge and conformal weights of $M_{2}$ satisfy the generalized $\T_2$ condition of $U(1)_3$ theory, we will show that their degeneracies do not match. Therefore they are \emph{not} in generalized Hecke relation. 
We determine the four distinct characters of $M_{2}$ to be
\be
\ba
\chi_0^{M_2}=&\,q^{-\frac{1}{12}}( 1 + 2 q + 11 q^2 + 22 q^3 + 50 q^4 + 96 q^5  +\dots),\\
\chi_{\frac{1}{6}}^{M_2}=&\,q^{\frac{1}{12}}( 1 + 4 q + 11 q^2 + 26 q^3 + 55 q^4 + 110 q^5   +\dots),\\
\chi_{\frac{1}{2}}^{M_2}=&\,q^{\frac{5}{12}}( 2 + 6 q + 14 q^2 + 34 q^3 + 70 q^4 + 136 q^5  +\dots),\\
\chi_{\frac{2}{3}}^{M_2}=&\,q^{\frac{7}{12}}( 3 + 6 q + 18 q^2 + 36 q^3 + 81 q^4 + 150 q^5  +\dots).
\ea
\ee
Clearly, the conductor $N=12$. We find this theory can be realized by the coset $
M_2=(D_4)_1/(A_2)_1.$ Indeed, we checked the following character relations
\be
\chi^{(D_4)_1}_0=\chi^{M_2}_0\chi^{(A_2)_1}_0+2\chi^{M_2}_{\frac23}\chi^{(A_2)_1}_{\frac13},\qquad \chi^{(D_4)_1}_{\frac12}=\chi^{M_2}_{\frac12}\chi^{(A_2)_1}_0+2\chi^{M_2}_{\frac16}\chi^{(A_2)_1}_{\frac13}.
\ee
Therefore, the degeneracies of the four characters of $M_2$ are $(1,6,3,2)$ respectively. 

Consider the Hecke images of $M_2$ theory. We find for $p=1, 11\textrm{ mod } 12$, $\rho(\sigma_{p})=\mathrm{Id},$ while for $p=5, 7\textrm{ mod } 12$, $\rho(\sigma_{p})=-\mathrm{Id}$. We summarize all admissible Hecke images $\T_p$ for $p<12$ in $c=24$ pairs in Table \ref{tb:HeckeM2}. Note the sums of spin-1 currents $m_1+\tm_1$ for all  pairs in Table \ref{tb:HeckeM2} are divisible by the conductor 12. The $\T_{11}$ image has made an appearance in holomorphic modular bootstrap (\cite{Kaidi:2021ent}, Table 5). 

\begin{table}[ht]
\def\arraystretch{1.1}
	\centering
	\begin{tabular}{|c|c|c|c|c|c|c|c|c|c|c|c|c|c|}
		\hline
		 \multicolumn{4}{|c|}{$l=0$} & \multicolumn{4}{c|}{$l'=0$} & \multicolumn{2}{c|}{duality}\\
		\hline
		 $c$ & $h_i$  & $m_1$ & remark & $\tc$ & $\tilh_i$ &  $\tm_1$ & remark & $\!m_1\!+\!\tm_1\!$ & Sch\\
		\hline
$2$ & $\! (\frac16)_6,(\frac12)_3,(\frac23)_2 \!$ & $ 2$ & $\T_{1} $ & $ 22$ & $ \!(\frac{4}{3})_2,(\nb{\frac{3}{2}})_3,(\frac{11}{6})_6\!$ & $22 $ & $\T_{11}$   &  $  24 $  & $-$\\
		\hline
$10$ & $\! (\frac{1}{2})_3,(\frac{5}{6})_6,(\frac{4}{3})_2 \!$ & $ 110$ & $\T_{5},\star $ & $ 14$ & $ \!(\frac{2}{3})_2,(\frac{7}{6})_6,(\nb{\frac{3}{2}})_3 \!$ & $ 154$ & $\T_{7},\star$   &  $ 264  $  & $-$\\
		\hline	
		\end{tabular}
			\caption{Hecke images of $M_2$ theory.}
			\label{tb:HeckeM2}
		\end{table}

\subsection{Type $(A_1)_1^{\otimes 3}$ }
Consider the triple product theory $(A_1)_1^{\otimes 3}$ which has central charge
$c=3$ and weights with degeneracy $h_i=0,(\frac14)_3,(\frac12)_3,\frac34$. The conductor $N=8$. For the Hecke operation of $(A_1)_1^{\otimes 3}$, we find for $p$ modulo 8, the $\rho(\sigma_p)$ matrices are $\rho(\sigma_{1,7})=-\rho(\sigma_{3,5})=\mathrm{Id}$. We compute all admissible Hecke images for $p<8$ and list them in $c=24$ pairs in Table \ref{tb:HeckeA113}. Notably, we find the $\T_7$ Hecke image can describe a subtheory of WZW $(A_7)_4$ with weights $0,\frac{5}{4},(\frac{3}{2})_3,(\frac{7}{4})_3$. For example, the vacuum character can be decomposed as
\be
\ba
\chi^{\T_{7}}_0=&\,\chi_{0,1}^{(A_7)_4,4w_0}+2\chi_{2,14700}^{(A_7)_4,w_1+w_3+w_4+w_0}+\chi_{2,24255}^{(A_7)_4,w_1+w_2+w_6+w_7}\\
&\,+2\chi_{3,13860}^{(A_7)_4,4w_2}+\chi_{3,577500}^{(A_7)_4,w_2+w_3+w_5+w_6}+\chi_{4,232848}^{(A_7)_4,4w_4}\\
=&\,q^{-\frac{7}{8}}(1 + 63 q + 55734 q^2 + 3714697 q^3 +106849134 q^4 +\dots).
\ea
\ee
This subtheory of $(A_7)_4$ can pair with $(A_1)_1^{\otimes 3}$ to form a $c=24$ theory in the Schellekens' list No.8, of which the
associated holomorphic VOA was constructed in \cite{lam2011}.

\begin{table}[ht]
\def\arraystretch{1.1}
	\centering
	\begin{tabular}{|c|c|c|c|c|c|c|c|c|c|c|c|c|c|}
		\hline
		 \multicolumn{4}{|c|}{$l=0$} & \multicolumn{4}{c|}{$l'=0$} & \multicolumn{2}{c|}{duality}\\
		\hline
		 $c$ & $h_i$  & $m_1$ & remark & $\tc$ & $\tilh_i$ &  $\tm_1$ & remark & $\!\!m_1\!+\!\tm_1\!\!$ & \!Sch\!\\
		\hline
$3$ & $( \frac14)_3,(\frac12)_3,\frac34$ & $ 9$ & $\!\T_{1},(A_1)_1^{\otimes 3}\! $ & $ 21$ & $ \frac{5}{4},(\nb{\frac{3}{2}})_3,(\frac{7}{4})_3 $ & $63 $ & $\T_{7}$   &  $  72$  & $\!15-18\!$\\
		\hline
$9$ & $ (\frac{1}{2})_3,(\frac{3}{4})_3,\frac{5}{4}$ & $ 117$ & $\T_{3},\star $ & $ 15$ & $  \frac{3}{4},(\frac{5}{4})_3,(\nb{\frac{3}{2}})_3$ & $ 195$ & $\T_{5},\star$   &  $  312  $  & $-$\\
		\hline	
		\end{tabular}
			\caption{Hecke images of $(A_1)_1^{\otimes 3}$ theory.}
			\label{tb:HeckeA113}
		\end{table}

There exist one more theory from holomorphic modular bootstrap (\cite{Kaidi:2021ent}, Table 6) falling into type $(A_1)_1^{\otimes 3}$. We find the $c=21$ theory with $h_i=0,\frac{3}{4},\frac{3}{2},\frac{9}{4}$ and $m_1= 399$ is just $(E_7)_1^{\otimes 3}$, which can pair with $(A_1)_1^{\otimes 3}$, and the bilinear relation of characters gives $J+ 744 $.

\subsection{Type $(A_6)_1$ }
WZW model $(A_6)_1$ has central charge $c=6$ and conformal weights with degeneracy as $h_i=0,(\frac{3}{7})_2,(\frac{5}{7})_2,(\frac{6}{7})_2.$
The four distinct characters are
\be
\ba
\chi_0^{w_0}=&\,q^{-\frac{1}{4}}( 1 + 48 q + 489 q^2 + 2842 q^3 + 13083 q^4+\dots),\\
\chi_{3/7}^{w_1}=&\,q^{\frac{5}{28}}( 7 + 147 q + 1071 q^2 + 5628 q^3 + 23709 q^4 +\dots),\\
\chi_{5/7}^{w_2}=&\,q^{\frac{13}{28}}( 21 + 273 q + 1764 q^2 + 8652 q^3 + 34790 q^4 +\dots),\\
\chi_{6/7}^{w_3}=&\,q^{\frac{17}{28}}( 35 + 357 q + 2268 q^2 + 10619 q^3 + 42042 q^4 +\dots).
\ea
\ee
Here $w_{0,1,2,\dots,6}$ denote the affine weights of $\hat{A}_6^{(1)}$. Clearly the conductor $N=28$.  
Consider the Hecke images $\T_p$ of WZW $(A_6)_1$. We find the $\T_3$ image has $c=18$, $l=0$, weights 
$h_i=0,(\frac{8}{7})_2,(\frac{9}{7})_2,(\frac{11}{7})_2$ and spin-1 current $m_1=144.$
This theory can be regarded as a subtheory of $(A_6)_1^{\otimes 3}$ and forms a $c=24$ pair with $(A_6)_1$ itself, which appeared in the Schellekens' list No.46. Besides, for $p\ge 5$, we notice that the Hecke images $\T_p(A_6)_1$ have $l\ge 6$ MLDEs.

\section{RCFTs with five characters}\label{sec:5chi}
The potential RCFTs with five characters, non-integral weights and $l=0$ MLDEs have been recently classified by modular bootstrap in  (\cite{Kaidi:2021ent}, Table 7). There are in total 23 theories. We propose that all these 23 theories can be merely generated by two simple types $(LY)_4$ and $(LY)_1^{\otimes 4}$ which have 5 primaries and 16 primaries respectively. 
We will discuss the Hecke images and cosets of these two types individually.
We also discuss the $U(1)_4$ theory which has 
five characters and integral weights. Though degenerate, this theory also has many interesting Hecke images.
There exists more degenerate $l=0$ RCFTs such as $(A_1)_4$ which has $c=2,h_i=0,\frac{1}{8},\frac{1}{3},\frac{5}{8},1$ and $N=24$.  We do not intend to discuss the Hecke images of $(A_1)_4$ in detail, instead just comment that $(A_1)_4$ and its $\T_{11}$ image form a $c=24$ theory in the Schellekens' list No.2.

\subsection{Type $(LY)_4$ }
Minimal model $M(11,2)$ is a non-unitary theory with real central charge $c=-\frac{232}{11}$ and conformal weights $h=\{0,-\frac{4}{11},-\frac{7}{11},-\frac{9}{11},-\frac{10}{11}\}$. Consider the effective theory $M_{\rm eff}(11,2)$ i.e., $(LY)_4$ which has $c_{\rm eff}=\frac{8}{11}$ and $h_{\rm eff}=\{0,\frac{1}{11},\frac{3}{11},\frac{6}{11},\frac{10}{11}\}$. The conductor is $N=33$. The $\rho(\sigma_p)$ matrices for various Hecke operations can be found in Appendix D.2.2 of \cite{Wuthesis}. We compute its Hecke images for all admissible $p< 33$ and collect the results in Table \ref{tb:HeckeM112}. 
Some Hecke images can be identified as WZW models such as $\T_{13}=(F_4)_2$ and $\T_{20}$ as a subtheory of $(A_8)_2$ \cite{Wuthesis}. Together $\T_{13}$ and $\T_{20}$ Hecke images form a $c=24$ theory in the Schellekens' list No.36, of which the associated holomorphic VOA was constructed in \cite{lam2012}.
Besides, \cite{Wuthesis} also points out that the $\T_{31}$ image gives the characters of RCFT associated to the Thompson group \cite{Bae:2020pvv}, while the $\T_{2}$ image describes the 3C conjugacy class of the Monster group. The only unitary theories in Table \ref{tb:HeckeM112} are $\T_{13},\T_{31},\T_{20},\T_2$ belonging to the MTC classes $5^B_{\pm 16/11}$ in Table II of \cite{Schoutens:2015uia}, and all others are non-unitary. For $c=24$ pairs $(\T_p,\T_{p'})$, we have $p+p'=33$ and $(l,l')=(0,2)$. We summarize all admissible pairs in Table \ref{tb:HeckeM112}. It is easy to check the sums of spin-1 currents $m_1+\tm_1$ for all pairs are divisible by the conductor 33. There also exist other $c=8k$ pairs. For example, we notice $\T_4$ and $\T_7$ are dual w.r.t WZW $(E_8)_1$.

 \begin{table}[ht]
 \def\arraystretch{1.1}
	\centering
	\begin{tabular}{|c|c|c|c|c|c|c|c|c|c|c|c|c|}
		\hline
		 \multicolumn{4}{|c|}{$l=0$} & \multicolumn{4}{c|}{$l'=2$} & \multicolumn{2}{c|}{duality}\\
		\hline
		 $c$ & $h_i$  & $m_1$ & \!\!\!\!remark\!\!\!\! & $\tc$ & $\tilh_i$ &  $\tm_1$ & \!remark\! & $\!\!m_1\!+\!\tm_1\!\!\!$ & \!\!Sch\!\!\\
		\hline
$\!\!-\frac{232}{11}\!\!$ & $\!\!\!-\{\frac{4}{11},\frac{7}{11},\frac{9}{11},\frac{10}{11}\}\!\!\! $ & $0$ & $\T_{-29}$ & $\frac{496}{11}$ & $\frac{26}{11},\frac{29}{11},\frac{31}{11},\frac{32}{11} $ & $0$ & $\mathsf{T}_{62}$ & $0$ & $0$\\
		  $\frac{8}{11}$ & $\frac{1}{11},\frac{3}{11},\frac{6}{11},\frac{10}{11}$ & $1$ & $\T_1$ & $\frac{256}{11}$ & $\frac{12}{11},\frac{16}{11},\frac{19}{11},\frac{21}{11}$ & $32$ & $\mathsf{T}_{32}$ & $33$ & $-$\\
		  $\frac{32}{11}$ & $\frac{2}{11},\frac{4}{11},\frac{7}{11},\frac{12}{11}$ & $8$ & $\mathsf{T}_{4}$  & $\frac{232}{11}$ & $\frac{10}{11},\frac{15}{11},\frac{18}{11},\frac{20}{11}$ & $58$ & $\mathsf{T}_{29}$ & $66$ & $-$\\
		 	  $56 \over 11$  & ${4\over 11},{ 7\over 11},{ 9\over 11}, {10\over 11}$ & $28$ & $\mathsf{T}_{7}$ & $\frac{208}{11}$ & $\frac{12}{11},\frac{13}{11},\frac{15}{11},\frac{18}{11}$ & $104$ & $\mathsf{T}_{26}$ &  $132$ & $-$ \\
		   $\frac{80}{11}$ & ${5\over 11},{ 8\over 11},{ 10\over 11}, {12\over 11}$ & $120$ & $\mathsf{T}_{10}$ & $\frac{184}{11}$ & $\frac{10}{11},\frac{12}{11},\frac{14}{11},\frac{17}{11}$ & $276$ & $\mathsf{T}_{23}$ & $396$ & $-$ \\
		  $\frac{104}{11}$ & $\frac{6}{11},\frac{9}{11},\frac{12}{11},\frac{13}{11}$ & $52$ & $\!\!\T_{13},(F_4)_2\!\!$ & $\frac{160}{11}$ & $\frac{9}{11},\frac{10}{11},\frac{13}{11},\frac{16}{11}$ & $80$ & $\!\!\T_{20},(A_8)_2\!\!$ & $132$ & $36$ \\
		 		$\frac{128}{11}$ &  ${\frac{6}{11},\frac{8}{11},\frac{15}{11},\frac{16}{11}}$ & ${224}$ & $\mathsf{T}_{16}$ & $\frac{136}{11}$ & ${\frac{6}{11},\frac{7}{11},\frac{14}{11},\frac{16}{11}}$ & $238$ & $\mathsf{T}_{17}$ & $462$ & $-$ \\
 ${152 \over 11}$ & $\frac{8}{11},\frac{13}{11},\frac{14}{11},\frac{15}{11}  $ & $ 304 $ & $\T_{19},\star $ & $112\over 11$ & $\frac{7}{11},\frac{8}{11},\frac{9}{11},\frac{14}{11}$ & $224$ &  $\T_{14},\star$ & $528$ & $-$  \\
 ${200 \over 11}$ & $ \frac{9}{11},\frac{14}{11},\frac{18}{11},\frac{19}{11} $ & $ 225 $ & $\T_{25},\star $ & $64\over 11$ & $\frac{3}{11},\frac{4}{11},\frac{8}{11},\frac{13}{11} $ & $72$ &  $\T_{8},\star$ & $297$ & $-$  \\
		  $\frac{224}{11}$  &  ${\frac{14}{11},\frac{16}{11},\frac{17}{11},\frac{18}{11}} $ & $112$ & $\mathsf{T}_{28}$ & $\frac{40}{11}$ & ${ \frac{4}{11},\frac{5}{11},\frac{6}{11},\frac{8}{11}}$ & $20$ & $\mathsf{T}_{5}$ & $132$ & $-$\\
		    ${248 \over 11}$ & ${13\over 11} , {16 \over 11} , {20\over 11},{21 \over 11}$ & $0$  & $\T_{31},Th$ & $16\over 11$ & ${1\over 11},{2\over 11},{6\over 11},{9\over 11}$ & $0$ & $\mathsf{T}_{2},D_{3\rm C}$ & $0$ & $0$\\
		\hline
	
		\end{tabular}
		\caption{Hecke images of $(LY)_4$.}
			\label{tb:HeckeM112}
		\end{table}
Consider the $c=24$ dual of the original $M(11,2)$ theory which we formally denote as $\T_{-29}$. The dual theory has central charge $c=\frac{496}{11}$ and weights $h_i=0,\frac{26}{11},\frac{29}{11},\frac{31}{11},\frac{32}{11}$. We notice it can be realized as Hecke image $\T_{62}$ and has an $l=2$ MLDE. We list this pair $(M(13,2),\T_{62})$ in the first row of Table \ref{tb:HeckeM112}. We also find this pair has the following bilinear relation of characters
\be
\ba
J(\tau)=&\, \chi_{0}^{\T_{-29}}\chi_0^{\T_{62}}-\chi_{-\frac{4}{11}}^{\T_{-29}}\chi_{\frac{26}{11}}^{\T_{62}}+\chi_{-\frac{7}{11}}^{\T_{-29}}\chi_{\frac{29}{11}}^{\T_{62}}-\chi_{-\frac{9}{11}}^{\T_{-29}}\chi_{\frac{31}{11}}^{\T_{62}}+\chi_{-\frac{10}{11}}^{\T_{-29}}\chi_{\frac{32}{11}}^{\T_{62}}.
\ea
\ee

Now let us examine our main proposal and explain how all potential RCFTs from holomorphic modular bootstrap -- the right side of Table 7 in \cite{Kaidi:2021ent} -- are related to $(LY)_4$ by Hecke and $c=8k$ coset operations.
Apart from the $l=0$ Hecke images in the left side of Table \ref{tb:HeckeM112}, there still remains five theories. We discuss them individually as follows.
\bitem
\item We identify the following $c=24$ pair:
\be
l=0,\ (c,h,m_1)=(\frac{32}{11},\frac{1}{11},\frac{4}{11},\frac{7}{11},\frac{13}{11},10),\Longleftrightarrow
l=2,\T_{29},
\ee
the characters of which satisfy the following bilinear relation
\be
\sum\chi^{c=\frac{32}{11}}\chi^{\mathsf{T}_{29}}=J+126.
\ee
\item We identify the following $c=24$ pair:
\be
l=0,\ (c,h,m_1)=(\frac{128}{11},\frac{5}{11},\frac{8}{11},\frac{15}{11},\frac{17}{11},248),\Longleftrightarrow
l=2,\T_{17},
\ee
the characters of which satisfy the following bilinear relation
\be
\sum\chi^{c=\frac{128}{11}}\chi^{\mathsf{T}_{17}}=J+894.
\ee
\item  We identify the following $c=24$ pair:
\be
l=0,\ (c,h,m_1)=(\frac{224}{11},\frac{5}{11},\frac{14}{11},\frac{18}{11},\frac{28}{11},528),\Longleftrightarrow
l=2,\T_{5},
\ee
the characters of which satisfy the following bilinear relation
\be
\sum\chi^{c=\frac{224}{11}}\chi^{\mathsf{T}_{5}}=J+708.
\ee
\item WZW $(E_8)_3$ is characterized by
$
(c,h,m_1)=(\frac{248}{11},\frac{10}{11},\frac{16}{11},\frac{20}{11},\frac{24}{11},248).
$
We identify the following $c=24$ pair:
\be
\sum\chi^{(E_8)_3}\chi^{\mathsf{T}_{2}}=J+744.
\ee
Note this is an $(l,l')=(0,2)$ pair and $n_i=1,2,2,3$.
\item We identify the following $c=32$ pair:
\be
\sum\chi^{c=\frac{344}{11}}\chi^{\mathsf{T}_{1}}=j^{1/3}(J+1000).
\ee
Note this is an $(l,l')=(0,0)$ pair and $n_i=1,2,3,4$. This satisfies the relation \eqref{llrelation}.
\eitem

\subsection{Type $(YL)_1^{\otimes 4}$ }
Consider the product theory $(YL)_1^{\otimes 4}$ which has central charge $c=\frac{8}{5}$ and conformal weights with degeneracy $h_i=0,(\frac15)_4,(\frac25)_6,(\frac35)_4,\frac45$. The conductor is $N=15$. 
We find there exist two classes for its Hecke operation. 
For $p=1, 4, 11, 14\textrm{ mod } 15$, i.e., $p^2\equiv 1 \textrm{ mod } 15$, $
\rho(\sigma_{p})=\mathrm{Id}.$ 
For $p=2, 7,8 ,13 \textrm{ mod } 15$, i.e., $p^2\equiv 4 \textrm{ mod } 15$, the $\rho(\sigma_{p})$ can be easily derived from those of $(YL)_1$. We compute all admissible Hecke images $\T_p$ of $(YL)_1^{\otimes 4}$ for $p<15$. Notably, we find the $\T_7$ image describes a subtheory of WZW model $(D_4)_4$, while the $\T_8$ image describes a subtheory of WZW model $(A_2)_2^{\otimes 4}$. 
For example, we have the following relation between the $\T_{7}$ Hecke image and $(D_4)_4$ characters:
\be
\ba
\chi^{\T_{7}}_0=&\,\chi_{0,1}^{(D_4)_4,4w_0}+3\chi_{2,294}^{(D_4)_4,4w_1}=q^{-\frac{7}{15}}( 1 + 28 q + 1316 q^2 + 18480 q^3 +\dots),\\
\chi^{\T_{7}}_{\frac75}=&\,\chi_{\frac75,300}^{(D_4)_4,2w_2}=q^{\frac{14}{15}}( 300 + 6475 q + 76300 q^2 +\dots),\\
\chi^{\T_{7}}_{\frac45}=&\,\chi_{\frac45,35}^{(D_4)_4,2w_1+2w_0}+\chi_{\frac95,840}^{(D_4)_4,2w_1+2w_3}=q^{\frac{1}{3}}( 35 + 1820 q + 29800 q^2 +\dots),\\
\chi^{\T_{7}}_{\frac65}=&\,\chi_{\frac65,350}^{(D_4)_4,w_1+w_3+w_4+w_0}=q^{\frac{11}{15}}( 350 + 9800 q + 126175 q^2 +\dots),\\
\chi^{\T_{7}}_{\frac{3}{5}}=&\,\chi_{\frac35,28}^{(D_4)_4,w_2+2w_0}+3\chi_{\frac85,567}^{(D_4)_4,2w_1+w_2}=q^{\frac{2}{15}}( 28 + 2485 q + 46844 q^2+\dots).\\
\ea
\ee
The five characters of $\T_{7}$ in the above order have degeneracy $(1,4,6,4,1)$. 
Together $(\T_7,\T_8)$ are glued as a $c=24$ theory appearing in the Schellekens' list No.13, of which the associated holomorphic VOA was constructed in \cite{lam2012}. 
In total 
there exist 4 pairs of Hecke images w.r.t $c=24$ which satisfy $p+p'=15$ and $(l,l')=(0,2)$. We summarize the relevant information in Table \ref{tb:HeckeYL4}. Note all theories in Table \ref{tb:HeckeYL4} have degeneracy $(1,4,6,4,1)$. Clearly, the sums of spin-1 currents $m_1+\tm_1$ for all pairs are divisible by the conductor 15.

\begin{table}[ht]
\def\arraystretch{1.1}
	\centering
	\begin{tabular}{|c|c|c|c|c|c|c|c|c|c|c|c|c|c|}
		\hline
		 \multicolumn{4}{|c|}{$l=0$} & \multicolumn{4}{c|}{$l'=2$} & \multicolumn{2}{c|}{duality}\\
		\hline
		 $c$ & $h_i$  & $m_1$ & remark & $\tc$ & $\tilh_i$ &  $\tm_1$ & remark & $m_1+\tm_1$ & Sch\\
		\hline
   $\frac{8}{5}$  &  ${\frac{1}{5},\frac{2}{5},\frac{3}{5},\frac{4}{5}} $ & $4 $ & $\T_{1}$ & $\frac{112}{5}$ & ${ \frac{6}{5},\frac{7}{5},\frac{8}{5},\frac{9}{5}}$ & $56 $ & $\T_{14} $ & $ 60$ & $-$\\
       $\frac{32}{5}$  &  ${\frac{2}{5},\frac{3}{5},\frac{4}{5},\frac{6}{5}} $ & $80 $ & $\T_{4}$ & $\frac{88}{5}$ & ${ \frac{4}{5},\frac{6}{5},\frac{7}{5},\frac{8}{5}}$ & $ 220$ & $\T_{11} $ & $300 $ & $-$ \\
         $\frac{56}{5}$  &  ${\frac{3}{5},\frac{4}{5},\frac{6}{5},\frac{7}{5}} $ & $28 $ & $\T_{7},(D_4)_4$ & $\frac{64}{5}$ & ${\frac{3}{5},\frac{4}{5},\frac{6}{5},\frac{7}{5} }$ & $32 $ & $\T_{8},(A_2)_2^{\otimes 4} $ & $ 60$ & $ 13$\\
     $\frac{104}{5}$ & ${\frac{6}{5},\frac{7}{5},\frac{8}{5},\frac{9}{5} }$ & $0 $ & $\T_{13} $ &         $\frac{16}{5}$  &  ${\frac{1}{5},\frac{2}{5},\frac{3}{5},\frac{4}{5}} $ & $ 0$ & $\T_{2}$ &  $ 0$ & $-$\\
		\hline
	
		\end{tabular}
			\caption{Hecke images of $(YL)_1^{\otimes 4}$. We omit the degeneracy here.}
			\label{tb:HeckeYL4}
		\end{table}
		
Let us check our main proposal and explain how all potential RCFTs from holomorphic modular bootstrap -- the left side of Table 7 in \cite{Kaidi:2021ent} -- are related to $(LY)_1^{\otimes 4}$ by Hecke and $c=8k$ coset operations. Apart from the $l=0$ Hecke images in Table \ref{tb:HeckeYL4}, there still lefts with six $l=0$ admissible theories from holomorphic modular bootstrap. We now discuss them individually.
\bitem
\item We identify the following pair w.r.t $c=24$:
\be
l=0,\ (c,h,m_1)=(\frac{32}{5},\frac{1}{5},\frac{3}{5},\frac{4}{5},\frac{7}{5},82),\Longleftrightarrow
\T_{11},\quad n_i=1,2,2,3.
\ee
The bilinear form of characters is
\be
\chi^{c=\frac{32}{5}}_0\chi^{\mathsf{T}_{11}}_0+\chi^{c=\frac{32}{5}}_{\frac{1}{5}}\chi^{\mathsf{T}_{11}}_{\frac{4}{5}}+6\chi^{c=\frac{32}{5}}_{\frac{3}{5}}\chi^{\mathsf{T}_{11}}_{\frac{7}{5}}+4\chi^{c=\frac{32}{5}}_{\frac{4}{5}}\chi^{\mathsf{T}_{11}}_{\frac{6}{5}}+4\chi^{c=\frac{32}{5}}_{\frac{7}{5}}\chi^{\mathsf{T}_{11}}_{\frac{8}{5}}=J+324.
\ee
\item 
We identify the following pair w.r.t $c=24$:
\be
l=0,\ (c,h,m_1)=(\frac{56}{5},\frac{2}{5},\frac{4}{5},\frac{6}{5},\frac{8}{5},56),\Longleftrightarrow
\T_8,\quad n_i=1,2,2,3.
\ee
The bilinear form of characters is
\be
\chi^{c=\frac{56}{5}}_0\chi^{\mathsf{T}_{8}}_0+4\chi^{c=\frac{56}{5}}_{\frac{2}{5}}\chi^{\mathsf{T}_{8}}_{\frac{3}{5}}+6\chi^{c=\frac{56}{5}}_{\frac{4}{5}}\chi^{\mathsf{T}_{8}}_{\frac{6}{5}}+4\chi^{c=\frac{56}{5}}_{\frac{6}{5}}\chi^{\mathsf{T}_{8}}_{\frac{4}{5}}+\chi^{c=\frac{56}{5}}_{\frac{8}{5}}\chi^{\mathsf{T}_{8}}_{\frac{7}{5}}=J+312.
\ee
\item 
We identify the following pair w.r.t $c=24$:
\be
l=0,\ (c,h,m_1)=(\frac{104}{5},\frac{3}{5},\frac{6}{5},\frac{9}{5},\frac{12}{5},208),\Longleftrightarrow
\T_2,\quad n_i=1,2,2,3.
\ee
The bilinear form of characters is
\be
\chi^{c=\frac{104}{5}}_0\chi^{\mathsf{T}_{2}}_0+4\chi^{c=\frac{104}{5}}_{\frac{3}{5}}\chi^{\mathsf{T}_{2}}_{\frac{2}{5}}+6\chi^{c=\frac{104}{5}}_{\frac{6}{5}}\chi^{\mathsf{T}_{2}}_{\frac{4}{5}}+4\chi^{c=\frac{104}{5}}_{\frac{9}{5}}\chi^{\mathsf{T}_{2}}_{\frac{1}{5}}+\chi^{c=\frac{104}{5}}_{\frac{12}{5}}\chi^{\mathsf{T}_{2}}_{\frac{3}{5}}=J+312.
\ee
\item We identify the following pair w.r.t $c=24$:
\be
l=0,\ (c,h,m_1)=(\frac{104}{5},\frac{4}{5},\frac{7}{5},\frac{8}{5},\frac{11}{5},520),\Longleftrightarrow
\T_2,\quad n_i=1,2,2,3.
\ee
The bilinear form of characters is
\be
\chi^{c=\frac{104}{5}}_0\chi^{\mathsf{T}_{2}}_0+4\chi^{c=\frac{104}{5}}_{\frac{4}{5}}\chi^{\mathsf{T}_{2}}_{\frac{1}{5}}+6\chi^{c=\frac{104}{5}}_{\frac{7}{5}}\chi^{\mathsf{T}_{2}}_{\frac{3}{5}}+4\chi^{c=\frac{104}{5}}_{\frac{8}{5}}\chi^{\mathsf{T}_{2}}_{\frac{2}{5}}+\chi^{c=\frac{104}{5}}_{\frac{11}{5}}\chi^{\mathsf{T}_{2}}_{\frac{4}{5}}=J+1560.
\ee
\item We identify the following pair w.r.t $c=32$:
\be
l=0,\ (c,h,m_1)=(\frac{128}{5},\frac{4}{5},\frac{7}{5},\frac{11}{5},\frac{13}{5},28),\Longleftrightarrow
\T_4,\quad n_i=2,2,3,3.
\ee
The bilinear form of characters is
\be
\sum\chi^{c=\frac{128}{5}}\chi^{\mathsf{T}_{4}}=j^{1/3}(J-140).
\ee
\item The $c=\frac{152}{5}$ theory is just $(E_{7\frac12})_1^{\otimes 4}$. Obviously, it is dual to $\T_1$ w.r.t $c=32$.
\be
l=0,\ (c,h,m_1)=(\frac{152}{5},\frac{4}{5},\frac{8}{5},\frac{12}{5},\frac{16}{5},760),\Longleftrightarrow
\T_1,\quad n_i=1,2,3,4.
\ee
\eitem

\subsection{Type $U(1)_4$ }
Consider the $c=1$ RCFT $U(1)_{4}$ model. It has conformal weights and degeneracy $h_i=0,(\frac{1}{16})_2,(\frac{1}{4})_2,(\frac{9}{16})_2,1$. Note this theory is degenerate due to the presence of weight-1 primary. It is easy to compute the five distinct characters to be
\be
\ba
\chi_0=\,&q^{-\frac{1}{24}} (1 + q + 2 q^2 + 3 q^3 + 7 q^4 + 9 q^5 + 15 q^6  +\dots),\\
\chi_{\frac{1}{16}}=\,&q^{\frac{1}{48}} (1 + q + 2 q^2 + 4 q^3 + 6 q^4 + 10 q^5 + 15 q^6   +\dots),\\
\chi_{\frac{1}{4}}=\,&q^{\frac{5}{24}} ( 1 + q + 3 q^2 + 4 q^3 + 7 q^4 + 10 q^5 + 17 q^6  +\dots),\\
\chi_{\frac{9}{16}}=\,&q^{\frac{25}{48}} ( 1 + 2 q + 3 q^2 + 5 q^3 + 8 q^4 + 12 q^5 + 18q^6  \dots),\\
\chi_{1}=\,&q^{\frac{23}{24}} ( 2 + 2 q + 4 q^2 + 6 q^3 + 10 q^4 + 14 q^5 + 22 q^6 +\dots).
\ea
\ee
Clearly the conductor is $N=48$. The $S$-matrix of the five characters is 
\be
\rho(S)= \frac{1}{\sqrt{8}}\left(
\begin{array}{ccccc}
 1 & 2 & 2 & 2 & 1 \\
 1 & \sqrt{2} & 0 & -\sqrt{2} & -1 \\
 1 & 0 & -2 & 0 & 1 \\
 1 & -\sqrt{2} & 0 & \sqrt{2} & -1 \\
 1 & -2 & 2 & -2 & 1 \\
\end{array}
\right).
\ee
As in Ising model, there are two classes for the Hecke operation: for $p^2\equiv 1 \textrm{ mod } 48$, $
\rho(\sigma_{p})=\mathrm{Id},$ while for $p^2\equiv 25 \textrm{ mod } 48$, 
\be
\rho(\sigma_p)= \left(
\begin{array}{ccccc}
 -1 & 0 & 0 & 0 & 0 \\
 0 & 0 & 0 & -1 & 0 \\
 0 & 0 & -1 & 0 & 0 \\
 0 & -1 & 0 & 0 & 0 \\
 0 & 0 & 0 & 0 & -1 \\
\end{array}
\right).
\ee
We compute all admissible Hecke images $\T_p$ for $p<48$ and collect them in $c=24$ pairs in Table \ref{tb:HeckeU1R4}. We find Hecke relations $\T_{7}U(1)_4=(A_7)_1$, and $\T_{17}U(1)_4$ can describe a subtheory of $(D_9)_2$. Together they form a $c=24$ pair in Schellekens' list No.50. The characters relations between $\T_{17}$ and WZW $(D_9)_2$  are for example:
\be
\ba
\chi_0^{\T_{17}}=&\chi^{(D_9)_2,2w_0}_0+\chi^{(D_9)_2,w_6}_2=q^{-\frac{17}{24}}(1 + 153 q + 30498 q^2 + 1078939 q^3+\dots),\\
\chi_{1}^{\T_{17}}=&\chi^{(D_9)_2,2w_1}_1+\chi^{(D_9)_2,w_6}_2=q^{\frac{7}{24}}(170 + 30362 q + 1079636 q^2+\dots).
\ea
\ee
We also compute Hecke images $\T_p$ for $25\le p\le 47$, however those have $l\ge 6$ MLDEs, thus are not of our main interest here.

\begin{table}[ht]
\def\arraystretch{1.1}
	\centering
	\begin{tabular}{|c|c|c|c|c|c|c|c|c|c|c|c|c|c|}
		\hline
		 \multicolumn{4}{|c|}{$l=0$} & \multicolumn{4}{c|}{$l'=2$} & \multicolumn{2}{c|}{duality}\\
		\hline
		 $c$ & $h_i$  & $m_1$ & remark & $\tc$ & $\tilh_i$ &  $\tm_1$ & remark & $m_1+\tm_1$ & Sch\\
		\hline
$1$ & $\frac{1}{16},\frac{1}{4},\frac{9}{16},1 $ & $ 1$ & $\T_{1} $ & $23$ & $ 1,\frac{23}{16},\frac{7}{4},\frac{31}{16}$ & $ 23$ & $\T_{23}$   &  $ 24  $  & $1 $\\
$7 $ & $\frac{7}{16},\frac{3}{4},\frac{15}{16},1  $ & $ 63$ & $\T_{ 7},(A_7)_1 $ & $17 $ & $1,\frac{17}{16},\frac{5}{4},\frac{25}{16} $ & $ 153 $ & $\T_{17 }$   &  $ 216  $  & $49,50 $\\
$13 $ & $ \frac{13}{16},1,\frac{5}{4},\frac{21}{16} $ & $ 195$ & $\T_{ 13},\star $ & $ 11$ & $ \frac{11}{16},\frac{3}{4},1,\frac{19}{16}$ & $ 165 $ & $\T_{11 },\star$   &  $ 360  $  & $ -$\\
$ 19$ & $1,\frac{19}{16},\frac{27}{16},\frac{7}{4}  $ & $ 133$ & $\T_{ 19},\star $ & $5 $ & $ \frac{1}{4},\frac{5}{16},\frac{13}{16},1$ & $ 35 $ & 
$\T_{5 },\star$   &  $ 168  $  & $- $\\
		\hline
	
		\end{tabular}
			\caption{Hecke images of $U(1)_{4}$. All non-integral weights have degeneracy two.}
			\label{tb:HeckeU1R4}
		\end{table}

\section{RCFTs with six characters}\label{sec:6chi}
RCFTs with six characters are not yet classified or studied from holomorphic modular bootstrap. Nevertheless, the rank 6 MTC with $N_{k}^{ij}\le 3$ are very recently classified in \cite{2022arXiv220314829N}, see Table 3 and 4 there. We choose several interesting initial theories with six characters to study their Hecke images and discover lots of new Hecke relations. In the following, we discuss the Hecke operation of type $(LY)_5$, $M(5,4)$, $M_{\rm sub}(7,6)$, $(LY)_1\otimes (LY)_2$, $(LY)_2^{\otimes 2}$ and $\mathrm{Ising}^{\otimes 2}$ individually. There of course exist more RCFTs with six characters not included in these Hecke types. For example, the Hecke images of $M_{\rm eff}(7,3)$ have been discussed in Appendix D.2.7 of \cite{Wuthesis} which reveals WZW $(C_5)_1=\T_{11}M_{\rm eff}(7,3)$. Besides, these also exist some RCFTs with integral central charge such as WZW $(B_2)_2$ and $(G_2)_3$ which seem to be a bit isolated in the Hecke program.

\subsection{Type $(LY)_5$ }
Non-unitary minimal model $M(13,2)$ has central charge $c=-\frac{350}{13}$ and conformal weights $h_i=0,-\frac{5}{13},-\frac{9}{13},-\frac{12}{13},-\frac{14}{13},-\frac{15}{13}$.
The effective theory $M_{\rm eff}(13,2)$, i.e., $(LY)_5$ has central charge
$c_{\rm eff}=\frac{10}{13}$ and conformal weights $h_{\rm eff}=0,\frac{1}{13},\frac{3}{13},\frac{6}{13},\frac{10}{13},\frac{15}{13}$. It is easy to compute the five characters to be
\be
\ba
\chi_0=\,&q^{-\frac{5}{156}} (1 + q + 2 q^2 + 3 q^3 + 5 q^4 + 7 q^5 + 10 q^6 + 13 q^7 +\dots),\\
\chi_{\frac{1}{13}}=\,&q^{\frac{7}{156}} (1 + q + 2 q^2 + 3 q^3 + 5 q^4 + 6 q^5 + 10 q^6 + 13 q^7 +\dots),\\
\chi_{\frac{3}{13}}=\,&q^{\frac{31}{156}} (1 + q + 2 q^2 + 3 q^3 + 4 q^4 + 6 q^5 + 9 q^6 + 12 q^7 +\dots),\\
\chi_{\frac{6}{13}}=\,&q^{\frac{67}{156}} (1 + q + 2 q^2 + 2 q^3 + 4 q^4 + 5 q^5 + 8 q^6 + 10 q^7 +\dots),\\
\chi_{\frac{10}{13}}=\,&q^{\frac{115}{156}} (1 + q + q^2 + 2 q^3 + 3 q^4 + 4 q^5 + 6 q^6 + 8 q^7 +\dots),\\
\chi_{\frac{15}{13}}=\,&q^{\frac{175}{156}} (1 + q^2 + q^3 + 2 q^4 + 2 q^5 + 4 q^6 + 4 q^7 +\dots).
\ea
\ee
Clearly the conductor $N=156$. The classes of Hecke operations and various $\rho(\sigma_p)$ matrices have been discussed in Appendix D.2.3 of \cite{Wuthesis}. We compute the Hecke images for all admissible $p< 156$ and collect the results for $p<38$ in Table \ref{tb:HeckeM132}. The $\T_{37}$ image belongs to the MTC class $6_{-46/13}^B$ in the Table III of \cite{Schoutens:2015uia}. Meanwhile, a unitary $c=\frac{46}{13}$ theory as the root of WZW $(A_1)_{11}$ belonging to the MTC class $6_{46/13}^B$ \cite{Schoutens:2015uia} is dual to $\T_{37}$ w.r.t $c=32$. Most Hecke images of type $(LY)_5$ appear to be non-unitary.

\begin{table}[ht]
\def\arraystretch{1.1}
	\centering
	\begin{tabular}{|c|c|c|c|c|c|c|c|c|c|c|c|c|c|}
		\hline
		 $c$ & $h_i$  & $m_1$ & $l$ & remark  \\
		\hline
$\frac{10}{13}$  & $\frac{1}{13},\frac{3}{13},\frac{6}{13},\frac{10}{13},\frac{15}{13}$ & $1$ & $0$  & $\T_1$  \\
$\frac{50}{13}$  & $\frac{2}{13},\frac{4}{13},\frac{5}{13},\frac{10}{13},\frac{11}{13} $ & $20$ & $6$  & $\T_{5},\star$  \\
$\frac{70}{13}$  & $\frac{3}{13},\frac{5}{13},\frac{7}{13},\frac{8}{13},\frac{14}{13} $ & $35$ & $6$  & $\T_{7},\star$  \\
$\frac{110}{13}$  & $ \frac{6}{13},\frac{7}{13},\frac{9}{13},\frac{11}{13},\frac{14}{13}$ & $44$ &  $6$ & $\T_{11},\star$  \\
$\frac{170}{13}$  & $\frac{8}{13},\frac{11}{13},\frac{12}{13},\frac{14}{13},\frac{17}{13} $ & $136$ & $6$  & $\T_{17},\star$  \\
$\frac{190}{13}$  & $\frac{10}{13},\frac{12}{13},\frac{18}{13},\frac{19}{13},\frac{21}{13} $ & $190$ &  $0$ & $\T_{19},(E_{7\frac12})_{2}$  \\
$\frac{230}{13}$  & $\frac{10}{13},\frac{17}{13},\frac{20}{13},\frac{21}{13},\frac{22}{13} $ & $230$ &  $0$ & $\T_{23},\star$   \\
$\frac{250}{13}$  & $\frac{10}{13},\frac{11}{13},\frac{12}{13},\frac{16}{13},\frac{20}{13} $ & $175$ &  $12$ & $\T_{25},\star$   \\
$\frac{290}{13}$  & $\frac{16}{13},\frac{17}{13},\frac{18}{13},\frac{19}{13},\frac{22}{13} $ & $58$ & $6$  & $\T_{29},\star$   \\
$\frac{310}{13}$  & $\frac{15}{13},\frac{17}{13},\frac{18}{13},\frac{23}{13},\frac{24}{13} $ & $31$ & $6$  & $\T_{31}$   \\
$\frac{350}{13}$  & $ \frac{14}{13},\frac{15}{13},\frac{18}{13},\frac{22}{13},\frac{25}{13}$ & $0$ &  $12$ & $\T_{35},\star$   \\
$\frac{370}{13}$  & $\frac{19}{13},\frac{20}{13},\frac{22}{13},\frac{24}{13},\frac{27}{13} $ & $0$ & $6$  & $\T_{37}$  \\
		\hline
	
		\end{tabular}
			\caption{Hecke images $\mathsf{T}_p$ of $(LY)_5$ for $p<38$. }
			\label{tb:HeckeM132}
		\end{table}

As a fascinating example, let us consider the $\mathsf{T}_{19}$ Hecke image of $(LY)_5$. We find this Hecke image satisfies an $l=0$ MLDE and have conformal weights
$
h_i=0,\frac{10}{13},\frac{12}{13},\frac{18}{13},\frac{19}{13},\frac{21}{13}.
$
We regard this as exactly \emph{WZW $E_{7\frac12}$ model of level 2}! To the best of our knowledge, WZW models $(E_{7\frac12})_k$ for level $k>1$ have not been discussed in the literature yet. We compute the characters from Hecke operation to be
\be\label{chiE7half2}
\ba
\chi_0=\,&q^{-\frac{95}{156}} (1 + 190 q + 18335 q^2 + 448210 q^3 + 6264585 q^4 + 62455698 q^5 +\dots),\\
\chi_{\frac{10}{13}}=\,&q^{\frac{25}{156}} (57 + 10830 q + 321575 q^2 + 4979330 q^3 + 53025295 q^4+\dots),\\
\chi_{\frac{12}{13}}=\,&q^{\frac{49}{156}} (190 + 20596 q + 537890 q^2 + 7761500 q^3 + 79066030 q^4+\dots),\\
\chi_{\frac{18}{13}}=\,&q^{\frac{121}{156}} (1045 + 48070 q + 910955 q^2 + 10983690 q^3 + 99272435 q^4+\dots),\\
\chi_{\frac{19}{13}}=\,&q^{\frac{133}{156}} (2640 + 109155 q + 1979610 q^2 + 23245740 q^3 + 206319480 q^4+\dots),\\
\chi_{\frac{21}{13}}=\,&q^{\frac{157}{156}} (1520 + 51395 q + 860890 q^2 + 9606457 q^3 + 82347710 q^4+\dots),
\ea
\ee
with $S$-matrix
\be\label{SE7half2}
\rho(S)= \frac{2}{\sqrt{13}}\left(
\begin{array}{cccccc}
 \sin \left(\frac{3 \pi }{13}\right) & \cos \left(\frac{3 \pi }{26}\right) & \cos \left(\frac{\pi }{26}\right) & \sin \left(\frac{2 \pi }{13}\right) & \cos \left(\frac{5 \pi }{26}\right) & \sin \left(\frac{\pi }{13}\right) \\
 \cos \left(\frac{3 \pi }{26}\right) & -\cos \left(\frac{5 \pi }{26}\right) & \sin \left(\frac{3 \pi }{13}\right) & \sin \left(\frac{\pi }{13}\right) & -\sin \left(\frac{2 \pi }{13}\right) & -\cos \left(\frac{\pi }{26}\right) \\
 \cos \left(\frac{\pi }{26}\right) & \sin \left(\frac{3 \pi }{13}\right) & -\sin \left(\frac{\pi }{13}\right) & -\cos \left(\frac{5 \pi }{26}\right) & -\cos \left(\frac{3 \pi }{26}\right) & \sin \left(\frac{2 \pi }{13}\right) \\
 \sin \left(\frac{2 \pi }{13}\right) & \sin \left(\frac{\pi }{13}\right) & -\cos \left(\frac{5 \pi }{26}\right) & -\sin \left(\frac{3 \pi }{13}\right) & \cos \left(\frac{\pi }{26}\right) & -\cos \left(\frac{3 \pi }{26}\right) \\
 \cos \left(\frac{5 \pi }{26}\right) & -\sin \left(\frac{2 \pi }{13}\right) & -\cos \left(\frac{3 \pi }{26}\right) & \cos \left(\frac{\pi }{26}\right) & \sin \left(\frac{\pi }{13}\right) & \sin \left(\frac{3 \pi }{13}\right) \\
 \sin \left(\frac{\pi }{13}\right) & -\cos \left(\frac{\pi }{26}\right) & \sin \left(\frac{2 \pi }{13}\right) & -\cos \left(\frac{3 \pi }{26}\right) & \sin \left(\frac{3 \pi }{13}\right) & \cos \left(\frac{5 \pi }{26}\right) \\
\end{array}
\right) .
\ee
We find the six characters satisfy the following 6th order MLDE:
\be
\ba
&[{\cal D}^{6}+\mu_1E_4{\cal D}^{4}+\mu_2E_6{\cal D}^{3}+\mu_3E_4^2{\cal D}^{2}+\mu_4E_4E_6{\cal D}+(\mu_5E_4^3+\mu_6E_6^2)]\chi=0,\\[+2mm]
&\mu_1=-\frac{1225}{1872},\quad \mu_2=\frac{25205}{36504},\quad \mu_3=-\frac{1349885}{3504384} ,\quad \mu_4= \frac{36703535}{296120448} \\
&\mu_5= -\frac{57214927525}{4804258148352},\quad\mu_6=-\frac{3824637775}{450399201408}.
\ea
\ee
It was known $E_{7\frac12}$ as an intermediate Lie algebra has dimension $190$, dual Coxeter number 24 and fundamental representation $\bf 57$ with Casimir invariants $40$ \cite{Kawasetsu}. Assuming the central charge formula \eqref{WZWc} of WZW models still work, then for $(E_{7\frac12})_k$ model, central charge $c_k={190k}/({k+24})$. For example, $c_2=\frac{190}{13}$, $c_3= \frac{190}{9}$,  $c_4=\frac{190}{7}$... Thus $c_2$ matches with the central charge of $\T_{19}(LY)_5$. Besides, assuming the weight formula \eqref{WZWweights} of WZW models still work, then $(E_{7\frac12})_2$ should have a character of weights $\frac{10}{13}$ with leading Fourier coefficients $57$. Apparently, \eqref{chiE7half2} also satisfies this condition. One more evidence is that the character $\chi_{\frac{12}{13}}$ in \eqref{chiE7half2} has leading Fourier coefficients $190$ which is exactly the dimension of adjoint representation of $E_{7\frac12}$. These support us to propose \eqref{chiE7half2} as the characters of generalized WZW $(E_{7\frac12})_2$. We remark that $(E_{7})_2$ also has six characters, the same number with $(E_{7\frac12})_2$.

We newly recognize from \eqref{chiE7half2} more “irreducible integrable representations" of $E_{7\frac12}$ which are $\bf 1045,2640$ and $\bf 1520$ with Casimir invariants $72,76,84$ respectively.\footnote{The Casimir invariants suggest that $(E_{7\frac12})_3$ should contain conformal weights $h_i=0,\frac{20}{27},\frac{8}{9},\frac{4}{3},\frac{38}{27},\frac{14}{9}$ with $\alpha_i$ as $-\frac{95}{108},-\frac{5}{36},\frac{1}{108},\frac{49}{108},\frac{19}{36},\frac{73}{108}$. Presumably, $(E_{7\frac12})_3$ has 12 characters and conductor 108.} 
We remark that the representation $\bf 2640$ appears in the defining property of Deligne exceptional series, see e.g. \cite{LM}:
\be
\mathrm{Sym}^2\mathbf{190}={\bf 1+ 15504+2640}, 
\ee
where ${\bf 15504}=2\bf Adj$. Similarly $\mathrm{Alt}^2\mathbf{190}=\mathbf{190}+\mathbf{17765}$. We also notice the $\bf 1520$ appears as the $\bf Y_3^*$ representation in the notion of (\cite{Beem:2019snk}, Appendix A.3). In fact, we are able to recognize the decomposition of representations in the flavored characters. For example, for the vacuum character, we find
\be
\ba
\chi_0=\,&q^{-\frac{95}{156}} (\mathbf{1} + \mathbf{190} q + ({\bf 1+190+2640+ 15504}) q^2  +({\bf 1} + 3\times{\bf 190} + \mathbf{1520}\\
&\qquad\ \  +\mathbf{2640}+{\bf 15504} + 
 2\times{\bf 17765 + 392445})q^3+\dots).
\ea
\ee
Here $\bf 392445$ is the representation $\bf A$ in the notion of (\cite{Beem:2019snk}, Appendix A.3). Besides, from the $S$-matrix \eqref{SE7half2}, it is easy to determine the ``fusion rule" for the non-vacuum primaries of $(E_{7\frac12})_2$ as follows
\be
\ba
&\phi_1\times \phi_1=\phi_0-\phi_1+\phi_2-\phi_3+\phi_4+\phi_5,\quad \phi_1\times \phi_2=\phi_1+\phi_3,\\
&\phi_1\times \phi_3=-\phi_1+\phi_2-\phi_3+\phi_4+\phi_5,\quad \phi_1\times \phi_4=\phi_1+\phi_3-\phi_5,\\
&\phi_1\times \phi_5=\phi_1+\phi_3-\phi_4-\phi_5,\quad \phi_2\times \phi_2=\phi_0+\phi_4,\quad \phi_2\times \phi_3=\phi_1-\phi_5,\\
&\phi_2\times \phi_4=\phi_2+\phi_5,\quad \phi_2\times \phi_5=-\phi_3+\phi_4,
\quad \phi_3\times \phi_3=\phi_0-\phi_1-\phi_3+\phi_4+\phi_5,\\
&\phi_3\times \phi_4=\phi_1+\phi_3-\phi_4,\quad \phi_3\times \phi_5=\phi_1-\phi_2+\phi_3-\phi_5,\quad  \phi_4\times \phi_4=\phi_0-\phi_3+\phi_4,\\
&\phi_4\times \phi_5=-\phi_1+\phi_2+\phi_5,\quad \phi_5\times \phi_5=\phi_0-\phi_1-\phi_3+\phi_4.
\ea
\ee

For $c=8k$ cosets, if we want both theories of a pair to be $\mathsf{T}_{p}(LY)_5$, then the first possibility is $c=40$ and $p+p'=52$. Unfortunately, there is no chance for both of them to avoid quasi-characters. Pairs with quasi-characters include
$
\sum\chi^{\mathsf{T}_{5}}\chi^{\mathsf{T}_{47}}=j^{2/3} (    J-516),
$
$
\sum\chi^{\mathsf{T}_{11}}\chi^{\mathsf{T}_{41}}=j^{2/3} (    J-540),
$ and 
$
\sum\chi^{\mathsf{T}_{23}}\chi^{\mathsf{T}_{29}}=j^{2/3} (    J-784).$
The next possibility is $c=80$ and $p+p'=104$. We notice the pairs including $(p,p')=(31,73)$ with $l=6,l'=12$, 
    $(p,p')=(37,67)$ with $l=6,l'=18$ and
 $(p,p')=(43,61)$ with $l=6,l'=12$.

\subsection{Type $M(5,4),$ tri-critical Ising }\label{sec:M54}
The tri-critical Ising model $M(5,4)$ is a unitary RCFT with central charge $c=\frac{7}{10}$ and conformal weights $h_i=0,\frac{3}{80},\frac{1}{10},\frac{7}{16},\frac{3}{5},{\frac{3}{2} }$. This is a well-known example which can be fermionized to the first unitary $\mathcal{N}=1$ supersymmetic minimal model $SM(l+2,l)$ for $l=3$ \cite{Friedan:1983xq}. The six characters of $M(5,4)$ have the following Fourier expansion
\be
\ba
\chi_0=\,&q^{-\frac{7}{240}} (1 + q^2 + q^3 + 2 q^4 + 2 q^5 + 4 q^6 + 4 q^7 +\dots),\\
\chi_{\frac{3}{80}}=\,&q^{\frac{1}{120}} (1 + q + 2 q^2 + 3 q^3 + 4 q^4 + 6 q^5 + 8 q^6 + 11 q^7 +\dots),\\
\chi_{\frac{1}{10}}=\,&q^{\frac{17}{240}} (1 + q + q^2 + 2 q^3 + 3 q^4 + 4 q^5 + 6 q^6 + 8 q^7 +\dots),\\
\chi_{\frac{7}{16}}=\,&q^{\frac{49}{120}} (1 + q + q^2 + 2 q^3 + 3 q^4 + 4 q^5 + 6 q^6 + 8 q^7 +\dots),\\
\chi_{\frac{3}{5}}=\,&q^{\frac{137}{240}} (1 + q + 2 q^2 + 2 q^3 + 4 q^4 + 5 q^5 + 7 q^6 + 9 q^7 +\dots),\\
\chi_{\frac{3}{2}}=\,&q^{\frac{353}{240}} (1 + q + 2 q^2 + 2 q^3 + 3 q^4 + 4 q^5 + 6 q^6 + 7 q^7 +\dots).
\ea
\ee
Clearly, the conductor $N=240$. The $S$-matrix is 
\be
\rho(S)=\left(
\begin{array}{cccccc}
 a_{-}& \sqrt{2}a_{+}& a_{+}& \sqrt{2}a_{-}& a_{+}& a_{-}\\
 \sqrt{2}a_{+}& 0 & \sqrt{2}a_{-}& 0 & -\sqrt{2}a_{-}& -\sqrt{2}a_{+}\\
 a_{+}& \sqrt{2}a_{-}& -a_{-}& -\sqrt{2}a_{+}& -a_{-}& a_{+}\\
 \sqrt{2}a_{-}& 0 & -\sqrt{2}a_{+}& 0 & \sqrt{2}a_{+}& -\sqrt{2}a_{-}\\
 a_{+}& -\sqrt{2}a_{-}& -a_{-}& \sqrt{2}a_{+}& -a_{-}& a_{+}\\
 a_{-}& -\sqrt{2}a_{+}& a_{+}& -\sqrt{2}a_{-}& a_{+}& a_{-}\\
\end{array}
\right)
.
\ee
Here $a_{\pm}=(\frac{1}{5}  (\frac{5}{8}\pm\frac{\sqrt{5}}{8} ))^{1/2}$. 

Let us now consider the Hecke images $\T_p$ of $M(5,4)$. There exist eight classes for the Hecke operation, each containing 8 $p$ coprime to 240. We summarize in the following table the classes of quadratic residue $p^2\equiv a \textrm{ mod } 240 $.
\begin{table}[ht]
	\centering
	\begin{tabular}{c|c|c}
		\hline
		 $a$ & $+$ & $-$   \\ \hline
1		 & $1, 41, 79, 119, 121, 161, 199, 239$ & $ 31, 49, 71, 89, 151, 169, 191, 209$ \\
49		 & $7, 47, 73, 113, 127, 167, 193, 233$ & $ 17, 23, 97, 103, 137, 143, 217, 223$\\
121		 & $ 11, 29, 91, 109, 131, 149, 211, 229$ & $19, 59, 61, 101, 139, 179, 181, 221 $\\
169		 &  $ 13, 53, 67, 107, 133, 173, 187, 227$ & $ 37, 43, 77, 83, 157, 163, 197, 203$\\
\hline
		\end{tabular}
		\end{table}
Here the signs mean that the $\rho(\sigma_{p})$ in the right column is the negative of the $\rho(\sigma_{p})$ in the left,  for example $ \rho(\sigma_{31})=-\rho(\sigma_{1})$. For each class, we compute the typical $\rho(\sigma_{p})$ matrices as
\be
\rho(\sigma_{1})=\mathrm{Id},\qquad
\rho(\sigma_{7})=\left(
\begin{array}{cccccc}
 0 & 0 & 0 & 0 & 1 & 0 \\
 0 & 0 & 0 & -1 & 0 & 0 \\
 0 & 0 & 0 & 0 & 0 & -1 \\
 0 & 1 & 0 & 0 & 0 & 0 \\
 -1 & 0 & 0 & 0 & 0 & 0 \\
 0 & 0 & 1 & 0 & 0 & 0 \\
\end{array}
\right).
\ee
\be
\rho(\sigma_{11})=\left(
\begin{array}{cccccc}
 0 & 0 & 0 & 0 & 0 & -1 \\
 0 & 1 & 0 & 0 & 0 & 0 \\
 0 & 0 & 0 & 0 & -1 & 0 \\
 0 & 0 & 0 & 1 & 0 & 0 \\
 0 & 0 & -1 & 0 & 0 & 0 \\
 -1 & 0 & 0 & 0 & 0 & 0 \\
\end{array}
\right),\qquad \rho(\sigma_{13})=\left(
\begin{array}{cccccc}
 0 & 0 & 1 & 0 & 0 & 0 \\
 0 & 0 & 0 & 1 & 0 & 0 \\
 -1 & 0 & 0 & 0 & 0 & 0 \\
 0 & -1 & 0 & 0 & 0 & 0 \\
 0 & 0 & 0 & 0 & 0 & -1 \\
 0 & 0 & 0 & 0 & 1 & 0 \\
\end{array}
\right).
\ee
Then all Hecke images $\T_p$ of $M(5,4)$ can be obtained from these $\rho(\sigma_{p})$ matrices. We collect the relevant information of Hecke images for all admissible $p\le 43$ in Table~\ref{tb:HeckeM54}. 

It is interesting that the WZW model $(E_7)_2$ emerges as the $\T_{19}$ image of $M(5,4)$. We compute the $\T_{19}M(5,4)$ image as
\be
\ba
\chi_{\frac32}=&\,q^{\frac{227}{240}}(1463+43757q+ 654664q^2+ 6593608q^3+ 51472463q^4+\dots) ,\\
\chi_{\frac{57}{80}}=&\,q^{\frac{19}{120}}(56+7448q+ 186352q^2+ 2512104q^3+ 23785720q^4+\dots)  ,\\
\chi_{\frac75}=&\,q^{\frac{203}{240}}(1539+52535q+ 824999q^2+ 8580229q^3+68438741q^4
\dots)  ,\\
\chi_{\frac{21}{16}}=&\,q^{\frac{91}{120}}(912+35112q+ 577752q^2+ 6183968q^3+50300600q^4+\dots),\\
\chi_{\frac{9}{10}}=&\,q^{\frac{83}{240}}(133+10318q+ 222509q^2+ 2768129q^3+ 24909931q^4+\dots) ,\\
\chi_0=&\,q^{-\frac{133}{240}}(1+133q+ 9044q^2+ 180215q^3+ 2158324q^4+\dots).
\ea
\ee
These are exactly the six characters of WZW $(E_7)_2$ theory! The $(E_7)_2$ theory can be fermionized to a supersymmetric theory satisfying a second order fermionic MLDE \cite{Bae:2020xzl}.

\begin{table}[ht]
\def\arraystretch{1.1}
	\centering
	\begin{tabular}{|c|c|c|c|c|c|c|c|c|c|c|c|c|c|}
		\hline
		 $c$ & $h_i$  & $m_1$ & $l$ & remark   \\
		\hline
$\frac{7}{10}$  & $\frac{3}{80},\frac{1}{10},\frac{7}{16},\frac{3}{5},\nb{\frac{3}{2} }$ & $0$ &  $0$ & $\T_{1}$   \\
$\frac{49}{10}$  & $\frac{1}{5},\frac{21}{80},\frac{1}{2},\frac{7}{10},\frac{17}{16} $ & $35$  &  $6$ & $\T_{7},\star$   \\
$\frac{77}{10}$  & $ \frac{33}{80},\frac{1}{2},\frac{3}{5},\frac{13}{16},\frac{11}{10}$ & $66$  & $6$  & $\T_{11},\star$  \\
$\frac{91}{10}$  & $\frac{39}{80},\frac{1}{2},\frac{11}{16},\frac{4}{5},\frac{13}{10} $ & $143$ &  $6$ & $\T_{13},\star$   \\
$\frac{119}{10}$  & $\frac{1}{2},\frac{51}{80},\frac{7}{10},\frac{6}{5},\frac{23}{16} $ & $221$ &  $6$ & $\T_{17}$  \\
$\frac{133}{10}$  & $\frac{57}{80},\frac{9}{10},\frac{21}{16},\frac{7}{5},\nb{\frac{3}{2}} $ & $133$ & $0$  & $\T_{19},(E_7)_2$   \\
$\frac{161}{10}$  & $\frac{4}{5},\frac{69}{80},\frac{17}{16},\frac{13}{10},\nb{\frac{3}{2}} $ & $207$ &  $6$ & $\T_{23}$  \\
$\frac{203}{10}$  & $\frac{9}{10},\frac{87}{80},\frac{7}{5},\nb{\frac{3}{2}},\frac{27}{16} $ & $58$ &  $6$ & $\T_{29},\star$   \\
$\frac{217}{10}$  & $\frac{11}{10},\frac{93}{80},\nb{\frac{3}{2}},\frac{25}{16},\frac{8}{5} $ & $31$  & $6$  & $\T_{31},\star$  \\
$\frac{259}{10}$  & $\frac{19}{16},\frac{6}{5},\frac{111}{80},\nb{\frac{3}{2}},\frac{17}{10} $ & $0$ &  $12$ & $\T_{37}$   \\
$\frac{287}{10}$  & $\nb{\frac{3}{2}},\frac{123}{80},\frac{8}{5},\frac{31}{16},\frac{21}{10} $ & $0$ &  $6$ & $\T_{41}$ \\
$\frac{301}{10}$  & $ \frac{13}{10},\nb{\frac{3}{2}},\frac{129}{80},\frac{9}{5},\frac{29}{16}$ & $0$ & $12$  & $\T_{43}$  \\
		\hline
		\end{tabular}
			\caption{Hecke images $\mathsf{T}_p$ of $M(5,4)$ for all admissible $p
			\le 43$. }
			\label{tb:HeckeM54}
		\end{table}

For $c=8k$ cosets, if we want both theories of a pair to be Hecke images of $M(5,4)$, then the first possibility is $c=56$ and $p+p'=80$. We find pairs such as
$(p,p')=(1,79)$, $l=0,l'=18$ and $(p,p')=(19,61)$, $l=0,l'=6$. Let us consider the first pair. The $\mathsf{T}_{79}M(5,4)$ theory has conformal weights  $h_i=0, \frac{12}{5},\frac{5}{2},\frac{41}{16},\frac{29}{10},\frac{237}{80} $. We notice this theory belongs to the MTC class $6_{-7/10}^B$ in Table III of \cite{Schoutens:2015uia}.
Note this pair has $n_i=3,3,3,3,4.$ It is easy to see this satisfies the relation \eqref{llrelation}.
We checked the bilinear relation of the characters of this pair to be
$
\sum\chi^{\mathsf{T}_{1}}\chi^{\mathsf{T}_{79}}=j^{1/3} (    J^2- 248 J-336387).
$
Consider the second pair $\mathsf{T}_{19}M(5,4)=(E_7)_2$ and $\mathsf{T}_{61}M(5,4)$ which have $l=0,6$ respectively. We notice $\mathsf{T}_{61}M(5,4)$ theory which has weights $h_i=0,\frac{21}{10},\frac{183}{80},\frac{5}{2},\frac{13}{5},\frac{43}{16} $ belongs to the MTC $6_{27/10}^B$ in Table III of \cite{Schoutens:2015uia}.
This pair has $n_i=3,3,4,4,4.$ It is easy to see this again satisfies the relation \eqref{llrelation}.
We checked the bilinear relation of the characters of this pair to be
$
\sum\chi^{\mathsf{T}_{19}}\chi^{\mathsf{T}_{61}}=j^{1/3} (    J^2- 115 J-358620).
$

\subsection{Type $M_{\rm sub}(7,6),$ tri-critical 3-states Potts model }
Unitary minimal model $M(7,6)$ has central charge $c=\frac{6}{7}$ and 15 primaries with conformal weights
$
h_i=0,\frac{1}{56},\frac{1}{21},\frac{5}{56},\frac{1}{7},\frac{3}{8},\frac{10}{21},\frac{33}{56},\frac{5}{7},\frac{4}{3},\frac{85}{56},\frac{12}{7},\frac{23}{8},\frac{22}{7},5.
$
It is well-known a subset of the 15 primaries can form a new theory $M_{\rm sub}(7,6)$ describing the tri-critical 3-states Potts model, also known as $(D_4,A_6)$ theory, which has partition function \cite{DiFrancesco:1997nk}
\be
Z=\sum_{s=1,2,3}|\chi_{1,s}+\chi_{5,s}|^2+2|\chi_{3,s}|^2.
\ee
Let us denote
\be
\ba
\chi^P_{0}=\,&\chi_{1,1}+\chi_{5,1},\quad \chi^P_{\frac17}=\chi_{1,2}+\chi_{5,2},\quad\chi^P_{\frac57}=\chi_{1,3}+\chi_{5,3},\\
&\chi^P_{\frac{1}{21}}=\chi_{3,3},\quad \chi^P_{\frac{10}{21}}=\chi_{3,2},\quad \chi^P_{\frac{4}{3}}=\chi_{3,1}.
\ea
\ee
Note the three primaries in the second line have degeneracy two. The six characters have the following Fourier expansion:
\be
\ba
\chi^P_{0}=\,&q^{-\frac{1}{28}} (1 + q^2 + q^3 + 2 q^4 + 3 q^5 + 5 q^6 + 6 q^7 + \dots)  ,\\
\chi^P_{\frac17}=\,& q^{\frac{3}{28}} (1 + q + q^2 + 3 q^3 + 4 q^4 + 6 q^5 + 9 q^6 + 13 q^7 + \dots),\\
\chi^P_{\frac57}=\,& q^{\frac{19}{28}}(1 + 2 q + 3 q^2 + 4 q^3 + 7 q^4 + 9 q^5 + 14 q^6 + 19 q^7 +\dots) ,\\
\chi^P_{\frac{1}{21}}=\,& q^{\frac{1}{84}}(1 + q + 2 q^2 + 3 q^3 + 5 q^4 + 7 q^5 + 11 q^6 + 15 q^7 +\dots)  ,\\
\chi^P_{\frac{10}{21}}=\,& q^{\frac{37}{84}}(1 + q + 2 q^2 + 3 q^3 + 5 q^4 + 7 q^5 + 10 q^6 + 14 q^7 +\dots) ,\\
\chi^P_{\frac{4}{3}}=\,& q^{\frac{109}{84}} (1 + q + 2 q^2 + 2 q^3 + 4 q^4 + 5 q^5 + 8 q^6 + 10 q^7 +\dots) .
\ea
\ee
Clearly, the conductor $N=84$.
There exist in total six classes for the Hecke operation $\T_p$, each contains four $p$ coprime to 84. We summarize in the following table the classes of quadratic residue $p^2\equiv a \textrm{ mod } 84$. 
\begin{table}[ht]
	\centering
	\begin{tabular}{c|c|c}
		\hline
		 $a$ & $+$ & $-$   \\ \hline
1		 & $1, 13,71, 83$ & $ 29, 41, 43, 55$ \\
25		 & $5, 19,65, 79$ & $ 23, 37, 47, 61$\\
121		 & $ 11, 25, 59, 73$ & $17,  31, 53,  67 $\\
\hline
		\end{tabular}
		\end{table}
		Again the $\rho(\sigma_{p})$ in the right column is the negative of the $\rho(\sigma_{p})$ in the left.
For each class, we compute the typical $\rho(\sigma_{p})$ matrices as $\rho(\sigma_{1})=\mathrm{Id}$,
\be
\rho(\sigma_{ 5})=\left(
\begin{array}{cccccc}
 0 & 0 & 0 & 0 & 1 & 0 \\
 0 & 0 & 0 & -1 & 0 & 0 \\
 1 & 0 & 0 & 0 & 0 & 0 \\
 0 & 0 & 0 & 0 & 0 & 1 \\
 0 & 0 & -1 & 0 & 0 & 0 \\
 0 & 1 & 0 & 0 & 0 & 0 \\
\end{array}
\right),\qquad \rho(\sigma_{11})=\left(
\begin{array}{cccccc}
 0 & 0 & -1 & 0 & 0 & 0 \\
 0 & 0 & 0 & 0 & 0 & -1 \\
 0 & 0 & 0 & 0 & 1 & 0 \\
 0 & 1 & 0 & 0 & 0 & 0 \\
 -1 & 0 & 0 & 0 & 0 & 0 \\
 0 & 0 & 0 & -1 & 0 & 0 \\
\end{array}
\right).
\ee
All Hecke images $\T_p$ can be obtained from these $\rho(\sigma_{p})$ matrices. We collect the relevant information of Hecke images for all admissible $p\le 30$ in Table~\ref{tb:HeckeM76}.

\begin{table}[ht]
\def\arraystretch{1.1}
	\centering
	\begin{tabular}{|c|c|c|c|c|c|c|c|c|c|c|c|c|c|}
		\hline
		 $c$ & $h_i$  & $m_1$ & $l$ & remark  \\
		\hline
$\frac{6}{7}$  & $ \frac{1}{21},\frac{1}{7},\frac{10}{21},\frac{5}{7},\frac{4}{3} $ & $ 0$ & $ 0$  & $\T_{1}$   \\
$\frac{30}{7}$  & $ \frac{8}{21},\frac{4}{7},\frac{2}{3},\frac{5}{7},\frac{26}{21} $ & $ 20$ & $0 $  & $\T_{5}$   \\
$\frac{66}{7}$  & $ \frac{11}{21},\frac{4}{7},\frac{2}{3},\frac{6}{7},\frac{26}{21} $ & $154 $ & $ 6$  & $\T_{11},\star$   \\
$\frac{78}{7}$  & $ \frac{13}{21},\frac{6}{7},\frac{25}{21},\frac{9}{7},\frac{4}{3} $ & $ 78$ & $ 0$  & $\T_{13},(E_6)_2$   \\
$\frac{102}{7}$  & $ \frac{2}{3},\frac{17}{21},\frac{23}{21},\frac{8}{7},\frac{10}{7} $ & $238 $ & $ 6$  & $\T_{17}$   \\
$\frac{114}{7}$  & $ \frac{5}{7},\frac{19}{21},\frac{22}{21},\frac{4}{3},\frac{11}{7} $ & $ 171$ & $ 6$  & $\T_{19}$   \\
$\frac{138}{7}$  & $ \frac{20}{21},\frac{23}{21},\frac{9}{7},\frac{10}{7},\frac{5}{3} $ & $ 92$ & $ 6$  & $\T_{23},\star$   \\
$\frac{120}{7}$  & $ \frac{19}{21},\frac{25}{21},\frac{4}{3},\frac{11}{7},\frac{13}{7} $ & $75$ & $ 6$  & $\T_{25},\star$   \\
$\frac{174}{7}$  & $ \frac{8}{7},\frac{29}{21},\frac{5}{3},\frac{12}{7},\frac{38}{21} $ & $ 0$ & $ 6$  & $\T_{29},\star$   \\
		\hline
	
		\end{tabular}
			\caption{Hecke images $\mathsf{T}_p$ of $M_{\rm sub}(7,6)$ for all admissible $p\le 30$. All weights with denominators $21$ and $3$ have degeneracy two.}
			\label{tb:HeckeM76}
		\end{table}

Interestingly, we find the $\mathsf{T}_{13}$ Hecke image gives exactly the characters of $(E_6)_2$ WZW theory which has Fourier expansion
\be
\ba
\chi_0^{2w_0}=&\,q^{-\frac{13}{28}}(1+78q+ 3159q^2+ 44564q^3+ 411372q^4+\dots) ,\\
\chi_{\frac{13}{21}}^{w_1+w_0}=&\,q^{\frac{13}{84}}(27+2106q+ 38961q^2+ 411723q^3+ 3172689q^4+\dots)  ,\\
\chi_{\frac{6}{7}}^{w_2}=&\,q^{\frac{11}{28}}(78+3654q+ 56862q^2+ 549796q^3+ 3997968q^4+\dots) ,\\
\chi_{\frac{25}{21}}^{w_3}=&\,q^{\frac{61}{84}}(351+9828q+ 126009q^2+ 1093716q^3+ 7389603q^4+\dots) ,\\
\chi_{\frac{9}{7}}^{w_1+w_6}=&\,q^{\frac{23}{28}}(650+15951q+ 195858q^2+ 1650961q^3+ 10943478q^4+
\dots) ,\\
\chi_{\frac{4}{3}}^{2w_1}=&\,q^{\frac{73}{84}}(351+8073q+ 97227q^2+ 807651q^3 +5304663q^4+\dots).
\ea
\ee
One can easily see the degeneracy is preserved upon Hecke operations from the affine Dynkin diagram of $\hat{E}_6^{(1)}$. 
Indeed, the characters with weights $\frac{13}{21},\frac{25}{21},\frac{4}{3}$ have degeneracy two, corresponding to $E_6$ irreducible representations $\bf 27,\overline{27},351,\overline{351},351',\overline{351'}$. The degeneracy is transferred by the $\rho(\sigma_{13})$ matrix.

\subsection{Type $(LY)_1\otimes(LY)_2$ }
Consider the product theory $M_{\rm eff}(5,2)\otimes M_{\rm eff}(7,2)$ which has effective central charge $c_{\rm eff}=\frac25+\frac47=\frac{34}{35}$ and conformal weights $h_{\rm eff}=0,\frac{1}{7},\frac{1}{5},\frac{12}{35},\frac{3}{7},\frac{22}{35}.$ It is easy to compute the six characters to be
\be
\ba
\chi_0=\,&q^{-\frac{17}{420}} (1 + 2 q + 4 q^2 + 6 q^3 + 10 q^4 + 15 q^5 + 24 q^6 + 34 q^7+\dots),\\
\chi_{\frac{1}{7}}=\,&q^{\frac{43}{420}} (1 + 2 q + 3 q^2 + 5 q^3 + 9 q^4 + 13 q^5 + 20 q^6 + 29 q^7+\dots),\\
\chi_{\frac{1}{5}}=\,&q^{\frac{67}{420}} (1 + q + 3 q^2 + 4 q^3 + 7 q^4 + 10 q^5 + 16 q^6 + 22 q^7+\dots),\\
\chi_{\frac{12}{35}}=\,&q^{\frac{127}{420}} (1 + q + 2 q^2 + 4 q^3 + 6 q^4 + 8 q^5 + 14 q^6 + 19 q^7+\dots),\\
\chi_{\frac{3}{7}}=\,&q^{\frac{163}{420}} (1 + q + 2 q^2 + 3 q^3 + 6 q^4 + 8 q^5 + 13 q^6 + 17 q^7+\dots),\\
\chi_{\frac{22}{35}}=\,&q^{\frac{247}{420}} (1 + 2 q^2 + 2 q^3 + 4 q^4 + 5 q^5 + 9 q^6 + 11 q^7+\dots).
\ea
\ee
Clearly, the conductor $N=420$. There exist 12 classes for the Hecke operation, each containing 8 $p$ coprime to 420. We summarize in the following table the classes of quadratic residue $p^2\equiv a \textrm{ mod } 420 $.
\begin{table}[ht]
	\centering
	\begin{tabular}{c|c|c}
		\hline
		 $a$ & $+$ & $-$   \\ \hline
$1 $		 & $1, 41, 139, 181, 239, 281, 379, 419$ & $29, 71, 169, 209, 211, 251, 349, 391$ \\
$121 $		 & $ 11, 31, 109, 151, 269, 311, 389, 409$ & $ 59, 101, 179, 199, 221, 241, 319, 361 $\\
$169 $		 &  $ 13, 113, 127, 167, 253, 293, 307, 407 $ & $ 43, 83, 97, 197, 223, 323, 337, 377 $\\
$ 289$		 & $ 17, 137, 143, 157, 263, 277, 283, 403 $ & $ 53, 67, 73, 193, 227, 347, 353, 367 $\\
$361 $		 &  $ 19, 61, 79, 121, 299, 341, 359, 401 $ & $ 89, 131, 149, 191, 229, 271, 289, 331 $\\
$109 $		 & $  23, 37, 103, 163, 257, 317, 383, 397$ & $ 47, 107, 173, 187, 233, 247, 313, 373 $\\
\hline
		\end{tabular}
		\end{table}
The $\rho(\sigma_p)$ matrices can be easily derived from those of $(LY)_1$ and $(LY)_2$. We compute many Hecke images $\T_p$ of $(LY)_1\otimes(LY)_2$ theory and summarize the results for $p\le 40$ in Table \ref{tb:HeckeLY1LY2}. Interestingly, we find the $\mathsf{T}_{23}$ image gives exactly the characters of the RCFT associated to the second largest Fischer group $Fi_{23}$ given in \cite{Bae:2020pvv}. The $Fi_{23}$ is a sporadic simple group of order 4089470473293004800. The $Fi_{23}$ RCFT has central charge $c=782/35$ and weights $h_i=0,\frac{9}{7},\frac{51}{35},\frac{8}{5},\frac{13}{7},\frac{66}{35}$. The characters were computed in \cite{Bae:2020pvv} as
\be
\ba
\chi_{0} &=  q^{-\frac{391}{420}}( 1 + 30889 q^2 + 2546974 q^3 + 85135558 q^4  + \cdots ), \\
\chi_{\frac{9}{7}} &= q^{\frac{149}{420}} ( 782 + 280347 q + 16687166 q^2 + 470844155 q^3 + \cdots ), \\
\chi_{\frac{51}{35}} &= q^{\frac{221}{420}} ( 3588 + 792948 q + 39982878 q^2 + 1031142072 q^3 + \cdots ), \\
\chi_{\frac85} &= q^{\frac{281}{420}}( 5083 + 817972 q + 36460359 q^2 + 877212478 q^3 + \cdots ), \\
\chi_{\frac{13}{7}} &= q^{\frac{389}{420}}( 25806 + 2622828 q + 96358822 q^2 + 2067752532 q^3 + \cdots ), \\
\chi_{\frac{66}{35}} &= q^{\frac{401}{420}}( 60996 + 5926778 q + 213547709 q^2 + 4527955950 q^3 + \cdots ).
\ea
\ee
We checked these are exactly the same with the Hecke image $\mathsf{T}_{23}$ of $(LY)_1\otimes(LY)_2$. Besides, four classes in the MTC classification in Table 4 of \cite{2022arXiv220314829N} are related to our type $(LY)_1\otimes(LY)_2$. 
The $\T_{23}$ Hecke image we discussed belongs to the MTC class $6_{-58/35}$, while its dual w.r.t $c=24$, i.e., the RCFT associated to the conjugacy class $D_{\mathrm{3A}}$ of the Monster group belongs to the MTC class $6_{58/35}$. Besides, the Hecke image $\T_{37}$ belongs to the MTC class $6_{138/35}$, while its dual w.r.t $c=8k$ belongs to the MTC class $6_{-138/35}$.

\begin{table}[ht]
\def\arraystretch{1.1}
	\centering
	\begin{tabular}{|c|c|c|c|c|c|c|c|c|c|c|c|c|c|}
		\hline
		 $c$ & $h_i$  & $m_1$ & $l$ & remark  \\
		\hline
$\frac{34}{35}$  & $\frac{1}{7},\frac{1}{5},\frac{12}{35},\frac{3}{7},\frac{22}{35}  $ & $2 $ & $6 $  & $\T_{1}$   \\
$\frac{374}{35}$  & $ \frac{4}{7},\frac{5}{7},\frac{27}{35},\frac{32}{35},\frac{6}{5} $ & $ 220$ & $6 $  & $\T_{11},\star$   \\
$\frac{442}{35}$  & $ \frac{4}{7},\frac{3}{5},\frac{6}{7},\frac{41}{35},\frac{51}{35} $ & $ 208$ & $6 $  & $\T_{13}$   \\
$\frac{578}{35}$  & $\frac{24}{35},\frac{29}{35},\frac{9}{7},\frac{7}{5},\frac{10}{7}  $ & $ 136$ & $ 6$  & $\T_{17},\star$   \\
$\frac{646}{35}$  & $\frac{4}{5},\frac{33}{35},\frac{8}{7},\frac{53}{35},\frac{12}{7} $ & $114 $ & $ 6$  & $\T_{19},\star$   \\
$\frac{782}{35}$  & $\frac{13}{7},\frac{9}{7},\frac{8}{5},\frac{51}{35},\frac{66}{35}  $ & $ 0$ & $ 0$  & $\T_{23},Fi_{23}$   \\
$\frac{986}{35}$  & $ \frac{10}{7},\frac{9}{5},\frac{68}{35},\frac{15}{7},\frac{78}{35} $ & $ 0$ & $ 0$  & $\T_{29},\star$   \\
$\frac{1054}{35}$  & $\frac{9}{7},\frac{10}{7},\frac{52}{35},\frac{57}{35},\frac{11}{5}  $ & $ 0$ & $12 $  & $\T_{31},\star$   \\
$\frac{1258}{35}$  & $ \frac{59}{35},\frac{13}{7},\frac{79}{35},\frac{16}{7},\frac{12}{5} $ & $ 0$ & $6 $  & $\T_{37}$   \\
		\hline
	
		\end{tabular}
			\caption{Hecke images of $(LY)_1\otimes(LY)_2$ for all admissible $p\le 40$. }
			\label{tb:HeckeLY1LY2}
		\end{table}

\subsection{Type $(LY)_2^{\otimes 2}$ }
Consider the double product of $(LY)_2$. Such product theory has central charge $c=8/7$ and conformal weights with degeneracy: 
$h_i=0,(\frac{1}{7})_2,\frac{2}{7},(\frac{3}{7})_2,(\frac{4}{7})_2,\frac{6}{7}.$ It is easy to compute the six distinct characters to be
\be
\ba
\chi_0=\,&q^{-\frac{1}{21}} (1 + 2 q + 5 q^2 + 8 q^3 + 14 q^4 + 22 q^5 + 36 q^6 + 54 q^7+\dots),\\
\chi_{\frac{1}{7}}=\,&q^{\frac{2}{21}} (1 + 2 q + 4 q^2 + 7 q^3 + 12 q^4 + 19 q^5 + 31 q^6 + 46 q^7+\dots),\\
\chi_{\frac{2}{7}}=\,&q^{\frac{5}{21}} (1 + 2 q + 3 q^2 + 6 q^3 + 11 q^4 + 16 q^5 + 26 q^6 + 40 q^7+\dots),\\
\chi_{\frac{3}{7}}=\,&q^{\frac{8}{21}} (1 + q + 3 q^2 + 4 q^3 + 8 q^4 + 12 q^5 + 20 q^6 + 28 q^7+\dots),\\
\chi_{\frac{4}{7}}=\,&q^{\frac{11}{21}} (1 + q + 2 q^2 + 4 q^3 + 7 q^4 + 10 q^5 + 17 q^6 + 24 q^7+\dots),\\
\chi_{\frac{6}{7}}=\,&q^{\frac{17}{21}} (1 + 2 q^2 + 2 q^3 + 5 q^4 + 6 q^5 + 11 q^6 + 14 q^7+\dots).
\ea
\ee
Clearly, the conductor $N=21$. The $S$-matrix can be easily deduced from the one of $(LY)_2$. We find there exist three classes for the Hecke operation of $(LY)_2^{\otimes 2}$: for $p=1,8,13,20 \textrm{ mod } 21$, i.e., $p^2\equiv 1 \textrm{ mod } 21,$ $\rho(\sigma_{p})=\mathrm{Id},$ while for $p^2\equiv 4$ and $16 \textrm{ mod } 21$, 
\be
\quad
\rho(\sigma_{2,5,16,19})= \left(
\begin{array}{cccccc}
 0 & 0 & 1 & 0 & 0 & 0 \\
 0 & 0 & 0 & 0 & -1 & 0 \\
 0 & 0 & 0 & 0 & 0 & 1 \\
 0 & -1 & 0 & 0 & 0 & 0 \\
 0 & 0 & 0 & 1 & 0 & 0 \\
 1 & 0 & 0 & 0 & 0 & 0 \\
\end{array}
\right)  ,\qquad 
\rho(\sigma_{4,10,11,17})=  \left(
\begin{array}{cccccc}
 0 & 0 & 0 & 0 & 0 & 1 \\
 0 & 0 & 0 & -1 & 0 & 0 \\
 1 & 0 & 0 & 0 & 0 & 0 \\
 0 & 0 & 0 & 0 & 1 & 0 \\
 0 & -1 & 0 & 0 & 0 & 0 \\
 0 & 0 & 1 & 0 & 0 & 0 \\
\end{array}
\right).
\ee
We then compute all the Hecke images $\T_p$ for admissible $p<21$ and summarize the results in $c=24$ pairs in Table \ref{tb:HeckeM27square}. Note the sums of spin-1 currents $m_1+\tm_1$ for all pairs in Table \ref{tb:HeckeM27square} are divisible by the conductor 21.

\begin{table}[ht]
\def\arraystretch{1.1}
	\centering
	\begin{tabular}{|c|c|c|c|c|c|c|c|c|c|c|c|c|c|}
		\hline
		 $c$ & $h_i$  & $m_1$   & remark  & $\tc$ & $\tilh_i$ &  $\tm_1$ &   remark  & $\!\!m_1\!+\!\tm_1\!\!$ & Sch\\
		\hline
$\frac{8}{7}$  & $ \frac{1}{7},\frac{2}{7},\frac{3}{7},\frac{4}{7},\frac{6}{7}$ & $2 $  & $\mathsf{T}_{1}$ & ${\frac{160 }{7}}$ &  $\frac{8}{7},\frac{10}{7},\frac{11}{7},\frac{12}{7},\frac{13}{7} $ & $ 40$  & $\mathsf{T}_{20}$ & $42$ & $-$\\
$\frac{16}{7}$  & $\frac{1}{7},\frac{2}{7},\frac{4}{7},\frac{5}{7},\frac{6}{7} $ & $0 $  & $\!\mathsf{T}_{2},D_{5\rm A}\!$ & ${\frac{152 }{7}}$ &  $ \frac{8}{7},\frac{9}{7},\frac{10}{7},\frac{12}{7},\frac{13}{7}$ & $ 0$  & $\!\mathsf{T}_{19},HN\!$ & $0$ & $0$\\
$\frac{32}{7}$  & $\frac{2}{7},\frac{3}{7},\frac{4}{7},\frac{5}{7},\frac{8}{7} $ & $24 $  & $\mathsf{T}_{4}$ & ${\frac{136 }{7}}$ &  $\frac{6}{7},\frac{9}{7},\frac{10}{7},\frac{11}{7},\frac{12}{7} $ & $102 $  & $\mathsf{T}_{17}$ & $126$& $-$\\
$\frac{40}{7}$  & $\frac{2}{7},\frac{3}{7},\frac{5}{7},\frac{6}{7},\frac{8}{7} $ & $ 10$  & $\mathsf{T}_{5}$ & ${\frac{128 }{7}}$ &  $\frac{6}{7},\frac{8}{7},\frac{9}{7},\frac{11}{7},\frac{12}{7} $ & $ 32$ & $\mathsf{T}_{16}$ & $42$& $-$\\
$\frac{64}{7}$  & $\frac{3}{7},\frac{4}{7},\frac{6}{7},\frac{8}{7},\frac{9}{7} $ & $176 $  & $\mathsf{T}_{8}$ & $\frac{104}{7}$  & $\frac{5}{7},\frac{6}{7},\frac{8}{7},\frac{10}{7},\frac{11}{7} $ & $286 $  & $\mathsf{T}_{13}$ & $462$& $-$\\
$\frac{80}{7}$  & $\frac{4}{7},\frac{5}{7},\frac{6}{7},\frac{9}{7},\frac{10}{7} $ & $ 160$ & $\mathsf{T}_{10}$ & $\frac{88}{7}$  & $\frac{4}{7},\frac{5}{7},\frac{8}{7},\frac{9}{7},\frac{10}{7} $ & $ 176$  & $\mathsf{T}_{11}$ & $336$ & $-$\\
		\hline
	
		\end{tabular}
			\caption{Hecke images $\mathsf{T}_p$ of $(LY)_2^{\otimes 2}$. All images have $l=3$ MLDEs. The degeneracies are omitted.}
			\label{tb:HeckeM27square}
		\end{table}

We find the $\mathsf{T}_{19}$ Hecke image of $(LY)_2^{\otimes 2}$ describes the RCFT associated to the Harada-Norton group ${HN}$ defined in \cite{Bae:2020pvv}. The Harada-Norton group is a simple sporadic simple group of order 273030912000000. The $HN$ RCFT has central charge $c=152/7$ and weights with degeneracy
$h_i=0,(\frac{8}{7})_2,\frac{9}{7},\frac{10}{7},(\frac{12}{7})_2,(\frac{13}{7})_2.$
We compute the $\T_{19}$ images of $(LY)_2^{\otimes 2}$ as
\be
\ba
&\chi_0 = q^{-\frac{19}{21}}(1 + 18316 q^{2}+1360096 q^3+42393826 q^4 + \cdots),\\
&\chi_{\frac87} = q^{\frac{5}{21}}(133 +65968 q+4172476 q^2+119360584 q^3 + \cdots), \\
&\chi_{\frac{12}{7}} = q^{\frac{17}{21}}(8778 +1003408 q+37866696 q^2 + \cdots ), \\
&\chi_{\frac97} = q^{\frac{8}{21}}(760 +231705 q+12595936 q^2+333082540 q^3 + \cdots), \\
&\chi_{\frac{13}{7}} = q^{\frac{20}{21}}(35112 +3184818 q+108781232 q^2 + \cdots), \\
&\chi_{\frac{10}{7}} = q^{\frac{11}{21}}(3344 +680504 q+32364068 q^2+795272512 q^3 + \cdots).
\ea
\ee
These are exactly the same characters computed in (3.115) of \cite{Bae:2020pvv}. The degeneracy is also matched. Besides, we find the $\mathsf{T}_{2}$ Hecke image describes the $D_{5\mathrm{A}}$ conjugacy class of the Monster group. The $D_{5\mathrm{A}}$ RCFT has an explicit construction in (\cite{lametal}, Theorem 3.19). We checked the characters there are exactly the same with our $\T_2$ Hecke image.

\subsection{Type $(LY)_1\otimes\mathrm{Ising}$ }
Consider product theory $(LY)_1\otimes\mathrm{Ising}$ which has central charge $c=\frac25+\frac12=\frac{9}{10}$ and conformal weights $h_i=0,\frac{1}{16},\frac{1}{5},\frac{21}{80},\frac{1}{2},\frac{7}{10}$. 
The conductor $N=80$. We compute the Hecke images $\T_p$ for all admissible $p<20$ and summarize relevant information in Table~\ref{tb:HeckeLY1Ising}. The Hecke images of type $(LY)_1\otimes\mathrm{Ising}$ contain some unitary theories. For example, the Hecke image $\T_7$ belongs to MTC class $6_{-17/10}^B=2^B_{14/5}\otimes 3^B_{7/2}$ in Table III of \cite{Schoutens:2015uia} which is represented by WZW $(G_2)_1\otimes (B_3)_1$, while the Hecke image $\T_{13}$ belongs to MTC class $6_{37/10}=2^B_{-14/5}\otimes 3^B_{-3/2}$ which is represented by WZW $(F_4)_1\otimes (B_6)_1$. Note the $\T_{19}$ image is not the naive guess $(E_{7\frac12})_1\otimes (B_9)_1$. It is interesting to consider whether it can be realized as a supersymmetric theory.

\begin{table}[ht]
\def\arraystretch{1.1}
	\centering
	\begin{tabular}{|c|c|c|c|c|c|c|c|c|c|c|c|c|c|}
		\hline
		 $c$ & $h_i$  & $m_1$ & $l$ & remark  \\
		\hline
$\frac{9}{10}$  & $ \frac{1}{16},\frac{1}{5},\frac{21}{80},\frac{1}{2},\frac{7}{10} $ & $1 $ & $6 $  & $\T_{1}$   \\
$\frac{27}{10}$  & $\frac{1}{10},\frac{3}{16},\frac{3}{5},\frac{63}{80},\nb{\frac{3}{2}}  $ & $ 6$ & $0 $  & $\T_{3},\star$   \\
$\frac{63}{10}$  & $ \frac{2}{5},\frac{7}{16},\frac{1}{2},\frac{67}{80},\frac{9}{10} $ & $ 35$ & $6 $  & $\T_{7}, 6_{-17/10}^B,(G_2)_1\otimes (B_3)_1$   \\
$\frac{81}{10}$  & $\frac{3}{10},\frac{29}{80},\frac{1}{2},\frac{9}{16},\frac{4}{5}  $ & $117 $ & $12 $  & $\T_{9},\star$   \\
$\frac{99}{10}$  & $ \frac{1}{2},\frac{11}{16},\frac{7}{10},\frac{71}{80},\frac{6}{5} $ & $ 165$ & $6 $  & $\T_{11},\star$   \\
$\frac{117}{10}$  & $\frac{1}{2},\frac{3}{5},\frac{13}{16},\frac{11}{10},\frac{113}{80}  $ & $ 130$ & $ 6$  & $\T_{13},6^B_{37/10},(F_4)_1\otimes (B_6)_1$   \\
$\frac{153}{10}$  & $ \frac{9}{10},\frac{17}{16},\frac{7}{5},\frac{117}{80},\nb{\frac{3}{2}} $ & $153 $ & $ 0$  & $\T_{17},\star$   \\
$\frac{171}{10}$  & $\frac{4}{5},\frac{79}{80},\frac{19}{16},\frac{13}{10},\nb{\frac{3}{2} } $ & $ 190$ & $ 6$  & $\T_{19}$   \\
		\hline
	
		\end{tabular}
			\caption{Hecke images $\mathsf{T}_p$ of $(LY)_1\otimes\mathrm{Ising}$ for all admissible $p<20$. }
			\label{tb:HeckeLY1Ising}
		\end{table}

\subsection{Type Ising$^{\otimes 2}$ }
Consider the double product of critical Ising model $M_{4,3}$. Such theory has central charge $1$ and nine primaries with conformal weights and degeneracy: $h_i=0,(\frac{1}{16})_2,\frac{1}{8},(\frac{1}{2})_2,(\frac{9}{16})_2,1.$ This theory is also known as the $\IZ_2$ orbifold of the $c=1$ $U(1)_2$ theory, see e.g. \cite{DiFrancesco:1997nk}.
The conductor $N=48$. Let us compute the Hecke images and bilinear relations of character of Ising$^{\otimes 2}$ for full nine characters without degeneracy. The nine primaries can be marked by $(0,0)_1,(0,\frac12)_2,(0,\frac{1}{16})_3,(\frac12,0)_4,(\frac12,\frac12)_5,(\frac12,\frac{1}{16})_6,(\frac{1}{16},0)_7,(\frac{1}{16},\frac12)_8,(\frac{1}{16},\frac{1}{16})_9$. We compute all Hecke images $\T_p$ for admissible $p<48$ and list the results for $p<24$ in Table \ref{tb:HeckeIsingsquare}. The theories in the right side of Table \ref{tb:HeckeIsingsquare} may be supersymmetric theories owing to the presence of weight $3/2$ primaries.
Notably it was found in \cite{Bae:2020pvv} section 3.2.8 that the $\mathsf{T}_{23}$ Hecke image of Ising$^{\otimes 2}$ gives exactly the characters of RCFT associated to the second Conway group $Co_2$, while Ising$^{\otimes 2}$ itself is associated to the conjugacy class 2B of the Monster group. We further find the pair $(\T_5,\T_{19})$ forms a $c=24$ theory in Schellekens' list No.25, while $(\T_7,\T_{17})$ forms a $c=24$ theory in Schellekens' list No.39. For example, the $\T_{17}$ image describes a sub-theory of WZW $(D_6)_2\otimes(C_4)_1$. 

\begin{table}[ht]
\def\arraystretch{1.1}
	\centering
	\begin{tabular}{|c|c|c|c|c|c|c|c|c|c|c|c|c|c|}
		\hline
		 $c$ & $h_i$  & $\!m_1\!$   & remark  & $\tc$ & $\tilh_i$ &  $\tm_1$ &   remark  & $\!\!m_1\!+\!\tm_1\!\!\!$ \\
		\hline
$1$ & $ \frac{1}{16},\frac{1}{8},\frac{1}{2},\frac{9}{16},1$ & $ 0$ & $\!\T_{1},D_{\mathrm{2B}}\!$ & $23 $ & $1,\frac{23}{16},\nb{\frac{3}{2}},\frac{15}{8},\frac{31}{16} $  & $ 0$ & $\!\!\T_{23},2.2^{1+22}.Co_2\!\!$ & $ 0 $ \\
$5 $ & $\frac{5}{16},\frac{1}{2},\frac{5}{8},\frac{13}{16},1 $ & $ 20$ & $\!\T_{5},(C_2)_1^{\otimes 2}\!$ & $19 $ & $ 1,\frac{19}{16},\frac{11}{8},\nb{\frac{3}{2}},\frac{27}{16}$  & $76 $ & $\T_{19}$ & $ 96 $ \\
$7 $ & $\frac{7}{16},\frac{1}{2},\frac{7}{8},\frac{15}{16},1 $ & $42 $ & $\!\T_{7},(B_3)_1^{\otimes 2}\!$ & $ 17$ & $ 1,\frac{17}{16},\frac{9}{8},\nb{\frac{3}{2}},\frac{25}{16}$  & $102 $ & $\T_{17}$ & $ 144 $ \\
$ 11$ & $\frac{1}{2},\frac{11}{16},1,\frac{19}{16},\frac{11}{8} $ & $\!110\! $ & $\!\!\T_{11},(B_5)_1^{\otimes 2}\!\!$ & $ 13$ & $\frac{5}{8},\frac{13}{16},1,\frac{21}{16},\nb{\frac{3}{2}} $  & $130 $ & $\T_{13}$ & $ 240 $ \\
		\hline
	
		\end{tabular}
			\caption{Hecke images $\mathsf{T}_p$ of Ising$^{\otimes 2}$ for $p<24$. All theories have $l=3$ MLDE. All weights with denominators $16$ and $2$ have degeneracy two.}
			\label{tb:HeckeIsingsquare}
		\end{table}

We find the characters of all four pairs w.r.t $c=24$ in Table \ref{tb:HeckeIsingsquare} satisfy the following bilinear relation
\be
\sum\chi^{\mathsf{T}_{p}}\cdot M\cdot\chi^{\mathsf{T}_{24-p}}=J+m_1+\tm_1,
\ee
where all nonzero elements of $M$ are $M_{ii}=1,i=1,2,4,5,7,8,9$ and $M_{36}=M_{63}=1$. If directly taking $M$ to be an identity matrix, the bilinear relation still gives $J+\mathcal{N}$ but the $\mathcal{N}$ could be different from $m_1+\tm_1$.

\section{RCFTs with seven characters}\label{sec:7chi}
RCFTs with seven characters are not yet classified or studied from MLDEs. Nevertheless, the rank 7 MTC with $N_{k}^{ij}\le 1$ has been classified in \cite{Schoutens:2015uia}, see Table IV therein. Here we choose four interesting theories with seven characters to discuss their Hecke images and cosets which are $U(1)_6$, $M_{\rm eff}(8,3)$, $(A_1)_1^{\otimes 6}$ and $(LY)_6$. The first three types could involve many interesting fermionic or supersymmetric RCFTs.

\subsection{Type ${U(1)_{6}}$ }
Consider the compact Boson theory ${U(1)_{6}}$ which has $c= 1$ and conformal weights with degeneracy
$
h_i = 0, (\frac{1}{24})_2, (\frac{1}{6})_2, (\frac{3}{8})_2, (\frac{2}{3})_2, (\frac{25}{24})_2, \frac{3}{2}.
$
This theory has a non-anomalous $\mathbb{Z}_2$ symmetry, which can be fermionized to give a supersymmetric RCFT \cite{Bae:2020xzl}. The seven characters are given as follows,
\be
\begin{aligned}
\chi_0 &= q^{-\frac{1}{24}}(1+q+2 q^2+3 q^3+5 q^4+7 q^5+13 q^6+17 q^7 + \cdots),\\
\chi_{1/24} &= q^0(1+q+2 q^2+3 q^3+5 q^4+8 q^5+12 q^6+18 q^7 +\cdots), \\
\chi_{1/6} &= q^{\frac{1}{8}} (1+q+2 q^2+3 q^3+6 q^4+8 q^5+13 q^6+18 q^7+ \cdots),\\
\chi_{3/8} &= q^{\frac13}(1+q+2 q^2+4 q^3+6 q^4+9 q^5+14 q^6+20 q^7 + \cdots),\\
\chi_{2/3} &= q^{\frac{5}{8}} (1+q+3 q^2+4 q^3+7 q^4+10 q^5+16 q^6+22 q^7 + \cdots), \\
\chi_{25/24} &=q (1+2 q+3 q^2+5 q^3+8 q^4+12 q^5+18 q^6+26 q^7 + \cdots),\\
\chi_{3/2} &=q^{\frac{35}{24}}(2+2 q+4 q^2+6 q^3+10 q^4+14 q^5+22 q^6+30 q^7+  \cdots).
\end{aligned}
\ee
Note that $\chi_{1/24} = \chi_{25/24} + 1$, which signifies the unbroken supersymmetry. This phenomenon also happens for $M(8,3)$ and $(A_1)_1^{\otimes 6}$ theories. The conductor of ${U(1)_{6}}$ is $N = 24$. The $S$-matrix is 
\be
\rho(
S)=\frac{1}{2\sqrt{3}}\left(
\begin{array}{ccccccc}
 1 & 2 & 2 & 2 & 2 & 2 & 1 \\
 1 & \sqrt{3} & 1 & 0 & -1 & -\sqrt{3} & -1 \\
 1 & 1 & -1 & -2 & -1 & 1 & 1 \\
 1 & 0 & -2 & 0 & 2 & 0 & -1 \\
 1 & -1 & -1 & 2 & -1 & -1 & 1 \\
 1 & -\sqrt{3} & 1 & 0 & -1 & \sqrt{3} & -1 \\
 1 & -2 & 2 & -2 & 2 & -2 & 1 \\
\end{array}
\right) .
\ee

Consider the Hecke images $\T_p$ of ${U(1)_{6}}$. We find for $p=1, 5,19,23\textrm{ mod } 24$, $\rho(\sigma_{p})=\mathrm{Id},$ while for $p=7, 11,13,17\textrm{ mod } 24$, $\rho(\sigma_{p})$ has non-vanishing elements $(11),(33),(44),(55)$, $(77),(26),(62)$ as $-1$. We compute all admissible  Hecke images for $p<24$ and summarize the results in Table \ref{tb:HeckeU1R6}.
In particular, we find the $\T_{11}$ images gives exactly the characters of $(A_{11})_1$ WZW model. We remark that $(A_{11})_1$ can be fermionized into a supersymmetric RCFT \cite{Bae:2020xzl}. 
We checked the direct bilinear products of the characters of pairs $(\T_1,\T_{23})$ and $(\T_{11},\T_{13})$ give $J+72$ and $J+648$ respectively. However, the bilinear products of the quasi-characters of pairs $(\T_5,\T_{19})$ and $(\T_{7},\T_{17})$ are more complicated, and require non-integral intermediate matrices. This phenomenon can happen when two characters of the initial theory differ only by a constant. We discuss more about this in the next subsection.
\begin{table}[ht]
\def\arraystretch{1.1}
	\centering
	\begin{tabular}{|c|c|c|c|c|c|c|c|c|c|c|c|c|c|}
		\hline
		 $c$ & $h_i$  & $m_1$ & $l$ & remark  & $\tc$ & $\tilh_i$ &  $\tm_1$ & $l'$ & remark   \\
		\hline
$1$  & $\frac{1}{24},\frac{1}{6},\frac{3}{8},\frac{2}{3},\frac{25}{24},\nb{\frac{3}{2}}$ & $1$ & $0$ & $\mathsf{T}_1$ & $23$ & $\frac{23}{24},\frac{4}{3},\nb{\frac{3}{2}},\frac{13}{8},\frac{11}{6},\frac{47}{24}$ & $23$ & $6$ & $\mathsf{T}_{23}$    \\
$5$  & $\frac{5}{24},\frac{1}{3},\frac{1}{2},\frac{5}{6},\frac{7}{8},\frac{29}{24} $ & $ 25$ & $6$ &  $\mathsf{T}_{5}, \star$ & $19$ & $ \frac{19}{24},\frac{9}{8},\frac{7}{6},\nb{\frac{3}{2}},\frac{5}{3},\frac{43}{24}$ & $95$ & $6$ & $\mathsf{T}_{19}, \star$    \\
$7$  & $\frac{7}{24},\frac{1}{2},\frac{5}{8},\frac{2}{3},\frac{7}{6},\frac{31}{24} $ & $49  $ & $6$ & $\mathsf{T}_{7},\star$ & $17$ & $\frac{17}{24},\frac{5}{6},\frac{4}{3},\frac{11}{8},\nb{\frac{3}{2}},\frac{41}{24} $ & $119$ & $6$ & $\mathsf{T}_{17},\star$    \\
$11$  & $\frac{11}{24},\frac{5}{6},\frac{9}{8},\frac{4}{3},\frac{35}{24},\nb{\frac{3}{2}} $ & $ 143$ & $0$ & $\mathsf{T}_{11},(A_{11})_1$ & $13$ & $\frac{13}{24},\frac{2}{3},\frac{7}{8},\frac{7}{6},\nb{\frac{3}{2}},\frac{37}{24} $ & $169$ & $6$ & $\mathsf{T}_{13}$    \\
\hline
	
		\end{tabular}
			\caption{Hecke images $\mathsf{T}_p$ of $U(1)_{6}$ theory. All weights with denominators $24,8,6$ and $3$ have degeneracy two.} 
			\label{tb:HeckeU1R6}
		\end{table}

\subsection{Type $M_{\rm eff}(8,3)$ }
Non-unitary minimal model $M(8,3)$ has central charge $c=-\frac{21}{4}$ and conformal weights
$h_i=-\frac{1}{4},-\frac{7}{32},-\frac{3}{32},0,\frac{1}{4},\frac{25}{32},\frac{3}{2}$. With weight-$\frac32$ primary, this becomes a supersymmetric theory. The effective theory $M_{\rm eff }(8,3)$ has $c_{\rm eff}=\frac34$ and $h_i^{\rm eff}=0,\frac{1}{32},\frac{5}{32},\frac{1}{4},\frac{1}{2},\frac{33}{32},\frac{7}{4}$. The seven characters have the following Fourier expansion
\be
\ba
\chi_0=\,&q^{-\frac{1}{32}} (1 + q + 2 q^2 + 2 q^3 + 4 q^4 + 5 q^5 + 8 q^6 + 10 q^7  +\dots),\\
\chi_{\frac{1}{32}}=\,& q^0(1 + q + q^2 + 2 q^3 + 3 q^4 + 4 q^5 + 6 q^6 + 8 q^7 +\dots),\\
\chi_{\frac{5}{32}}=\,&q^{\frac{1}{8}} (1 + q + 2 q^2 + 3 q^3 + 4 q^4 + 6 q^5 + 9 q^6 + 12 q^7 +\dots),\\
\chi_{\frac{1}{4}}=\,&q^{\frac{7}{32}} (1 + q^2 + q^3 + 2 q^4 + 2 q^5 + 4 q^6 + 4 q^7 +\dots),\\
\chi_{\frac{1}{2}}=\,&q^{\frac{15}{32}} (1 + q + 2 q^2 + 3 q^3 + 5 q^4 + 6 q^5 + 9 q^6 + 12 q^7 +\dots),\\
\chi_{\frac{33}{32}}=\,& q(1 + q + 2 q^2 + 3 q^3 + 4 q^4 + 6 q^5 + 8 q^6 + 11 q^7 +\dots),\\
\chi_{\frac{7}{4}}=\,&q^{\frac{55}{32}} (1 + q + q^2 + 2 q^3 + 3 q^4 + 4 q^5 + 6 q^6 + 7 q^7+\dots).
\ea
\ee
Note $\chi_{{1}/{32}}=\chi_{{33}/{32}}+1$, which often happens for supersymmetric theory.
Clearly the conductor $N=32$. The $S$-matrix of $M_{\rm eff }(8,3)$ is 
\be
\rho(S)=\frac{1}{2} \left(
\begin{array}{ccccccc}
 \cos \left(\frac{\pi }{8}\right) & \frac{1}{\sqrt{2}} & 1 & \sin \left(\frac{\pi }{8}\right) & \cos \left(\frac{\pi }{8}\right) & \frac{1}{\sqrt{2}} & \sin \left(\frac{\pi }{8}\right) \\
 \frac{1}{\sqrt{2}} & 1 & 0 & \frac{1}{\sqrt{2}} & -\frac{1}{\sqrt{2}} & -1 & -\frac{1}{\sqrt{2}} \\
 1 & 0 & 0 & -1 & -1 & 0 & 1 \\
 \sin \left(\frac{\pi }{8}\right) & \frac{1}{\sqrt{2}} & -1 & -\cos \left(\frac{\pi }{8}\right) & \sin \left(\frac{\pi }{8}\right) & \frac{1}{\sqrt{2}} & -\cos \left(\frac{\pi }{8}\right) \\
 \cos \left(\frac{\pi }{8}\right) & -\frac{1}{\sqrt{2}} & -1 & \sin \left(\frac{\pi }{8}\right) & \cos \left(\frac{\pi }{8}\right) & -\frac{1}{\sqrt{2}} & \sin \left(\frac{\pi }{8}\right) \\
 \frac{1}{\sqrt{2}} & -1 & 0 & \frac{1}{\sqrt{2}} & -\frac{1}{\sqrt{2}} & 1 & -\frac{1}{\sqrt{2}} \\
 \sin \left(\frac{\pi }{8}\right) & -\frac{1}{\sqrt{2}} & 1 & -\cos \left(\frac{\pi }{8}\right) & \sin \left(\frac{\pi }{8}\right) & -\frac{1}{\sqrt{2}} & -\cos \left(\frac{\pi }{8}\right) \\
\end{array}
\right).
\ee

Consider the Hecke images of $M_{\rm eff }(8,3)$. There exist four classes for the Hecke operation, see Appendix D.2.7 in \cite{Wuthesis}. 
We compute the Hecke images $\T_p$ for all admissible $p<32$ and summarize the results in $c=24$ pairs in Table \ref{tb:HeckeM38}. There are eight pairs w.r.t $c=24$ which satisfy $p+p'=32$. The $\mathsf{T}_3$ and $\mathsf{T}_{13}$ images describe exactly the WZW  $(A_1)_6$ and $(C_6)_1$ theories \cite{Wuthesis}, which are both renowned supersymmetric theories. Besides, we find the pair $(\T_{13},\T_{19})$ form a $c=24$ theory in the Schellekens' list No.48, where $\T_{19}$ describe a subtheory of $(C_6)_1\otimes (B_4)_1$. Note the sums of spin-1 currents $m_1+\tm_1$ for all pairs in Table \ref{tb:HeckeM38} are divisible by the conductor 32.

\begin{table}[ht]
\def\arraystretch{1.1}
	\centering
	\begin{tabular}{|c|c|c|c|c|c|c|c|c|c|c|c|c|c|}
		\hline
		 $\!\!c\!\!$ & $h_i$  & $\!\!m_1\!\!$ & $l$ & remark  & $\tc$ & $\tilh_i$ &  $\tm_1$ & $l'$ & \!\!remark\!\! & $\!+\!$  \\
		\hline
$\frac34$  & $\frac{1}{32},\frac{5}{32},\frac{1}{4},\frac{1}{2},\frac{33}{32},\frac{7}{4}$ & $1$ & $0$ & $\mathsf{T}_1$ & ${\frac{93}{4}}$ & $\frac{31}{32},\frac{5}{4},\nb{\frac{3}{2}},\frac{7}{4},\frac{59}{32},\frac{63}{32}$ & $31$ & $0$ & $\mathsf{T}_{31}$ & $32$  \\
$\frac{9}{4}$  & $\frac{3}{32},\frac{1}{4},\frac{15}{32},\frac{3}{4},\frac{35}{32},\nb{\frac{3}{2}} $ & $ 3$ & $0$ &  $\!\!(A_1)_6,\mathsf{T}_{3}\!\!$ & ${\frac{87}{4}}$ & $ \frac{29}{32},\frac{5}{4},\nb{\frac{3}{2}},\frac{49}{32},\frac{7}{4},\frac{61}{32}$ & $29$ & $6$ & $\mathsf{T}_{29}$ & $32$  \\
$\frac{15}{4}$  & $\frac{5}{32},\frac{1}{4},\frac{1}{2},\frac{3}{4},\frac{25}{32},\frac{37}{32} $ & $10  $ & $6$ & $\mathsf{T}_{5},\star$ & ${\frac{81}{4}}$ & $\frac{27}{32},\frac{27}{32},\frac{39}{32},\frac{5}{4},\nb{\frac{3}{2}},\frac{7}{4} $ & $54$ & $\!\!12\!\!$ & $\!\!\mathsf{T}_{27},\star\!\!$ & $64$  \\
$\frac{21}{4}$  & $\frac{7}{32},\frac{1}{4},\frac{1}{2},\frac{3}{4},\frac{35}{32},\frac{39}{32} $ & $ 42$ & $6$ & $\mathsf{T}_{7},\star$ & ${\frac{75}{4}}$ & $\frac{25}{32},\frac{25}{32},\frac{29}{32},\frac{5}{4},\nb{\frac{3}{2}},\frac{7}{4} $ & $150$ & $\!\!12\!\!$ & $\!\!\mathsf{T}_{25},\star\!\!$ & $192$  \\
$\frac{27}{4}$  & $\frac{9}{32},\frac{9}{32},\frac{13}{32},\frac{1}{2},\frac{3}{4},\frac{5}{4} $ & $ 81$ & $\!\!12\!\!$ & $\mathsf{T}_{9},\star$ & ${\frac{69}{4}}$ & $\frac{23}{32},\frac{5}{4},\nb{\frac{3}{2}},\frac{51}{32},\frac{55}{32},\frac{7}{4} $ & $207$ & $0$ & $\!\!\mathsf{T}_{23},\star\!\!$ & $288$  \\
$\frac{33}{4}$  & $\frac{11}{32},\frac{1}{2},\frac{23}{32},\frac{3}{4},\frac{5}{4},\frac{43}{32} $ & $44 $ & $6$ & $\mathsf{T}_{11},\star$ & ${\frac{63}{4}}$ & $\frac{21}{32},\frac{3}{4},\frac{5}{4},\frac{41}{32},\nb{\frac{3}{2}},\frac{53}{32} $ &  $84$ & $6$ & $\!\!\mathsf{T}_{21},\star\!\!$ & $128$  \\
$\frac{39}{4}$  & $\frac{13}{32},\frac{3}{4},\frac{33}{32},\frac{5}{4},\frac{45}{32},\nb{\frac{3}{2}} $ & $ 78$ & $0$ & $\!\!(C_6)_1,\mathsf{T}_{13}\!\!$ & ${\frac{57}{4}}$ & $\frac{19}{32},\frac{3}{4},\frac{31}{32},\frac{5}{4},\nb{\frac{3}{2}},\frac{51}{32} $ & $114$ & $6$ & $\mathsf{T}_{19}$ & $192$  \\
$\frac{45}{4}$  & $\frac{15}{32},\frac{1}{2},\frac{3}{4},\frac{5}{4},\frac{43}{32},\frac{47}{32} $ & $\!\!225 \!\!$ & $6$ & $\mathsf{T}_{15}$ & ${\frac{51}{4}}$ & $\frac{17}{32},\frac{21}{32},\frac{3}{4},\frac{5}{4},\nb{\frac{3}{2}},\frac{49}{32} $ & $255$ & $6$ & $\mathsf{T}_{17}$ & $480$  \\
		\hline
	
		\end{tabular}
			\caption{Hecke images $\mathsf{T}_p$ of $M_{\rm eff}(8,3)$. Only $c=\frac{9}{4},\frac{39}{4},\frac{57}{4},\frac{87}{4}$ theories are unitary. The $+$ is short for $m_1+\tm_1$.
			}
			\label{tb:HeckeM38}
		\end{table}

We find the bilinear relations of characters w.r.t $c=24$ are complicated for most of the pairs.  
Due to the relation $\chi_{{1}/{32}}=\chi_{{33}/{32}}+1$ of $M_{\rm eff}(8,3)$, the intermediate $M$ matrices could have fractions for the $(22),(26),(62),(66)$ elements. Nevertheless the summation of the four elements is still 2. For example, for the four pairs with ordinary characters, we determine the following $M$ matrices and bilinear relations:
\be 
\sum\chi^{\mathsf{T}_{1}}\cdot M\cdot\chi^{\mathsf{T}_{31}}=J+32(1+a),\qquad M=\left(
\begin{array}{cc}
  a  & 1-a \\
  1-a  & a \\
\end{array}
\right) .
\ee
\be
\sum\chi^{\mathsf{T}_{3}}\cdot M\cdot\chi^{\mathsf{T}_{29}}=J+87a,\qquad M=\left(
\begin{array}{ccccccc}
  a-\frac{32}{87}  & \frac{30}{29}-a \\
  \frac{4}{3}-a  & a \\
\end{array}
\right) .
\ee
\be
\sum\chi^{\mathsf{T}_{13}}\cdot M\cdot\chi^{\mathsf{T}_{19}}=J+247a+160,\qquad M=\left(
\begin{array}{ccccccc}
  a-\frac{32}{247}  & \frac{20}{19} - a   \\
  \frac{14}{13}-a  & a   \\
\end{array}
\right) .
\ee
\be
\sum\chi^{\mathsf{T}_{15}}\cdot M\cdot\chi^{\mathsf{T}_{17}}=J+270a+354,\qquad M=\left(
\begin{array}{ccccccc}
  a-\frac{7}{15}  & 1-a   \\
  \frac{22}{15}-a  & a   \\
\end{array}
\right) .
\ee
Here $M$ matrices are short for just the $(22),(26),(62),(66)$ elements. All other non-vanishing elements are $(11),(33),(44),(55),(77)$ as 1.

\subsection{Type $(LY)_6$ }
Non-unitary minimal model $M(15,2)$ has central charge $c=-\frac{164}{5}$ and conformal weights
$
h_i=-\frac{7}{5},-\frac{4}{3},-\frac{6}{5},-1,-\frac{11}{15},-\frac{2}{5},0. 
$ While the effective theory
$M_{\rm eff }(15,2)$, i.e., $(LY)_6$ has $c_{\rm eff }=\frac45$ and  $h_{\rm eff }=0,\frac{1}{15},\frac{1}{5},\frac{2}{5},\frac{2}{3},1,\frac{7}{5}$. The conductor $N=30$. 
The $\rho(\sigma_p)$ matrices for various classes of the Hecke operation can be found in the Appendix D.2.4 of \cite{Wuthesis}. We compute all admissible Hecke images $\T_p$ for $p<30$ and summarize the relevant information in $c=24$ pairs in Table \ref{tb:HeckeM152}. We can see the sums of spin-1 currents $m_1+\tm_1$ for all pairs in Table \ref{tb:HeckeM152} are divisible by the conductor 30. We also study the dual theory of original $M(15,2)$ w.r.t $c=24$ and find it can be realized as Hecke image $\T_{71}(LY)_6$. On the other hand, the original $M(15,2)$ can be formally denoted as $\T_{-41}(LY)_6$.
\begin{table}[ht]
\def\arraystretch{1.1}
	\centering
	\begin{tabular}{|c|c|c|c|c|c|c|c|c|c|c|c|c|c|}
		\hline
		 $c$ & $h_i$  & $\!\!m_1\!\!$ & $l$ & \!\!remark\!\!  & $\tc$ & $\tilh_i$ &  $\!\!\tm_1\!\!$ & $l'$ & \!\!remark\!\!  & $\!\!m_1\!+\!\tm_1\!\!\!$  \\
		\hline
		$\!\!\!-\frac{164}{5} \!\!\!$ & $\!\!\!-\{\frac{7}{5},\frac{4}{3},\frac{6}{5},1,\frac{11}{15},\frac{2}{5}\}\! \!\!$ & $ 0$ & $0 $ &  $\T_{-41} $ & $\!\!\!\frac{284}{5}\!\! $ & $\!\! \frac{12}{5},\frac{41}{15},3,\frac{16}{5},\frac{10}{3},\frac{17}{5}\!\!\! $ & $ 0$ & $\!\! 12\!\!$  & $\T_{71} $ & $0 $     \\
$\frac{4}{5} $ & $\frac{1}{15},\frac{1}{5},\frac{2}{5},\frac{2}{3},1,\frac{7}{5} $ & $ 1$ & $0 $ &  $\T_{1} $ & $\frac{116}{5} $ & $1,\frac{4}{3},\frac{8}{5},\frac{8}{5},\frac{9}{5},\frac{29}{15} $ & $ 29$ & $ 6$  & $\T_{29} $ & $30 $     \\
$\frac{28}{5} $ & $\frac{2}{5},\frac{7}{15},\frac{2}{3},\frac{4}{5},\frac{4}{5},1 $ & $42 $ & $ 6$ &  $\T_{7} $ & $\frac{92}{5} $ & $1,\frac{6}{5},\frac{6}{5},\frac{4}{3},\frac{23}{15},\frac{8}{5} $ & $138 $ & $ 6$  & $\T_{23} $ & $ 180$     \\
$\frac{44}{5} $ & $ \frac{2}{5},\frac{11}{15},1,\frac{6}{5},\frac{4}{3},\frac{7}{5}$ & $66 $ & $0 $ &  $\T_{11} $ & $\frac{76}{5} $ & $\frac{2}{3},\frac{4}{5},1,\frac{19}{15},\frac{8}{5},\frac{8}{5} $ & $114 $ & $ 6$  & $\T_{19} $ & $ 180$     \\
$\frac{52}{5} $ & $\frac{3}{5},\frac{2}{3},\frac{13}{15},1,\frac{6}{5},\frac{6}{5} $ & $52 $ & $6 $ &  $\T_{13},\star $ & $\frac{68}{5} $ & $ \frac{4}{5},\frac{4}{5},1,\frac{17}{15},\frac{4}{3},\frac{7}{5}$ & $ 68$ & $6 $  & $\T_{17},\star $ & $120 $     \\
		\hline
	
		\end{tabular}
			\caption{Hecke images $\mathsf{T}_p$ of $(LY)_6$. All theories here are non-unitary.}
			\label{tb:HeckeM152}
		\end{table}

\subsection{Type $(A_1)_1^{\otimes 6}$ }
The product theory $(A_1)_1^{\otimes 6}$ has central charge $c = 6$ and weights
$h_i =0,\frac{1}{4}, \frac{1}{2}, \frac{3}{4}, 1, \frac{5}{4}, \frac{3}{2}$ and degeneracy $1,6,15,20,15,6,1$ respectively. This theory has been discussed in \cite{Harvey:2020jvu} and is related to non-linear sigma model with target space being the K3 surface. We record the first few terms of the characters,
\be
\ba
\chi_0 &= q^{-\frac{1}{4}} (1+18 q+159 q^2+942 q^3+4323 q^4+16722 q^5  +\cdots),\\
\chi_{{1}/{4}} &= 2+32 q+256 q^2+1408 q^3+6144 q^4+22976 q^5  + \cdots,\\
\chi_{1/2} &= q^{\frac{1}{4}}(4+56 q+404 q^2+2072 q^3+8648 q^4+31360 q^5   + \cdots),\\
\chi_{3/4} &= q^{\frac{1}{2}}(8+96 q+624 q^2+3008 q^3+12072 q^4+42528 q^5  + \cdots),\\
\chi_{1} &= q^{\frac{3}{4}}(16+160 q+944 q^2+4320 q^3+16720 q^4+57312 q^5  + \cdots),\\
\chi_{5/4} &= q(32+256 q+1408 q^2+6144 q^3+22976 q^4+76800 q^5+  \cdots),  \\ 
\chi_{3/2} &= q^{\frac{5}{4}}(64+384 q+2112 q^2+8576 q^3+31488 q^4+102144 q^5+  \cdots).\\   
\ea
\ee
Clearly the conductor $N=4$. Note $\chi_{{1}/{4}}=\chi_{{5}/{4}}+1$. We find for all $p=2k+1,k\in\IZ$, $\rho(\sigma_p)=\mathrm{Id}$. We summarize the relevant information of the Hecke images for $p<4$ in Table \ref{tb:Heckec66}. The $(A_1)_1^{\otimes 6}$ model form a $c=24$ pair with its $\T_3$ image, which directly has conformal weights $ 0,\frac{3}{4},\frac{3}{2},\frac{5}{4},1,\frac{7}{4},\frac{3}{2}$ and multiplicities $1,8,1216,288,48,4224,1152$. The direct bilinear product of their characters gives $ J+168$. We also compute many higher $\T_{2k+1}$ Hecke images. For example, $\T_5$ and $\T_7$ images have good characters and $l=12$ MLDEs.
\begin{table}[ht]
\def\arraystretch{1.1}
	\centering
	\begin{tabular}{|c|c|c|c|c|c|c|c|c|c|c|c|c|c|}
		\hline
		 $c$ & $h_i$  & $m_1$ & $l$ & remark  & $\tc$ & $\tilh_i$ &  $\tm_1$ & $l'$ & remark     \\
		\hline
$6$  & $ \frac{1}{4}, \frac{1}{2}, \frac{3}{4}, 1, \frac{5}{4}, \nb{\frac{3}{2}}$ & $18$ & $0$ & $\mathsf{T}_1$ & $18$ & $\frac{3}{4},1,\frac{5}{4}, \nb{\frac{3}{2}},\nb{\frac{3}{2}},\frac{7}{4}$ & $54$ & $6$ & $\mathsf{T}_{3}$   \\
\hline
	
		\end{tabular}
			\caption{Hecke images $\mathsf{T}_p$ of $(A_1)_1^{\otimes6}$. The degeneracies are omitted.}
			\label{tb:Heckec66}
		\end{table}

\section{Summary and outlook}\label{sec:outlook}
The results of this work are twofold. On the one hand, we study the Hecke images of 2d RCFTs with up to seven characters beyond \cite{Harvey:2018rdc,Harvey:2019qzs,Wuthesis} and identify a large number of new Hecke relations among 2d RCFTs. For example, we find the Hecke image interpretations for WZW models $(E_{6})_2,(E_7)_2,(E_{7\frac12})_2$ and RCFTs associated to the second largest Fisher group $Fi_{23}$ and the Harada-Norton group $HN$. These should be interesting from the view point of both number theory and representation theory. On the other hand, we give an account of the holomorphic modular bootstrap results in \cite{Kaidi:2021ent} from the viewpoint of Hecke relations and $c=8k$ coset relations. We find all theories bootstrapped in \cite{Kaidi:2021ent} can be elegantly generated by these two operations from just a handful of initial theories. This new understanding also allows us to neatly determine the degeneracies and multiplicities of all theories bootstrapped in \cite{Kaidi:2021ent}. We also introduce the concept of generalized Hecke relation which sometimes allows $\T_p$ operation when $p$ is not coprime to the conductor $N$. We have observed this type of new relations exist in many examples. However, our current definition is still \emph{ad hoc}. It would be certainly desirable to find a direct definition of generalized Hecke operation.

Furthermore, bootstrap approach is also used to study 2d fermionic RCFTs recently in \cite{Bae:2020xzl,Bae:2021mej}. It is interesting to generalize the philosophy of the current work to the fermionic cases. In fact, we can define the \emph{fermionic Hecke operator} which map the characters of a fermionic RCFT to the characters of another fermionic RCFT. As a simple example, let us consider the tri-critical Ising model $M(5,4)$ and WZW model $(E_7)_2$ discussed in Section \ref{sec:M54}, both of which are well-known to preserve $\mathcal{N} = 1$ supersymmetry, but have Ramond ground state breaking supersymmetry spontaneously \cite{Friedan:1983xq}. We have declared the bosonic characters of the two theories satisfy an $\T_{19}$ Hecke relation, which directly maps the $M(5,4)$ weights $0,\frac{3}{80},\frac{1}{10},\frac{7}{16},\frac{3}{5},{\frac{3}{2} }$ to the $(E_7)_2$ weights in the order of $ \frac{3}{2},\frac{57}{80},\frac{7}{5},\frac{21}{16},\frac{9}{10},0$. The fermionic characters of supersymmetric minimal model $SM(5,3)$ are defined from the bosonic characters of $M(5,4)$ by
\be
\ba
    f^{\rm{NS}}_{0} &= \chi_{0} + \chi_{\frac{3}{2}},\quad
    f^{\rm{NS}}_{1}  = \chi_{\frac{1}{10}} + \chi_{\frac{3}{5}},\\
    f^{\widetilde{\rm{NS}}}_{0}  &= \chi_{0}  - \chi_{\frac{3}{2}} , 
    \quad
    f^{\widetilde{\rm{NS}}}_{1}  = \chi_{\frac{1}{10}}  - \chi_{\frac{3}{5}} ,  \\
    f^{\rm{R}}_{0} & =  \sqrt{2}\chi_{\frac{3}{80}} , 
    \qquad
    f^{\rm{R}}_{1}  =  \sqrt{2}\chi_{\frac{7}{16}}.
    \ea
\ee
The WZW $(E_7)_2$ theory has fermionic characters defined by
\be
\ba
f^{(E_7)_2,\rm{NS}}_{0} &= \chi^{(E_7)_2}_0 + \chi^{(E_7)_2}_{\frac{3}{2}},\quad
    f^{(E_7)_2,\rm{NS}}_{1}  = \chi^{(E_7)_2}_{\frac{9}{10}} + \chi^{(E_7)_2}_{\frac{7}{5}},\\
    f^{(E_7)_2,\widetilde{\rm{NS}}}_{0} &= \chi^{(E_7)_2}_0 - \chi^{(E_7)_2}_{\frac{3}{2}},\quad
    f^{(E_7)_2,\widetilde{\rm{NS}}}_{1} = \chi^{(E_7)_2}_{\frac{9}{10}} - \chi^{(E_7)_2}_{\frac{7}{5}}\\
    \tilde{f}^{(E_7)_2,\rm{R}}_{0}  &=  \sqrt{2}\chi^{(E_7)_2}_{\frac{21}{16}},
    \qquad\quad
    \tilde{f}^{(E_7)_2,\rm{R}}_{1}  =  \sqrt{2}\chi^{(E_7)_2}_{\frac{57}{80}}.
    \ea
\ee
Noticing  the $\rho(\sigma_{19})=-\rho(\sigma_{11})$ matrix in Section \ref{sec:M54}, we observe that the fact $\rho(\sigma_{19})$ matrix has elements $\rho_{16}=\rho_{61}=1$ and $\rho_{35}=\rho_{53}=1$ perfectly allows an uniform transformation of NS characters from $(f^{\rm{NS}}_{0},f^{\rm{NS}}_{1})$ to $ (f^{(E_7)_2,\rm{NS}}_{0},f^{(E_7)_2,\rm{NS}}_{1})$. We call such transformation as \emph{fermionic Hecke relation} $\T^{\rm F}_{19}$. Similar relations hold for $\widetilde{\rm{NS}}$ and R characters as well. 
Although in this case, both bosonic and fermionic theories are quite clear, 
for some potential NS characters bootstrapped from $\Gamma(2)$ MLDEs, e.g., in \cite{Bae:2020xzl,Bae:2021mej}, there could be many bosonic characters which may not be directly known for all of them. Therefore, we suggest that for fermionic theories, it could be more convenient to discuss fermionic Hecke relations directly.
We plan to systematically study the fermionic Hecke relations among fermionic/supersymmetric RCFTs in the future including their connection with the classification of $2+1$ dimensional fermionic topological orders \cite{Lan:2015sgw}.

At the physics level, it would be very interesting to understand if there is a physical meaning underlying the Hecke operator, and a possible answer may lead to an explanation of this huge zoo of Hecke relations. For example, it is well-known that 2d WZW RCFTs can be realized as the boundary theories of 3d Chern-Simons theory. It is intriguing to consider what the Hecke relations imply in 3d. Meanwhile, we find a lot of generalized coset relations with respect to single and modular invariant characters. For many of them it still remains to be understood whether those characters can be uplifted to holomorphic CFTs in a certain sense.

\section*{Acknowledgements}
We would like to thank Sungjay Lee, Ying-Hsuan Lin, Du Pei, Haowu Wang and Yuxiao Wu for useful discussions. We also thank Ying-Hsuan Lin and Yuxiao Wu for comments on the draft. ZH would like to thank LPENS, Paris and LPTHE, Paris for hospitality during the final stage of this project. Some results of this work have been presented by KS at Jeju workshop ``Advances in Theoretical Physics'' in January, 2022. ZD, KL and KS are supported by KIAS Grant PG076902, PG006904 and QP081001 respectively. KL is also supported in part by the National Research Foundation of Korea (NRF) Grant funded by the Korea government (MSIT) (No.2017R1D1A1B06034369).

\end{document}